\documentclass{IEEE-Con-Sys-mag}

\usepackage{cite}
\usepackage{amsmath,amssymb,amsfonts}
\usepackage{graphicx}
\usepackage{textcomp}
\usepackage{hyperref}
\usepackage{graphicx,color}
\usepackage{mathrsfs}
\usepackage{subfigure}
\usepackage{url}
\usepackage{booktabs}
\usepackage{array}
\usepackage[table]{xcolor}
\usepackage{mathtools} 		
\usepackage{upgreek}
\usepackage{dsfont}
\usepackage[most]{tcolorbox}
\usepackage{etoolbox}
\usepackage{tikz}  
\usepackage{wrapfig}

\usetikzlibrary{arrows.meta,positioning,fit,backgrounds}
\usetikzlibrary{calc}

\newcommand{\cL}{{\cal{L}}}
\newcommand{\cN}{{\cal N}}

\newcommand{\cG}{{\cal G}}

\newcommand{\cR}{{\cal R}}
\newcommand{\cT}{{\cal T}}

\newcommand{\cE}{{\cal E}}

\providecommand{\cGG}{\mathcal{GG}}
\providecommand{\cGL}{\mathcal{GL}}
\providecommand{\cLG}{\mathcal{LG}}
\providecommand{\cLL}{\mathcal{LL}}

\newcommand{\col}{\operatorname{col}}
\newcommand{\diag}{\operatorname{diag}}
\newcommand{\sat}{\operatorname{sat}}

\newcommand{\proj}{\operatorname{proj}}

\allowdisplaybreaks[3] 

\definecolor{noteblue}{RGB}{55,112,170}
\newtcolorbox{notebox}[1][NOTE]{
    enhanced,
    breakable,
    colback=white,
    frame hidden,
    boxrule=0pt,
    borderline north={1pt}{0pt}{noteblue},
    borderline south={1pt}{0pt}{noteblue},
    sharp corners,
    left=1mm,
    right=0mm,
    top=1.0mm,
    bottom=.5mm,
    before skip=9pt,
    after skip=8pt,
    title={#1},
    fonttitle=\sffamily\bfseries\footnotesize,
    coltitle=noteblue,
    attach boxed title to top left={
        xshift=0mm,
        yshift=-2mm
    },
    boxed title style={
        colback=white,
        colframe=white,
        boxrule=0pt,
        sharp corners,
        left=0mm,
        right=0mm,
        top=0.6mm,
        bottom=0.6mm
    }
}

\definecolor{notegreen}{RGB}{72,151,129}
\newtcolorbox{derivationbox}[1][]{
    enhanced,
    breakable,
    colback=white,
    frame hidden,
    boxrule=0pt,
    borderline north={1pt}{0pt}{notegreen},
    borderline south={1pt}{0pt}{notegreen},
    sharp corners,
    left=1mm,
    right=1mm,
    top=.5mm,
    bottom=.5mm,
    before skip=9pt,
    after skip=8pt,
    title={#1},
    fonttitle=\sffamily\bfseries\footnotesize,
    coltitle=notegreen,
    attach boxed title to top left={
        xshift=0mm,
        yshift=-2mm
    },
    boxed title style={
        colback=white,
        colframe=white,
        boxrule=0pt,
        sharp corners,
        left=1.2mm,
        right=1.2mm,
        top=0.6mm,
        bottom=0.6mm
    }
}

\definecolor{roadmapblue}{HTML}{176B87}
\definecolor{roadmaplight}{HTML}{3770AA}
\definecolor{roadmapgray}{HTML}{F2F2F2}
\definecolor{roadmaporange}{HTML}{CD8430}
\definecolor{roadmapgreen}{RGB}{232,243,233}
\definecolor{figurebg}{RGB}{221,229,242}

\newcommand{\blue}[1]{{\color{blue} #1}}

\newcommand{\load}{\mathrm{load}}
\newcommand{\inj}{\mathrm{inj}}
\newcommand{\net}{\mathrm{net}}

\newcommand{\real}{\mathbb{R}}
\newcommand{\complex}{\mathbb{C}}
\DeclareMathOperator*{\minimize}{minimize} 	
\DeclareMathOperator{\subto}{subject~to}

\jvol{}
\jnum{}
\paper{}
\jmonth{}
\jname{IEEE CONTROL SYSTEMS}
\pubyear{}

\begin{document}

\title{Multi-time-scale Control of Power Systems\stitle{From Classical Automatic Generation Control to Feedback Optimization}}

\author{EMILIANO DALL'ANESE, JOHN W. SIMPSON-PORCO, FLORIAN D\"{O}RFLER, SAVERIO BOLOGNANI, \\ AND SAIRAJ DHOPLE}

\maketitle

\begin{summary}
\summaryinitial{T}{he} power grid has been dominated by large synchronous generators and clearly separated control layers. Today,  power-electronics energy conversion interfaces are rapidly being deployed calling for another look at system-level control hierarchies. This article examines how multi-time-scale modeling, feedback control, and feedback optimization can help reinterpret the traditional organically grown hierarchy of primary control, automatic generation control, and economic dispatch, and enable new approaches for control of contemporary power systems. Beginning with an overview of power system models and conventional operational practices, the article develops connections to modern control architectures that can coordinate heterogeneous resources, account for operational constraints, and exploit faster actuation without overlooking first-principles physics. 
\end{summary}

\vspace{.5cm}

\begin{figure*}[t]
    \centering
    \hspace{-0.06\textwidth}
    \subfigure[Temporal scales of control.%
        \label{fig:temporal_control_scales}]{%
        \begin{minipage}[t]{0.45\textwidth}
            \centering
            \resizebox{!}{3.0cm}{%
                \includegraphics[width=265.48073pt,height=94.49867pt]{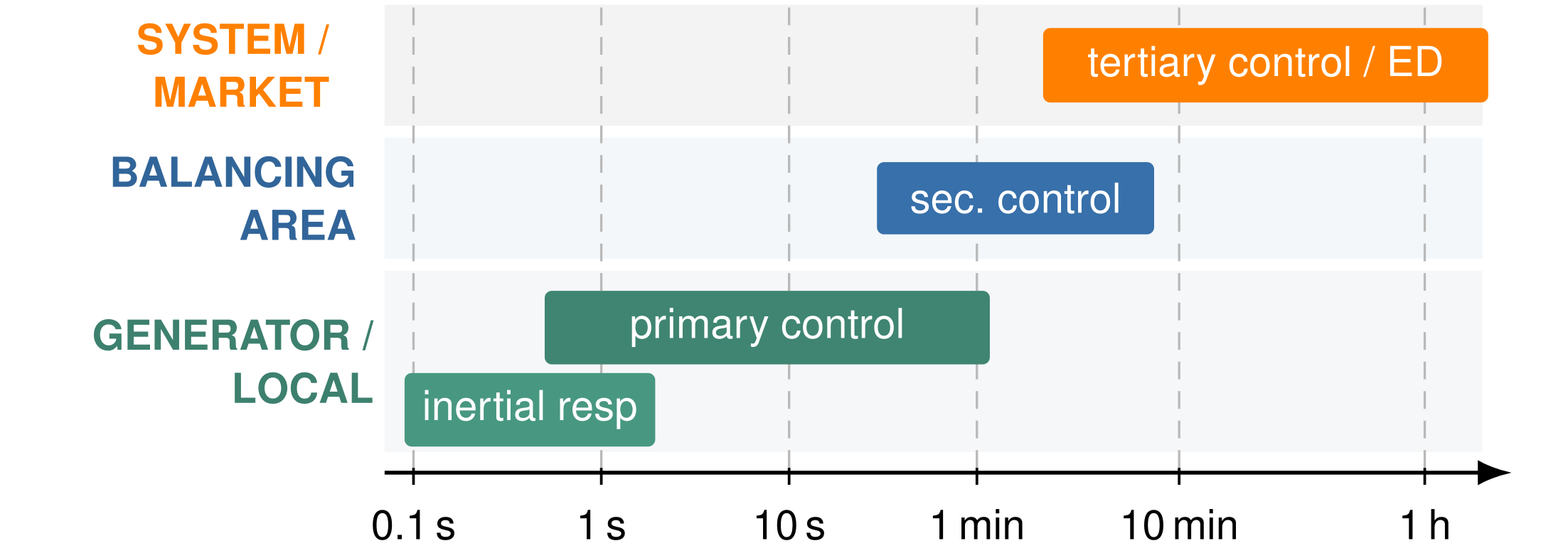}%
            }%
            \vspace{2mm}
        \end{minipage}%
    }
    \hspace{0.06\textwidth}
    \subfigure[Spatial organization into balancing areas.%
        \label{fig:balancing_areas}]{%
        \begin{minipage}[t]{0.45\textwidth}
            \centering
            \includegraphics[height=3.0cm]{%
                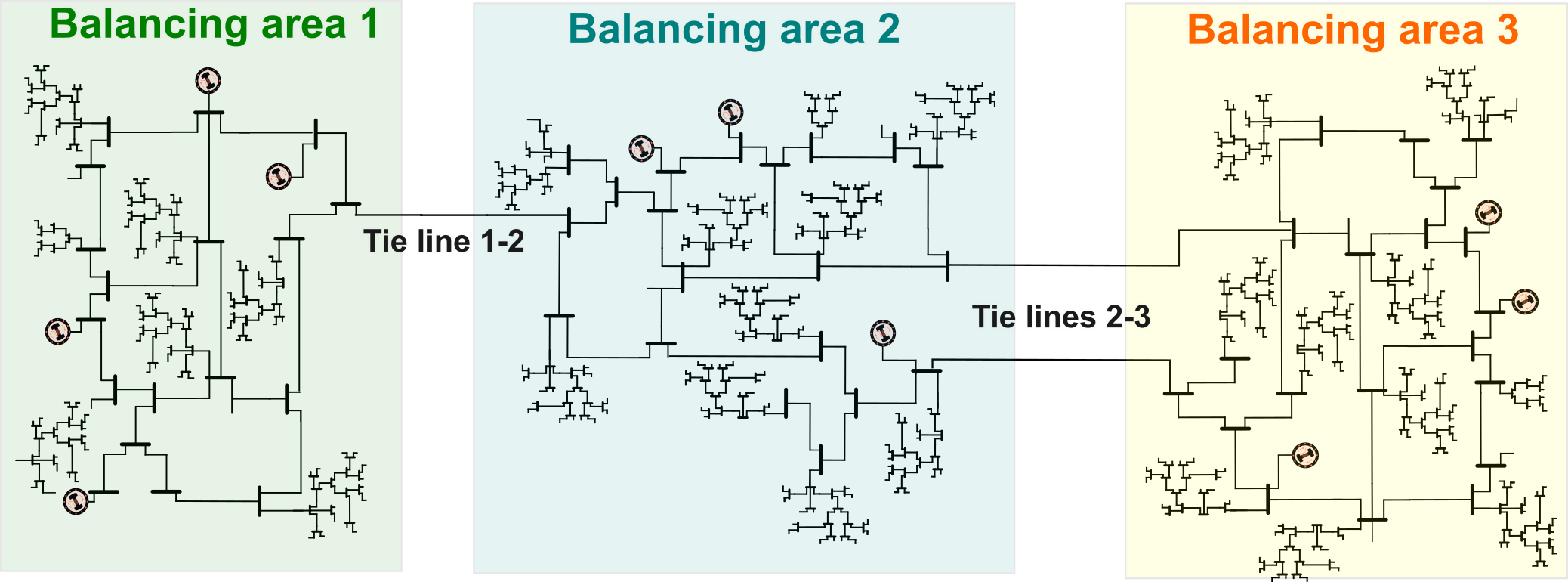}%
            \vspace{2mm}
        \end{minipage}%
    }

    \caption{Spatial and temporal organization of bulk power system operation:
    (a) Frequency-control and operational mechanisms act over progressively
    broader spatial regions and slower time scales. The highlighted time scales
    are representative and depend on the resources, implementation, and
    operating conditions. (b) An interconnected transmission system is partitioned into balancing areas linked by tie lines.}
    \label{fig:spatiotemporal_organization}
\end{figure*}

\chapterinitial{P}ower system operation and control has traditionally been organized across distinct spatial and temporal scales. From a spatial perspective, a large interconnected power system is partitioned into \emph{balancing areas} (Fig.~\ref{fig:balancing_areas}), which are electrically coupled through \emph{tie lines}. Within each balancing area, a \emph{balancing authority}, such as an independent system operator (ISO), regional transmission organization (RTO), or utility, is responsible for balancing generation and demand and maintaining operational constraints on key variables such as voltage and line power flows. Each balancing authority is also responsible for maintaining its net tie-line interchange near a scheduled value. Inter-area exchanges are established through interchange agreements, bilateral transactions, and, where wholesale markets exist, day-ahead and real-time market clearing.

\begin{figure}[h!]
    \centering
{\includegraphics[width=1.0\linewidth]
        {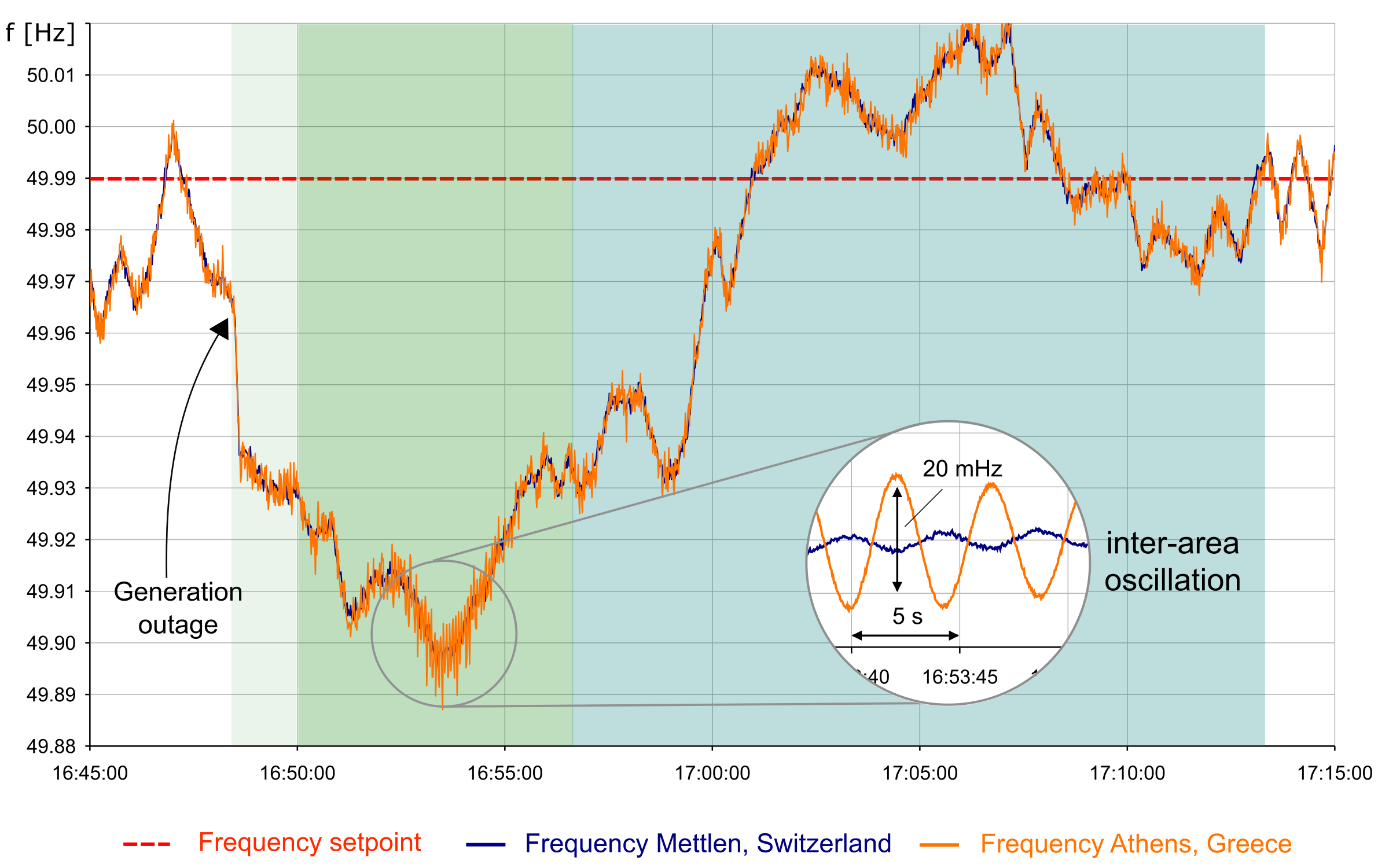}}
    \caption{The figure is based on data and an original graphic by Walter Sattinger from Swissgrid AG. Presented publicly at ETH Zürich Seminar “Stromdrehscheibe Schweiz – Netzbetrieb im Zentrum Europas”, 2007~\cite{sattinger2007stromdrehscheibe}. The time scales of the different responses after the outage are highlighted in shades of green. From left to right: inertial response, primary control (aptly also called frequency containment), and secondary frequency regulation to the setpoint. The inset highlights two distinct out-of-phase frequency measurements from Switzerland and Greece that showcase the (in this case excited) continental East-West inter-area oscillation mode. As a little known curiosity: the frequency setpoint is at 49.99 Hz (thus below the usual 50Hz) so that the myriads of  clocks synchronized to the AC waveform can compensate for an accumulated drift.}
    \label{fig:F_frequency_inter_area_oscillation}
\end{figure}

From a temporal perspective, power imbalances are managed through a hierarchy
of physical and control mechanisms (Fig.~\ref{fig:temporal_control_scales}). Following a sudden increase in demand or a loss of generation, the
stored kinetic energy in each synchronous generator provides an immediate inertial
response, causing a deceleration in the generators' rotation, and the network frequency begins to decline. Next, \emph{primary frequency control} acts over the coming seconds to arrest the
frequency excursion and stabilize the frequency at an off-nominal value. This is traditionally implemented through governors that regulate the flow of energy to the turbine driving the synchronous generator. \emph{Secondary
frequency control}, also known as automatic generation control
(AGC), then adjusts the generator power set-points over tens of seconds to minutes to restore nominal frequency and the
scheduled net interchanges. At still slower operational
intervals, (re)dispatch processes based on economic dispatch (ED)
and optimal power flow (OPF) update resource schedules, replenish regulation
reserves, and coordinate economic objectives. This
time-scale separation is the foundation of bulk
power system operation: it allows stabilization, frequency regulation, and
economic coordination to be addressed separately. From a control-theoretic standpoint, this layered structure can be formalized and analyzed using two-time-scale models and singular-perturbation methods
\cite{kokotovic1976singular,peponides1982singular,sauer2011time}. Figure~\ref{fig:F_frequency_inter_area_oscillation} shows an example of frequency evolution after a generation outage in the European transmission system (that occurred on December 8th, 2004); the figure exemplifies the actions of inertial response, primary control, and secondary control. The figure also highlights inter-area oscillations. For more information on this legacy layered architecture,  the reader is referred to the textbooks~\cite{kundur1994power,machowski1997power,sauer2017power,wood2013power} and to the \blue{sidebar ``History of Bulk Power System Control and Operations.''}

\begin{figure*}[t!]
    \centering
{\includegraphics[width=.95\linewidth]
        {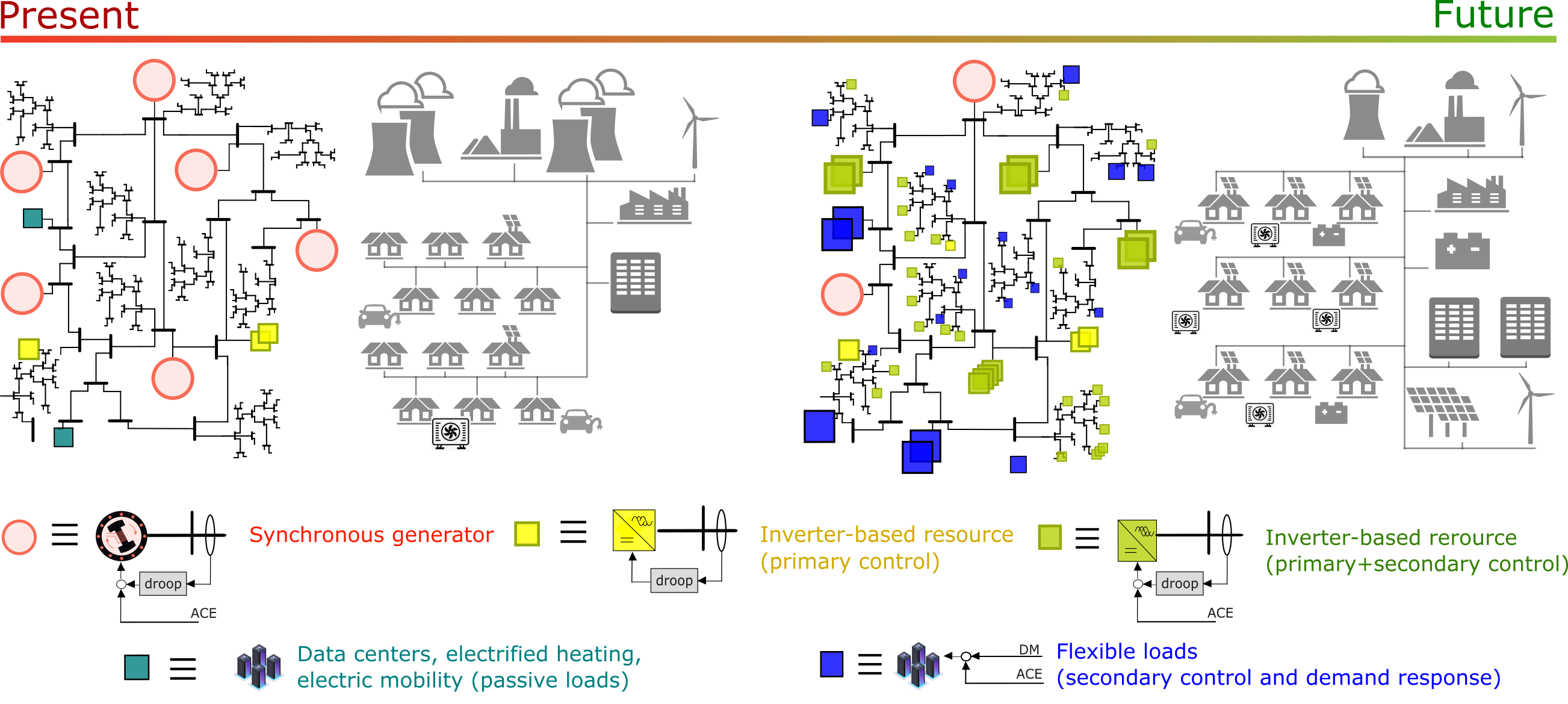}}
    \caption{Modern grids increasingly integrate inverter-based resources,    growing classes of demand, renewables, energy storage, electrified transportation, and data centers. The article highlights research questions related to integrating these resources within the control layers of
   Fig.~\ref{fig:spatiotemporal_organization} and within plausible new control
    architectures.}
    \label{fig:F_future}
\end{figure*}

\begin{sidebar}{History of Bulk Power System Control and Operations}\label{Side:History}

\setcounter{sfigure}{0}
\renewcommand{\thesfigure}{S\arabic{sfigure}}
\sdbarinitial{S}tories surrounding the genesis of electrification as they relate to the battle between Nikola Tesla's alternating- and Thomas Edison's direct-current based designs are well known. The history of the development of architectural layers examined in this article, i.e., primary, secondary, and tertiary control, does not enjoy the same level of familiarity and renown. In fact, the developments here are equally engaging and they feature equally prominent scientists and engineers. They are also dotted with victories attributable to innovations crossing disciplinary technical boundaries and fraught with several missteps with grave consequences given the scale and importance of electric infrastructure. To develop an appreciation for the innovations and interventions we outline in this article, it is insightful to ponder on how the technologies, timescales, and architectures of the contemporary grid as illustrated in Fig.~\ref{fig:spatiotemporal_organization} came to be. A necessarily abridged and incomplete history follows.

Primary control is tied to the action of governors that regulate frequency as (the inverse) function of active power within synchronous generators; a control action termed droop control. The enabling and underlying technology dates back to fly-ball governors invented by Christiaan Huygens in the 17th century and adopted for steam engines my James Watt in the 18th century. Droop control enables synchronization in interconnected energy-conversion systems. Early adopters and operators appear to have been curious about \textit{why it all worked}~\cite{Cohn-book-2017}; rigorous analytical justifications for synchronization came much later and are tied to inferences gleaned from significant research in the celebrated Kuramoto model~\cite{Dorfler-Synch-2010}. Grid-forming inverters adopt droop control at their core. Seminal research in this space appears only in the late 20th century with foundational work attributed to power-electronics experts,  Mukul Chandorkar, Deepak Divan, and Rambabu Adapa~\cite{Chandorkar-1993}. Early AC systems engaged parallel connected generators to serve load, but they required significant manual intervention and operator expertise to maintain stable operation. Large-scale interconnected grids were enabled by the development of automatic generation control in the early-to-mid 20th century by Nathan Cohn and Robert Brandt~\cite{Cohn-1966}. This distributed secondary-control scheme is sandwiched between real-time (primary) control and slower time-scale (tertiary) optimization. It appears to have organically emerged out of the need to balance economic optimality with steady-state performance and the concomitant desire to expand interconnections. Developments in tertiary control are not covered in this account since this layer is not examined in depth in our effort. Algorithmic innovations in this layer are tied to companion developments in control theory, linear algebra, and optimization methods, see, e.g.,~\cite{cohn2026calculating}.

\sdbarfig{%
  \resizebox{.95\linewidth}{!}{%
    \includegraphics[width=233.27957pt,height=304.75008pt]{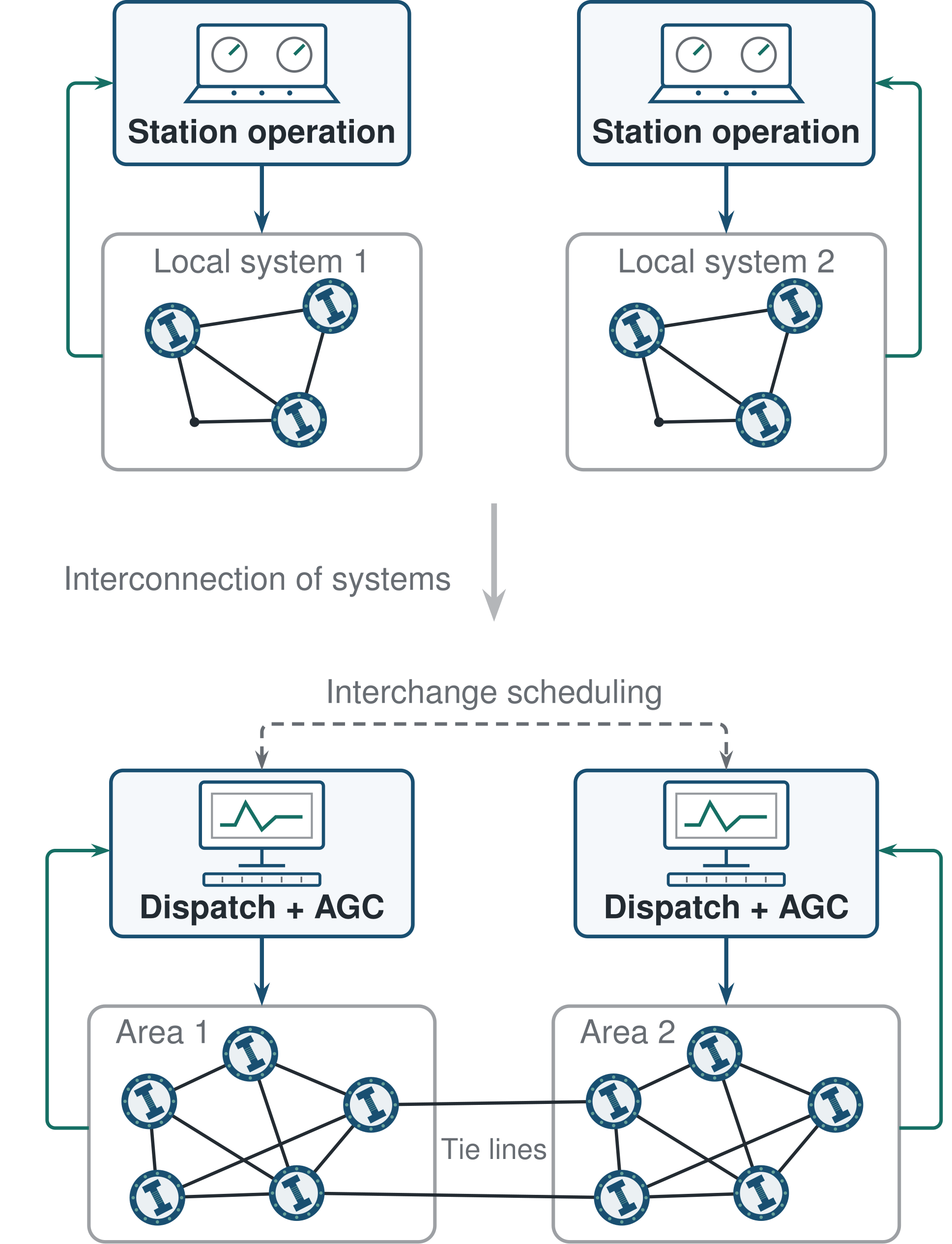}%
  }%
}{%
  Conceptual evolution from independent local systems (top)
to interconnected control areas (bottom). \label{fig:F_history_power}%
}
\end{sidebar}

The composition of present-day grids around the world is rapidly changing, (see Fig.~\ref{fig:F_future}), issuing substantial new operational challenges. In particular, operators are witnessing an unprecedented increase in the integration of inverter-based resources (IBRs) interfacing energy storage systems, photovoltaics, wind, electrified transportation fleets, and data center loads. While internal models and controllers for synchronous generators are based on Newtonian mechanics and established schemes, inverters are programmable energy-conversion interfaces offering unprecedented flexibility in operation. Two categories of control schemes for IBRs have emerged to be commonplace: grid-following (GFL) and grid-forming (GFM)~\cite{dorfler2023control}. The distinction (which the article will dwell on in detail) relates to whether the internal controls autonomously synthesize a terminal voltage much like synchronous generators (i.e., GFM) or rely on a network-induced voltage for operation (i.e., GFL). Modeling, analysis, and control of continental-scale systems with a heterogeneous mix of resources that are characterized by significantly varied dynamic characteristics is a daunting task. As a concrete example, the fast response speeds of IBRs and oftentimes unknown operational behavior has meant that they are not integrated into higher (i.e., secondary and tertiary) layer controls. At the same time, growing classes of demand, including data centers, heat pumps, and electric mobility, are being connected across transmission and distribution networks. For example, data centers can constitute large, concentrated loads whose consumption may exhibit rapid and uncertain changes associated with computational activity; this may drive the network towards undesirable/infeasible operating configurations or dynamic instability. From both the generation and demand side, present-day changes increase the variability of net demand and redistribute (increasingly controllable) injections and consumption across the network.

Reliable operation under this evolving resource mix requires
controls which take advantage of the capabilities of new resources while
respecting their limits and operational requirements. For example, GFL and GFM inverters respond rapidly to disturbances and participate in primary/secondary frequency control and economic dispatch. Data centers and other
flexible loads can likewise offer part of their consumption as a flexible
resource, subject to workload and quality-of-service requirements~\cite{wierman14ppportunities,colangelo2026ai,acun2026investigating,zhang2024data}. Integrating these resources into operational hierarchies must be accomplished systematically to ensure small-signal stability, large-disturbance stability, and satisfaction of operational constraints under uncertainty. 

These observations lead to two broad and complementary research directions: 
\begin{itemize}
    \item The first extends the existing layered operational paradigm by incorporating new resources and control methods within the  primary, secondary, and tertiary hierarchies.
    \item The second considers new control methods that reassign selected functions across layers and may compress conventional time-scale separations in search of improved dynamic performance.
\end{itemize}
The article examines both directions. It is important to stress that these are not mutually exclusive; different resources, services, and regions may adopt them to different degrees. However, both directions raise challenging control engineering questions centered around model approximations, network stability and  robustness, transient performance, constraint satisfaction, and distributed control implementations.

\section{Article Perspective and Scope}

This article offers a tutorial and perspective on multi-time-scale modeling and
control for frequency regulation in power systems with heterogeneous
resources. It develops a common model-based framework for
examining both incremental extensions of current practice and more substantial
changes to control architectures. A central theme is that architectural and dynamic
modeling choices cannot be separated from one another: a controller can safely exploit faster
actuation only when the design model acknowledges the physical dynamics, network
coupling, and constraints that are relevant to operation.

\begin{figure}[t]
    \centering
    \includegraphics[width=221.822998bp,height=431.527008bp]{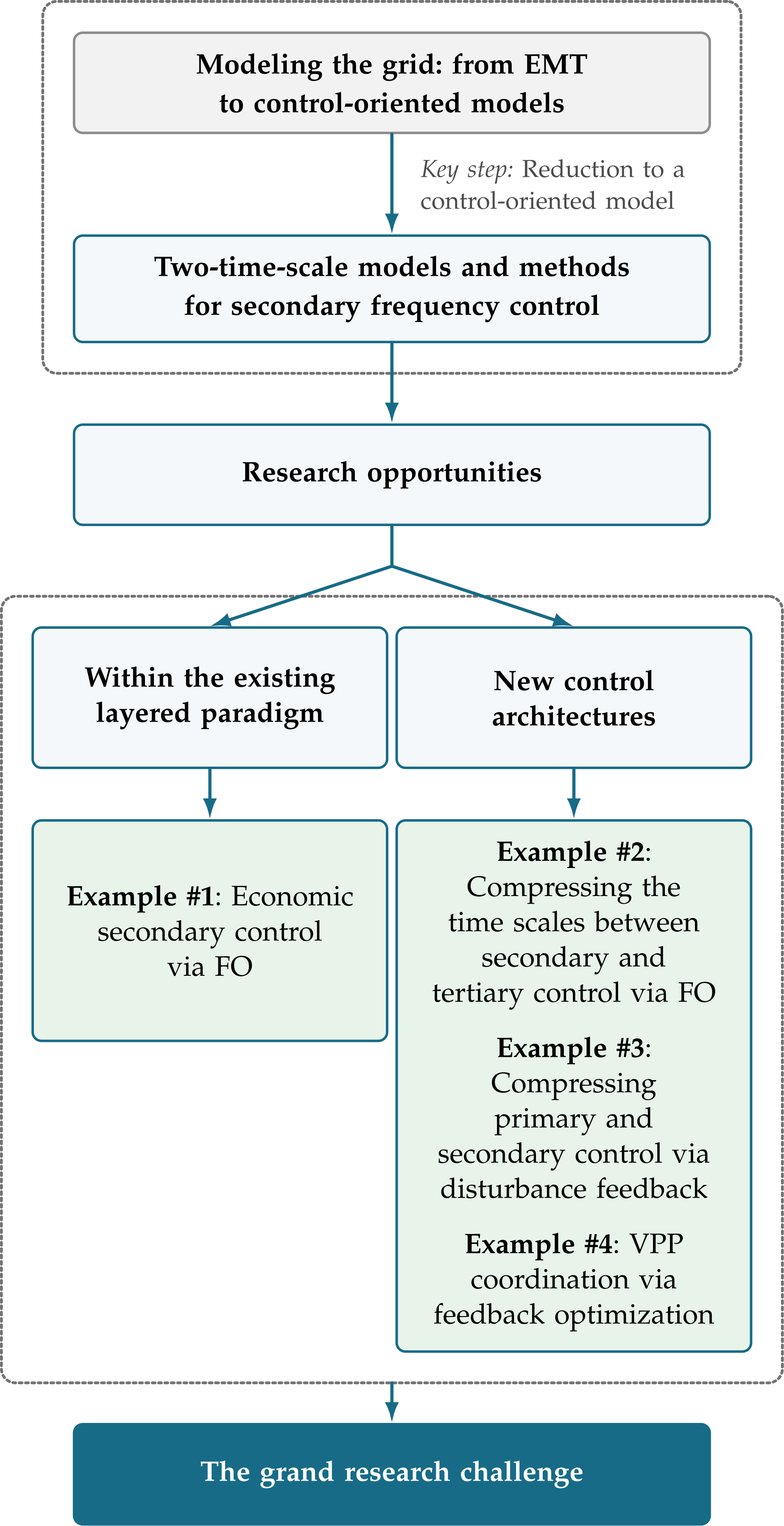}
    \caption{Organization of the article from the traditional power-system modeling
    hierarchy, to research opportunities, to illustrative new
    control methods and architectures.}
    \label{fig:article_roadmap}
\end{figure}

The organization of the article is summarized in Fig.~\ref{fig:article_roadmap}. The section ``Modeling the Grid: From EMT to Control-Oriented Models'' first describes a modeling hierarchy ranging from high-fidelity electromagnetic-transient (EMT) models down to a simplified linear ODE describing the network dynamics with primary control. The EMT model captures all network dynamics, including fast transients, while the simplified model retains only the aspects needed to describe the dynamics of primary frequency control. The section ``Two-time-scale Models and Methods for Secondary Frequency Control'' then introduces AGC, and uses singular-perturbation arguments to make explicit the assumptions under which AGC dynamics can be viewed independently of detailed primary-control dynamics.  

The sections ``Research Opportunities Within the Existing Layered Architecture'' and ``Research Opportunities Beyond the Existing Layered Architecture'' frame the two broad and complementary directions previously mentioned: extending the existing layered paradigm, and developing new architectures that reorganize selected control tasks. Their topics include AGC participation by IBRs and data centers, systematic design of primary frequency control, plant-level modeling and resource limits, transmission-distribution coordination, and compression of  time scales. A discussion on potential reverse engineering approaches, where feedback laws can sometimes be fruitfully reinterpreted as solving an optimization problem is provided; this is followed by a discussion on how ``feedback optimization'' methods~\cite{Jokic2009controller,brunner2012feedback,colombino2019online,hauswirth2020timescale,lawrence2021linear,tang2017real,bianchin2021time,hauswirth2024optimization} can provide principled forward engineering ways to design new architectures. 

The next section examines four concrete examples that arise from the research opportunities. The first replaces the conventional AGC allocation rule with feedback optimization while maintaining the separation between secondary and tertiary operation. The second uses feedback optimization to coordinate secondary regulation and economic redispatch. The third uses disturbance feedback to bring selected secondary actions onto the time scale of the system dynamics with primary control. The fourth one described how aggregations of resources forming virtual power plants can be coordinated. Remarks with a grand research challenge conclude the article. 

Power system modeling and control is a broad topic, and this article is, by no means, exhaustive and self contained. The technical development in this article focuses on transmission systems, active power, and frequency dynamics. Voltage and reactive power dynamics are, of course, considered in the higher-fidelity modeling hierarchy, but voltage-control architectures are not developed in the reduced models presented in the article. Detailed converter switching, protection, distribution-system operation, market design, and the procurement and compensation of flexibility are also outside the scope, although links to open research opportunities are provided. Distribution-connected resources are represented through aggregate injections or loads at transmission buses. 

\emph{\textbf{Reading notes}.} Two types of highlighted, self-contained notes support readers seeking different levels of technical detail. \textcolor{noteblue}{\textbf{Modeling notes}} record assumptions, limitations, and technical details, whereas \textcolor{notegreen}{\textbf{Derivations}} provide intermediate mathematical steps. Both can be skipped without breaking the main narrative. Sidebars provide longer, self-contained tutorials or technical developments and are referenced where they support the main text.

\emph{\textbf{Notation}}.
The sets of real and complex numbers are denoted by
$\mathbb R$ and $\mathbb C$, respectively;
$\mathbb R_+^n:=\{v\in\mathbb R^n:v_i\geq0,\ i=1,\ldots,n\}$,
and $\mathbb N:=\{0,1,\ldots\}$.
The superscript $^\top$ denotes transpose.
For column vectors $x_1,\ldots,x_m$,
$\col(x_1,\ldots,x_m):=[x_1^\top\ \cdots\ x_m^\top]^\top$
denotes vertical stacking. Indexed stacking follows the stated index ordering.
The operators $\diag$ and $\operatorname{blkdiag}$ construct
diagonal and block-diagonal matrices, respectively, and
$\otimes$ denotes the Kronecker product.
The symbols $I_n$, $\mathbf 1_n$, and $e_i$ denote the
$n$-dimensional identity matrix, the vector of ones, and the
$i$th standard basis vector; $\|\cdot\|$ denotes the Euclidean norm.
For real symmetric matrices, $A\succ B$ and $A\succeq B$
mean that $A-B$ is positive definite and positive semidefinite,
respectively. Vector inequalities are  component-wise.
For a complex scalar $v$, $v^*$ denotes its conjugate,
$\Re(v)$ and $\Im(v)$ denote its real and imaginary parts.

For a differentiable function $f$,
$\nabla f(x)$ denotes its column gradient.
For a differentiable map $g:\mathbb R^n\to\mathbb R^m$,
$\partial g/\partial x$ denotes its Jacobian, with
$(i,j)$th entry $\partial g_i/\partial x_j$.
For a nonempty closed convex set $\mathcal S\subseteq\mathbb R^n$,
the Euclidean projection is
$\proj_{\mathcal S}(v):=
\arg\min_{y\in\mathcal S}\frac12\|y-v\|^2$.
For $x\in\mathcal S$, its normal cone is
$N_{\mathcal S}(x):=
\{v\in\mathbb R^n:v^\top(y-x)\leq0
\text{ for all }y\in\mathcal S\}$.
For $a\leq b$, scalar saturation is defined by
$\sat_{[a,b]}(v):=\min\{b,\max\{a,v\}\}$;
saturation onto a box is applied componentwise.

\section{Modeling the Grid: From EMT to Control-Oriented Models} \label{sec:models}

Dynamical models used for analysis and control design must balance physical fidelity, analytical tractability, and operational relevance. At one extreme, EMT models capture fast switching dynamics, inverter controls, and detailed network effects, including line dynamics~\cite{lara2023revisiting,xiong2024open,wang2025co,filizadeh2025electromagnetic}. However, they are typically too complex for network-level stability analysis, and unnecessarily detailed for primary- and secondary-control timescales~\cite{dorfler2023control}. At the other extreme, simplified grid models (e.g., the ubiquitous \textit{swing equations}) are amenable to theoretical analysis, allowing one to  construct Lyapunov functions, design controllers, and draw detailed conclusions about stability and performance as a function of device and network parameters. However, such models over-approximate or outright omit essential phenomena on time scales of interest.

The tension between model fidelity and tractability is especially relevant for the multi-time-scale analysis and control problems studied in this article. The analysis and design of primary and secondary frequency controllers rely on the availability of multi-time-scale grid models, with a partition of states into ``fast'' and ``slow'' - as in classical singular-perturbation approaches~\cite{subotic2020lyapunov,chow1990singular,peponides1982singular}- as well as approximations of electrical models~\cite{ajala2022model,singh2022towards}. Electromagnetic transients, inverter inner controls, synchronization loops, electromechanical frequency dynamics, turbine-governor dynamics, primary control, AGC, economic dispatch, and online optimization all evolve on different, and sometimes overlapping, time scales in modern power grids~\cite{taylor2016power,subotic2020lyapunov}. Therefore, a  \emph{modeling hierarchy} is required before any control design and/or analysis can be attempted. Which dynamic effects should be considered depends on the phenomenon under study and on how stability properties propagate through the modeling hierarchy. As concrete examples:
\begin{itemize}
    \item a study of inverter-current limiting or synchronization may need to keep averaged converter dynamics, while omitting switching dynamics, and treating primary control action as constant.
    \item a study of AGC may need  modeling of the primary frequency control dynamics, while  omitting faster converter modes and network transients, and holding economic dispatch routines constant.
\end{itemize}

\begin{pullquote}
\emph{Design models} capture the dynamics essential for control synthesis, and \emph{validation models} are used to assess the controllers in simulation.
\end{pullquote}

A useful workflow involves separating \emph{design models} from \emph{validation models}. Design models are necessarily lower order; they capture dominant electromechanical and power-flow dynamics, and can be used to develop intuition, facilitate stability analysis, and guide control design. Validation of stability criteria or control designs can then be performed with higher-fidelity EMT simulations that include detailed device, control, and network dynamics.  During validation, one may find that certain omitted dynamics were more important than originally thought; these dynamics can then be accounted for in the lower-fidelity models, and the design and analysis steps can be repeated.

\subsection{High-fidelity Models} \label{subsec:high_fidelity_model}
A high-fidelity power transmission system model includes dynamics for synchronous generators and their associated control systems, device-level models for IBRs, loads, ancillary control equipment such as voltage compensators, and network dynamics describing how terminal voltages, currents, and power injections are coupled. Abstractly, such a model can often be written as a
system of nonlinear differential-algebraic equations (DAEs) of the form
\begin{subequations}
\label{eq:high_fidelity_dae}
\begin{align}
    \dot x_{\rm dev} &= f_{\rm dev}(x_{\rm dev},x_{\rm net},z_{\rm a},u,d),
    \label{eq:hf_devices} \\
    \dot x_{\rm net} &= f_{\rm net}(x_{\rm dev},x_{\rm net},z_{\rm a},u,d),
    \label{eq:hf_network} \\
    0 &= g(x_{\rm dev},x_{\rm net},z_{\rm a},u,d),
    \label{eq:hf_algebraic} \\
    y &= h(x_{\rm dev},x_{\rm net},z_{\rm a},u,d), 
    \label{eq:hf_output}
\end{align}
\end{subequations}
where the vector $x_{\rm dev}$ collects device and controller states, such as generator rotor angles and frequencies, turbine-governor states, excitation-system states, inverter internal-controller states (including, e.g., PLL states, voltage- and current-loop controller states), and dynamic load states. Depending on the modeling framework, network variables
may appear either among the dynamic states $x_{\rm net}$ or the algebraic variables $z_{\rm a}$. In the former case,  $x_{\rm net}$ collects dynamic network
states such as line currents, filter currents, and bus-capacitor voltages; see sidebar ``Dynamic Network Models.'' In the latter case, $z_{\rm a}$ collects terminal voltage phasors, algebraic current injections,
algebraic power-flow variables, and network constraints in phasor-domain or
quasi-steady models; see sidebar ``The Power Flow Equations and DC Power Flow.'' The vector $u$ in \eqref{eq:high_fidelity_dae} denotes control inputs or setpoints,
including commands issued by AGC, economic dispatch, or other controllers, $d$ denotes
exogenous disturbances and operating condition changes, such as load variations
and contingencies, and $y$ denotes available measurements. 

The model \eqref{eq:high_fidelity_dae} should be thought of as capturing the point of view of an (arguably idealized) omniscient grid operator, for whom the lower-level dynamics are fixed, with control inputs to be used for higher-level operational objectives; this is the view we take throughout this article. Additionally, the representation~\eqref{eq:high_fidelity_dae} is deliberately broad: its
purpose here is not to prescribe any \emph{particular} high-fidelity model, but to
emphasize that bulk grid dynamics integrate together device-level dynamics and network models. In general, \eqref{eq:high_fidelity_dae} is a large-scale nonlinear DAE, whose state dimension can easily reach tens of thousands for a realistic transmission network. Detailed network and device-level models for IBRs, forming the dynamics
in~\eqref{eq:hf_devices}, are described in,
e.g.,~\cite{ilic2007hierarchical,dorfler2023control,lara2023powersimulationsdynamics,ajala2022model,pogaku2007modeling,yazdani2010voltage},
while standard textbooks provide detailed models for conventional
generators, inverters, and networks~\cite{kundur1994power,bergen2009power,machowski1997power,yazdani2010voltage,sauer2017power}.

\subsection{Control-oriented Models} \label{subsec:control_oriented_model}
Control-oriented models for primary control can be derived from~\eqref{eq:high_fidelity_dae} through successive elimination of dynamic and algebraic variables. A standard procedure consists of three steps: 1) network approximation, 2) device model reduction, and, 3) network-device interconnection and (when desired) reduction. We develop these steps below, culminating in a reduced-order linear ODE model for primary frequency-control dynamics on a time scale of seconds.

\subsubsection{Network approximation and DC power flow} The electromagnetic network dynamics (described in detail in the sidebar ``Dynamic Network Models'') are much faster than the classical electromechanical and control dynamics of primary and secondary control in synchronous generator-dominated networks. The dynamic network equations can, therefore, be replaced by their equilibrium conditions leading to a purely algebraic network model. The most commonly used algebraic model is the \emph{power flow equations}; see sidebar ``The Power Flow Equations and DC Power Flow.''  The  DC power flow is a linearized, reduced, and decoupled version of the more general power flow equations enabling the examination of
transmission-level frequency dynamics. It specifies the linear relationship
\begin{equation}
\label{eq:dc_pf}
    p_i^{\inj}
    =
    \sum_{j \in \cN_i} b_{ij}(\theta_i-\theta_j),
\end{equation}
between active power injections and voltage angles, where $b_{ij}>0$ is the (negative) susceptance of line $(i,j)$ and $\cN_i$ denotes the set
of neighbors of bus~$i$. In matrix form,
\begin{equation}
\label{eq:dc_power_flow}
    p^{\inj}
    =
    B_{\rm dc}\theta,
\end{equation}
where $B_{\rm dc} = B_{\rm dc}^{\top}$ is the susceptance matrix, also referred to as the susceptance-weighted network Laplacian or Kirchhoff matrix~\cite{dorfler2018electrical}. 

\begin{sidebar}{Dynamic Network Models}\label{Side:Network}
    \setcounter{sfigure}{1}
    \renewcommand{\thesfigure}{S\arabic{sfigure}}
    \sdbarinitial{H}igh-voltage transmission lines are commonly modelled using the so-called $\pi$-model, which consists of a resistive-inductive line with shunt capacitance at each bus~\cite{kundur1994power} (Fig.~\ref{fig:F_pi_line_model}).

\sdbarfig{%
  \resizebox{0.95\linewidth}{!}{%
    \includegraphics[width=394.87186pt,height=167.871pt]{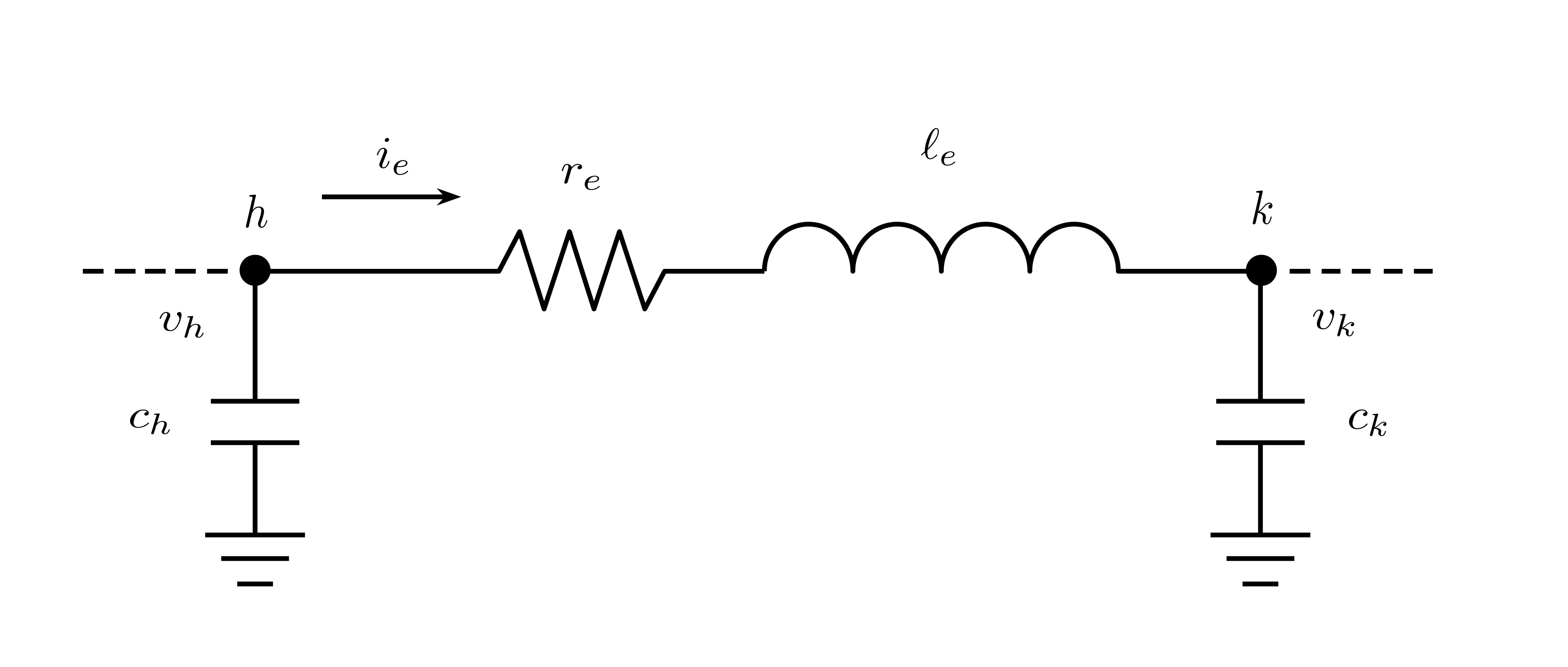}%
  }%
}{%
  Transmission line $\pi$-model. At node $h$, one may have current injection $i^{\inj}_{h}$ or current drawn from a load 
$i^{\load}_{h}(v_h)$. \label{fig:F_pi_line_model}%
}

The discussion here considers a balanced three-phase network in a so-called $dq$ frame
rotating at the nominal angular frequency~$\omega_0$. Let $\cE$ denote the set of transmission lines. Each line $e\in\cE$ connects two buses $h$ and $k$, and the reference direction for the current is arbitrarily assigned to be from $h$ to $k$. In $dq$ coordinates, the current $i_e = \mathrm{col}(i_{e,\mathrm{d}},i_{e,\mathrm{q}}) \in \real^2$ of line $e$ is governed by
\begin{equation}
\label{eq:line_dynamic_dq}
    \ell_e \tfrac{\mathrm{d}}{\mathrm{d}t} i_e
    =
    -(r_e I_2 + J\omega_0 \ell_e)i_e
    +
    v_h - v_k,
\end{equation}
where $r_e, \ell_e > 0$ are the line resistance and inductance,  $v_h = \col(v_{h,\mathrm{d}},v_{h,\mathrm{q}}) \in \real^2$ is the $dq$
voltage at bus $h$, $I_2$ is the $2\times 2$ identity matrix, and
$
    J =
    \left[\begin{smallmatrix}
        0 & -1 \\
        1 & \phantom{-}0
    \end{smallmatrix}\right]
$
 is the $90^\circ$ rotation matrix arising from the rotating frame. The voltage $v_h$ at bus $h$ satisfies
 $$
c_h \tfrac{\mathrm{d}}{\mathrm{d}t}v_h
=
-\omega_0 c_h Jv_h
-\sum_{e \in \cE} a_{ie}i_e
+i^{\inj}_{h}
-i^{\load}_{i}(v_h),
 $$
where $c_h > 0$ is the shunt capacitance, $i^{\inj}_{h}$ is the  current injection, $i^{\load}_{h}$ is the current drawn by the (possibly voltage-dependent) load, and $a_{ie}$ is the signed bus--branch incidence coefficient associated with bus $i$ and line $e$.  Stacking  currents and voltages, the network model can be written  as
\begin{subequations}
\label{eq:full_dq_network_compact}
\begin{align}
    L \tfrac{\mathrm{d}}{\mathrm{d}t} i &= -Z i + \mathcal{A}^\top v,
    \label{eq:dq_lines}\\
    C \tfrac{\mathrm{d}}{\mathrm{d}t} v &= -\omega_0 C\mathcal Jv-\mathcal{A} i + i^{\rm inj} - i^{\rm load}(v),
    \label{eq:dq_buses}
\end{align}
\end{subequations}
where $L:=\operatorname{blkdiag}_{e\in\cE}(\ell_e I_2)$,
$Z:=\operatorname{blkdiag}_{e\in\cE}(r_e I_2+\omega_0\ell_eJ)$, and $C$
are the block-diagonal matrices of line inductances, line
impedances, and bus capacitances, respectively; $\mathcal{J}:=I_n\otimes J$;
and $\mathcal{A}$ is the $dq$-lifted bus-branch incidence matrix.

The model \eqref{eq:full_dq_network_compact} retains line currents and bus voltages as dynamic
states \cite{lara2023revisiting}, and in this setting $(i,v)$ would be part of the network dynamics in \eqref{eq:hf_network}. If these dynamics are neglected and approximated as being in equilibrium, the bus currents and voltages become algebraic variables, and \eqref{eq:full_dq_network_compact} reduces to the algebraic equation
\begin{equation}\label{eq:full_dq_network_compact_algebraic}
i^{\rm inj} - i^{\rm load}(v) = Yv, ~~~~ i = Z^{-1}\mathcal{A}^{\top}v,
\end{equation}
where
$Y := \mathcal{A}Z^{-1}\mathcal{A}^\top+\omega_0 C\mathcal J$
is the $dq$-lifted bus admittance matrix, including the shunt-capacitance
contribution. Thus, under this approximation, $(i,v)$ are instead algebraic variables, and \eqref{eq:full_dq_network_compact_algebraic} would be included in \eqref{eq:hf_algebraic}.
\end{sidebar}

\begin{sidebar}{The Power Flow Equations and DC Power Flow}
    \sdbarinitial{T}he power flow equations are a static network model, describing the steady-state relationship between complex voltage and complex power in a synchronous AC power system. They are the most commonly used network model in power systems analysis, operations, and planning; see \cite{DKM-IAH:19} for further information and \cite{JWSP:17b} for recent theoretical analysis. This sidebar derives the power flow equations, along with a commonly used approximation known as the DC power flow.

Consider the static network model \eqref{eq:full_dq_network_compact_algebraic}. The $2n$ dimensional real vectors in \eqref{eq:full_dq_network_compact_algebraic}, representing $dq$-frame voltages and currents, are instead re-expressed in complex form, and the equations rewritten with complex matrices. With an obvious abuse of notation, write this complex version of the equilibrium equation as
\begin{equation}\label{Eq:NetworkComplex}
i^{\rm inj} - i^{\rm load}(v) = Yv, 
\end{equation}
where $v = \mathrm{col}(v_i) \in \complex^n$ with $v_i = V_i e^{j \theta_i}$ is the complex ``phasor'' voltage, $i^{\rm inj} - i^{\rm load} \in \complex^n$ the net complex current injection vector, and $Y \in \complex^{n \times n}$ the bus admittance matrix \cite{kundur1994power,sauer2017power}. 

The inputs and outputs of device-level models are often expressed in terms of power, not current, and thus \eqref{Eq:NetworkComplex} must be further rewritten to facilitate construction of the interconnected model. The complex power injected at bus $i$ is given by $s_i^{\inj} := v_i (i_i^{\inj})^*$, with real and imaginary parts $s_i^{\inj} = p_i^{\inj} + j q_i^{\inj}$ the \emph{active and reactive power} injections. Conjugating the $i$th equation in \eqref{Eq:NetworkComplex} and multiplying by $v_i$, one finds that
$$
p_i^{\inj} + jq_i^{\inj} - v_i i^{\mathrm{load}}_{i}(v_i)^*  =  \sum_{j=1}^{n}\nolimits v_i Y_{ij}^* v_j^*,
$$
which upon separating real and imaginary yields the classical \emph{power flow equations}
\begin{subequations}\label{eq:ac_power}
\begin{align}
\label{eq:ac_active_power}
    p_i^{\net}
    &=
    \sum_{j=1}^{n}
    V_i V_j
    \left(
        G_{ij}\cos(\theta_i-\theta_j)
        +
        B_{ij}\sin(\theta_i-\theta_j)
    \right)\\
    \label{eq:ac_reactive_power}
        q_i^{\net}
    &=
    \sum_{j=1}^{n}
    V_i V_j
    \left(
        G_{ij}\sin(\theta_i-\theta_j)
        -
        B_{ij}\cos(\theta_i-\theta_j)
    \right)
\end{align}
\end{subequations}
where $G_{ij}$ and $B_{ij}$ are the real and imaginary parts of $Y_{ij}$, $p_i^{\rm net} = p_i^{\rm inj} - \Re(v_ii^{\mathrm{load}}_{i}(v_i)^*)$, and $q_{i}^{\rm net} = q_{i}^{\rm inj} - \Im(v_ii^{\mathrm{load}}_{i}(v_i)^*)$.

An important linear approximation of \eqref{eq:ac_power} is known as the \emph{DC power flow}. This reduction is obtained under
the following assumptions: (i) transmission lines are predominantly inductive $|B_{ij}| >\!\!> |G_{ij}|$, (ii) voltage
magnitudes $V_i$ remain close to their nominal values of $1$ per-unit, (iii) voltage-angle differences across lines are sufficiently small, and (iv) reactive power is not of interest. Under these
assumptions, \eqref{eq:ac_active_power} reduces to \eqref{eq:dc_pf}.
\end{sidebar}

\subsubsection{Reduced-order device models}

Detailed device dynamics are now replaced by reduced-order models that capture
 the dominant electromechanical states relevant for primary frequency dynamics. This is described for (i) synchronous generators, (ii) grid-forming inverters, (iii) grid-following inverters, and (iv) controllable and non-controllable loads, modeled as active power demands. Hereafter, the deviation from the electrical radian synchronous frequency~$\omega_0$ at bus~$i$ is denoted as
\begin{equation}
\label{eq:Omega_def}
\Omega_i := \omega_i - \omega_0. 
\end{equation}
In a later section, the article will cover primary and secondary frequency control; within this control framework,~$\omega_0$ is the nominal frequency  to
which the network is expected to settle with secondary control (for example, see the dotted line in Fig.~\ref{fig:F_frequency_inter_area_oscillation} for that particular example of transmission system).

\vspace{.2cm}

\emph{\textbf{Synchronous generators.}}
Each synchronous generator is represented by the classical swing equation coupled with a first-order turbine-governor model:
\begin{subequations}
\label{eq:sg_model}
\begin{align}
\dot\theta_i &= \Omega_i, \label{eq:sg_angle}\\
M_i^{\rm sg} \dot\Omega_i &= -D_i^{\rm sg} \Omega_i + p^{\rm sg}_i - p_i^{\rm inj} , \label{eq:sg_swing}\\
\tau^{\rm sg}_i \dot p^{\rm sg}_i &= -p^{\rm sg}_i + \bar p^{\rm sg}_i - K_{i}^{\rm sg}\,\Omega_i , \label{eq:sg_governor}
\end{align}
\end{subequations}
where $\theta_i$ is the voltage angle, $M_i^{\rm sg} := 2H_i^{\rm sg}/\omega_0 > 0$ is the inertia coefficient, $H_i^{\rm sg}$ is the inertia constant, $D_i^{\rm sg} \ge 0$ is the damping coefficient, $p^{\rm sg}_i$ is the mechanical power, $\tau^{\rm sg}_i > 0$ is the turbine-governor time constant, $\bar p^{\rm sg}_i$ is the mechanical-power setpoint (subsequently, we discuss how this is tied to secondary controllers such as AGC and feedback optimization), and $K_{i}^{\rm sg} \ge 0$ is the governor gain. The active-power injection $p_i^{\rm inj}$ is determined by the DC power-flow equation~\eqref{eq:dc_pf_block}. This model captures inertial response through~$M_i^{\rm sg}$, damping through~$D_i^{\rm sg}$, and delayed primary-frequency response through the governor state~$p^{\rm sg}_i$.  Some remarks are provided in sidebar ``Model Approximation for Generators and Inverters''.

\vspace{.2cm}

\emph{\textbf{Grid-forming inverters.}}
On the time scales of primary and secondary frequency control, each GFM inverter can be represented by the following swing-equivalent droop model, as summarized in the sidebar ``Model Approximation for Generators and Inverters'': 
\begin{subequations}
\label{eq:gfm_model}
\begin{align}
\dot\theta_i &= \Omega_i, \label{eq:gfm_angle}\\
M_{i}^{\rm gfm} \dot\Omega_i &= -D_{i}^{\rm gfm}\,\Omega_i + \bar p_i^{\rm gfm} - p_i^{\rm inj} , \label{eq:gfm_swing}
\end{align}
\end{subequations}
where $M_{i}^{\rm gfm} > 0$ is the virtual inertia coefficient, $D_{i}^{\rm gfm} > 0$ is the virtual damping coefficient, $\bar p_i^{\rm gfm}$ is the active-power setpoint, and $p_i^{\rm inj}$ is determined by the network. The setpoint $\bar p_i^{\rm gfm}$ is typically the dispatch determined by the ISO/RTO, balancing authority, or vertically integrated utility through economic dispatch or optimal power flow (later, the article will comment on how this setpoint can be modified to include a correction from the secondary layer). Note that $M_{i}^{\rm gfm}$ is not representing inertia; it is just a time constant coming from the model reduction. As a convention, $\bar p_i^{\rm gfm} > 0$ models generation. In~\eqref{eq:gfm_model}, the filter, modulation, and inner voltage-controller dynamics are neglected, as is any saturation. Therefore,~\eqref{eq:gfm_model} derives from similar approximations applied to a high-fidelity model as generators (discussed previously). It nonetheless captures dynamics most relevant for the primary and secondary frequency control layers. 

\begin{sidebar}{Model Approximation for Generators and Inverters}
    \setcounter{sfigure}{2}
    \renewcommand{\thesfigure}{S\arabic{sfigure}}
    
\sdbarinitial{T}his sidebar comments on the approximations underlying models synchronous generators, GFL inverters, and GFM inverters. Figure~\ref{fig:F_device_models} illustrates their control
structure.

\textbf{Synchronous generator model}.   The classical swing equations
\eqref{eq:sg_angle}--\eqref{eq:sg_swing} describe rotor-angle
and rotor-speed dynamics, with mechanical input and
electrical output determining the accelerating power. The model neglects neglects stator electromagnetic transients,
field- and damper-winding dynamics, and excitation-system
dynamics~\cite{sauer2017power,machowski1997power,kundur1994power}.
Together with the fixed-voltage network approximation,
it describes aggregate frequency motion and relative
electromechanical oscillations among generators.
Accurate prediction of modal frequencies and damping may,
however, require the omitted machine and excitation dynamics.
The first-order turbine-governor model~\eqref{eq:sg_governor}
represents the delay in mechanical-power response and
the governor's proportional response to frequency deviation:
a decrease in frequency commands an increase in mechanical
power. It does not represent individual turbine stages,
reheaters, valve-position and rate limits, governor deadbands,
or the nonminimum-phase response of hydraulic turbines.
Detailed models of these mechanisms are provided
in~\cite{IEEEPES-TGM:13}. The first-order approximation
is adopted here to make the interaction between primary
and secondary control analytically tractable.

\textbf{Grid-forming inverters}. 
A GFM controller generates the angle of an internal
voltage reference. Detailed converter models describe
the DC link, AC filters, modulation, inner control loops,
and outer power and voltage controls
\cite{yazdani2010voltage,pogaku2007modeling}. The representation~\eqref{eq:gfm_model} assumes that
the inner loops provide sufficiently fast and stable voltage tracking, that voltage-magnitude variations can
be neglected, and that DC-side power availability does
not constrain the requested response.
The angle and frequency dynamics have
a swing-equation structure for suitable virtual
synchronous-machine and filtered-droop implementations~\cite{dorfler2023control}.
The coefficients $M_i^{\rm gfm}$ and $D_i^{\rm gfm}$
are determined by the control law and its parameters;
they do not represent mechanical inertia and damping. The assumptions leading from detailed converter models to
controlled-voltage-source representations are discussed in, e.g.,~\cite[Sec.~5]{schiffer2016survey}.
Detailed models and systematic model reduction procedures,
including those that account for current limiting are developed
in~\cite{ajala2022model}.

\textbf{Grid-following inverters}. 
A GFL controller regulates injected current using
a reference frame synchronized to the terminal voltage,
typically through a phase-locked loop (PLL).
An outer frequency--watt controller adjusts the
active power command using the estimated frequency
deviation~\cite{yazdani2010voltage,dorfler2023control}.
Equation~\eqref{eq:gfl_static_droop} is obtained by
approximating the PLL frequency estimate by the local
terminal frequency and the delivered active power
by its command. These approximations remove the dynamics of the
PLL, current-control, and power-tracking states,
leaving an algebraic relationship between power
injection and frequency.
The approximation requires the corresponding
closed-loop dynamics to be stable and sufficiently
fast relative to the dynamics under study.
It also assumes adequate DC-side power availability
and operation within current and power limits.
Consequently, it cannot describe synchronization
transients, control interactions in weak grids,
or responses involving limit activation; such studies
require more detailed converter and network
models~\cite{dorfler2023control}.

\sdbarfig{%
  \resizebox{\linewidth}{!}{%
    \includegraphics[width=249.88725pt,height=273.98947pt]{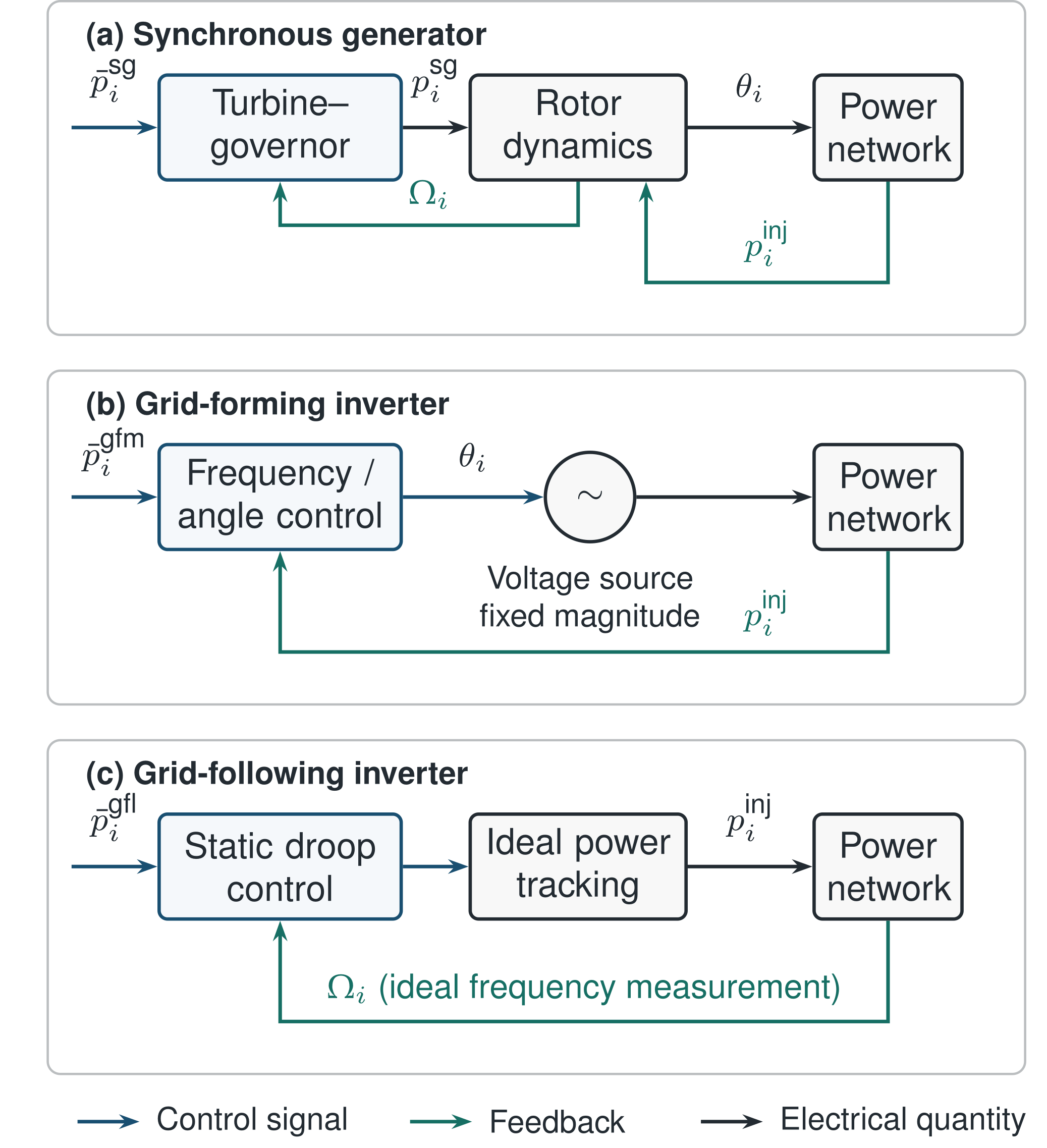}%
  }%
}{Reduced models adopted in this article:
(a) synchronous generator with turbine--governor dynamics;
(b) GFM inverter with swing-type frequency and angle dynamics;
and (c) GFL inverter with a static frequency--watt response.
\label{fig:F_device_models}
}

\end{sidebar}

\vspace{.2cm}

\emph{\textbf{Grid-following inverters.}}
When synchronization and power-tracking dynamics are
sufficiently fast relative to the dynamics of interest~\cite{dorfler2023control},
a GFL inverter with frequency--watt control can be
approximated by the static response:
\begin{equation}
\label{eq:gfl_static_droop}
p_i^{\rm inj}
=
\bar p_i^{\rm gfl} - D_{i}^{\rm gfl}\,\Omega_i,
\end{equation}
where $\bar p_i^{\rm gfl}$ is the active-power setpoint set at the tertiary layer, and $D_{i}^{\rm gfl} \ge 0$ is the frequency-watt droop gain; see the  discussion in the sidebar ``Model Approximation for Generators and Inverters''. As a convention, $\bar p_i^{\rm gfl} > 0$ models generation (e.g., from a photovoltaic system, or from a battery discharging). In \eqref{eq:gfl_static_droop}, the PLL, current controller, and power filter dynamics have been neglected, as is any saturation, leading to a quasi-static linear frequency-responsive power injection model. 

\vspace{.2cm}

\emph{\textbf{Buses with only loads.}} In the DC active-power model, a load-only bus can be modeled as $p_i^{\rm inj} = -p_i^{\rm load}$, where $p_i^{\rm load} > 0$ is the active-power demand. For frequency-sensitive loads, one may use the model \begin{equation}\label{eq:load_model}
    p_i^{\rm inj} = - p_i^{\rm load} - D_{i}^{\rm load}\,\Omega_i, 
\end{equation}
with $D_{i}^{\rm load} \ge 0$ a load-damping coefficient. Both GFL and load-only buses are algebraic.

\subsubsection{Network-device interconnection and reduction to state-space model}

The device models \eqref{eq:sg_model}, \eqref{eq:gfm_model}, \eqref{eq:gfl_static_droop}, \eqref{eq:load_model} are now interconnected with the DC power flow model \eqref{eq:dc_pf} through the power variables and an appropriate partitioning of buses. Let $\cG^{\rm sg}$ denote the set of buses containing synchronous generators, $\cG^{\rm gfm}$ grid-forming inverters, and let $\cG = \cG^{\rm sg} \cup \cG^{\rm gfm}$; these are buses that will be referred to as \emph{swing-type} buses. Similarly, let $\cL^{\rm gfl}$ denote the set of buses containing grid-following inverters and $\cL^{\rm l}$ the ones with loads only, and define $\cL = \cL^{\rm gfl} \cup \cL^{\rm l}$; these buses are referred to as \emph{algebraic} buses. Let $n_G := |\cG|$ and $n_L := |\cL|$ denote the number of swing-type and algebraic buses, with $n_G + n_L = n$. 

\begin{notebox}[Modeling note: co-located and aggregated resources]
This article assumes that each bus has loads and either a generator, a GFM inverter, or a GFL inverter. This is  for simplicity of exposition; the model can be obviously extended to cases where multiple types or aggregations of resources are connected at a single bus.
\end{notebox}

\vspace{.2cm} 

Partitioning the DC power-flow equation \eqref{eq:dc_power_flow} according to the bus sets $\cG$ and $\cL$ yields the block equation
\begin{equation}
\label{eq:dc_pf_block}
\begin{bmatrix}
p_\cG^{\rm inj} \\
p_\cL^{\rm inj}
\end{bmatrix}
=
\begin{bmatrix}
B_\cGG & B_\cGL \\
B_\cLG & B_\cLL
\end{bmatrix}
\begin{bmatrix}
\theta_\cG \\
\theta_\cL
\end{bmatrix},
\end{equation}
with matrices $B_\cGG$, $B_\cGL = B_\cLG^\top$, and $B_\cLL$ of appropriate dimensions, and where $\theta_\cG \in \mathbb{R}^{n_G}$ and $\theta_\cL \in \mathbb{R}^{n_L}$ are the voltage angles for the buses $\cG$ and $\cL$, respectively. 

Vectorizing the device models \eqref{eq:sg_model}, \eqref{eq:gfm_model}, \eqref{eq:gfl_static_droop}, \eqref{eq:load_model} and combining them with \eqref{eq:dc_pf_block} yields the linear DAE
\begin{subequations}
\label{eq:dae_model}
\begin{align}
\dot\theta_\cG
&=
\Omega_\cG,
\label{eq:dae_angle}
\\
M_\cG\dot\Omega_{\cG}
&=
-D_{\cG}\Omega_{\cG}
+
\begin{bmatrix}
p^{\rm sg}\\
\bar p^{\rm gfm}
\end{bmatrix}
-
\begin{bmatrix}
B_\cGG & B_\cGL
\end{bmatrix}
\begin{bmatrix}
\theta_\cG\\
\theta_\cL
\end{bmatrix},
\\
T\dot p^{\rm sg}
&=
-p^{\rm sg}
+\bar p^{\rm sg}
-K^{\rm sg}\Omega_{\cG^{\rm sg}},
\label{eq:dae_governor}
\\
0
&=
\begin{bmatrix}
\bar p^{\rm gfl}\\
-p^{\rm load}
\end{bmatrix}
-D_\cL\Omega_\cL
-
\begin{bmatrix}
B_\cLG & B_\cLL
\end{bmatrix}
\begin{bmatrix}
\theta_\cG\\
\theta_\cL
\end{bmatrix}, 
\label{eq:dae_algebraic_balance}
\end{align}
\end{subequations}
where $\bar p^{\rm sg}
=
\col(\bar p_i^{\rm sg})_{i\in\cG^{\rm sg}}$, $\bar p^{\rm gfm}
=
\col(\bar p_i^{\rm gfm})_{i\in\cG^{\rm gfm}}$,
$\bar p^{\rm gfl}
=
\col(\bar p_i^{\rm gfl})_{i\in\cL^{\rm gfl}}$, and $p^{\rm load}
=
\col(p_i^{\rm load})_{i\in\cL^{\rm l}}$, with $p^{\rm load}$ positive for consumption. The matrices
\[
M_\cG
=
\begin{bmatrix}
M^{\rm sg}&0\\
0&M^{\rm gfm}
\end{bmatrix},
~~~~
D_\cG
=
\begin{bmatrix}
D^{\rm sg}&0\\
0&D^{\rm gfm}
\end{bmatrix}
\]
collect the inertia and damping coefficients at synchronous-generator and
GFM buses, while
\[
D_\cL
=
\begin{bmatrix}
D^{\rm gfl}&0\\
0&D^{\rm load}
\end{bmatrix}
\]
collects the frequency--watt gains of GFL inverters and the
frequency-sensitivity coefficients of loads. The relation $\dot\theta_\cL
=
\Omega_\cL$ is left implicit. Finally,
\[
T
=
\diag(\tau_i^{\rm sg})_{i\in\cG^{\rm sg}},
\qquad
K^{\rm sg}
=
\diag(K_i^{\rm sg})_{i\in\cG^{\rm sg}}
\]
collect turbine-governor time constants and droop gains.

The DAE~\eqref{eq:dae_model} retains the distinction between buses in $\cG$,
which have swing-type dynamics, and buses in $\cL$, whose
active-power balances are algebraic. It is often more convenient for subsequent analysis and design to further reduce \eqref{eq:dae_model} to an explicit ODE model, and this is pursued next. In particular, the final reduced ODE model will be obtained by eliminating the algebraic-bus angles $\theta_{\cL}$, using a quasi-static approximation for their frequencies $\Omega_{\cL}$, and then finally specifying an arbitrary generator bus angle as a reference to resolve rotational symmetry.

To proceed, observe that if the network is connected and $\cG$ is nonempty, then the submatrix $B_{\cLL}$ satisfies $B_\cLL \succ 0$. Hence $B_\cLL$ is invertible \cite{dorfler2018electrical}, and in fact its inverse $B_{\cLL}^{-1}$ has nonnegative entries; this follows because $B_\cLL$ is a proper principal submatrix of the irreducible symmetric weighted Laplacian matrix $B_{\rm dc}$. Define the Kron-reduced network matrix
\begin{equation}
\label{eq:kron_reduced_matrix}
B_{\rm red}
:=
B_\cGG
-
B_\cGL B_\cLL^{-1}B_\cLG
\in\mathbb R^{n_G\times n_G}
\end{equation}
which is itself a positive-semidefinite Laplacian matrix satisfying $B_{\rm red}\mathbf 1_{n_G}= \mathbf 0_{n_G}$, and the matrix
\begin{equation}
\label{eq:frequency_divider_matrix}
\Phi
:=
-B_\cLL^{-1}B_\cLG
\in\mathbb R^{n_L\times n_G}.
\end{equation}
The matrix $\Phi$ is referred to as the \emph{frequency divider} matrix~\cite{milano2016frequency,ponce2024complex}; it is nonnegative, and satisfies $\Phi\mathbf 1_{n_G}=\mathbf 1_{n_L}$. As explained in ``Derivation: ODE via frequency divider'' below, the utility of $\Phi$ is that the  frequencies $\Omega_{\cL}$ can be approximated as
\begin{equation}
\label{eq:dc_frequency_divider_qs}
\Omega_\cL
\approx
\Phi\Omega_\cG.
\end{equation}
Since $\Phi$ is nonnegative, \eqref{eq:dc_frequency_divider_qs} states that each frequency at the buses $\cL$ is a weighted average of the frequencies at buses in $\cG$. Eliminating $\theta_{\cL}$ from \eqref{eq:dae_model} and using the frequency divider approximation \eqref{eq:dc_frequency_divider_qs}, one obtains the following reduced model:
\begin{tcolorbox}[
    enhanced,
    colback=noteblue!3,
    colframe=noteblue,
    boxrule=0.8pt,
    arc=2mm,
    left=1mm,
    right=1mm,
    top=-1mm,
    bottom=1mm,
    before skip=5pt,
    after skip=5pt
]
\begin{subequations}
\label{eq:representative_control_model}
\begin{align}
\dot\theta_\cG
&=
\Omega_\cG,
\label{eq:representative_angle}
\\
M_\cG\dot\Omega_\cG
&=
\!-\!D\Omega_\cG
\!+\!
\begin{bmatrix}
p^{\rm sg}\\
\bar p^{\rm gfm}
\end{bmatrix}
\!-\!B_{\rm red}\theta_\cG
+\Phi^\top \!
\begin{bmatrix}
\bar p^{\rm gfl}\\
-p^{\rm load}
\end{bmatrix},
\label{eq:representative_frequency}
\\
T\dot p^{\rm sg}
&=
-p^{\rm sg}
+\bar p^{\rm sg}
-K^{\rm sg}\Omega_{\cG^{\rm sg}}.
\label{eq:representative_governor}
\end{align}
\end{subequations}
\end{tcolorbox}

The first equation in \eqref{eq:representative_control_model} relates the angles and frequencies at buses with synchronous generators
and GFM inverters. The second describes their swing-type dynamics after eliminating
the buses in $\cL$, where
\begin{equation}
\label{eq:effective_damping}
D
:=
D_\cG+\Phi^\top D_\cL\Phi ,
\end{equation}
is an effective damping matrix which combines the damping at synchronous generator and GFM buses with the
frequency-responsive behavior of GFL inverters and loads. The third equation describes the turbine-governor dynamics of
the synchronous generators.

Because $B_{\rm red}\mathbf 1_{n_G}= \mathbf 0_{n_G}$, the model \eqref{eq:representative_control_model} is invariant under a
uniform shift of all voltage angles. This leads to a zero eigenvalue of the system dynamics, which can be eliminated by fixing a reference angle as follows. Select any bus $r\in\cG$ as
the angular reference and define
\[
\delta_i:=\theta_i-\theta_r,
\qquad
i\in\cG,\quad i\neq r.
\]
With the columns ordered so that the final column corresponds to bus $r$,
define
\[
C
:=
\begin{bmatrix}
I_{n_G-1}&-\mathbf 1_{n_G-1}
\end{bmatrix}
\in\mathbb R^{(n_G-1)\times n_G}.
\]
The relative-angle dynamics are then $\dot\delta=C\Omega_\cG$.
Moreover, by construction, $B_{\rm red}\theta_\cG=\hat B\delta$,
where
$\hat B\in\mathbb R^{n_G\times(n_G-1)}$ is obtained by removing the column
of $B_{\rm red}$ associated with the reference bus. Define the state
\begin{equation}
\label{eq:state_vector}
x
:=
\col(\delta,\Omega_\cG,p^{\rm sg})
\in\mathbb R^{n_x},
\end{equation}
with $n_x=(n_G-1)+n_G+n_{\rm sg}$. Let
$S_{\cG}^{\rm sg}\in\{0,1\}^{n_G\times n_{\rm sg}}$ and
$S_{\cG}^{\rm gfm}\in\{0,1\}^{n_G\times n_{\rm gfm}}$ embed synchronous
generators and GFM inverters in the $\cG$-bus ordering. Similarly, let
$S_{\cL}^{\rm gfl}\in\{0,1\}^{n_L\times n_{\rm gfl}}$ and
$S_{\cL}^{\rm load}\in\{0,1\}^{n_L\times n_{\rm load}}$ embed GFL-inverter
setpoints and load demands in the $\cL$-bus ordering. With this notation it holds that
\begin{align*}
\Omega_{\cG^{\rm sg}}
&=
(S_{\cG}^{\rm sg})^\top\Omega_\cG,
\\
\begin{bmatrix}
p^{\rm sg}\\
\bar p^{\rm gfm}
\end{bmatrix}
&=
S_{\cG}^{\rm sg}p^{\rm sg}
+
S_{\cG}^{\rm gfm}\bar p^{\rm gfm},
\\
\begin{bmatrix}
\bar p^{\rm gfl}\\
-p^{\rm load}
\end{bmatrix}
&=
S_{\cL}^{\rm gfl}\bar p^{\rm gfl}
-
S_{\cL}^{\rm load}p^{\rm load}.
\end{align*}

Routine calculations now show that
\begin{tcolorbox}[
    enhanced,
    colback=noteblue!3,
    colframe=noteblue,
    boxrule=0.8pt,
    arc=2mm,
    left=1mm,
    right=1mm,
    top=-1mm,
    bottom=1mm,
    before skip=5pt,
    after skip=5pt
]
\begin{equation}
\label{eq:state_space_form}
\dot x
=
Ax
+
B^{\rm sg}\bar p^{\rm sg}
+
B^{\rm gfm}\bar p^{\rm gfm}
+
B^{\rm gfl}\bar p^{\rm gfl}
+
F^{\rm load}p^{\rm load}
\end{equation}
\end{tcolorbox}
where the state matrix is
\begin{equation}
\label{eq:state_matrix}
A
=
\begin{bmatrix}
0
&
C
&
0
\\
-M_\cG^{-1}\hat B
&
-M_\cG^{-1}D
&
M_\cG^{-1}S_{\cG}^{\rm sg}
\\
0
&
-T^{-1}K^{\rm sg}(S_{\cG}^{\rm sg})^\top
&
-T^{-1}
\end{bmatrix},
\end{equation}
and the input matrices are
\begin{align}
B^{\rm sg}
&=
\begin{bmatrix}
0\\
0\\
T^{-1}
\end{bmatrix},
&
B^{\rm gfm}
&=
\begin{bmatrix}
0\\
M_\cG^{-1}S_{\cG}^{\rm gfm}\\
0
\end{bmatrix},
\label{eq:explicit_input_matrices_1}
\\[1mm]
B^{\rm gfl}
&=
\begin{bmatrix}
0\\
M_\cG^{-1}\Phi^\top S_{\cL}^{\rm gfl}\\
0
\end{bmatrix},
&
F^{\rm load}
&=
\begin{bmatrix}
0\\
-M_\cG^{-1}\Phi^\top S_{\cL}^{\rm load}\\
0
\end{bmatrix}.
\label{eq:explicit_input_matrices_2}
\end{align}
All zero blocks have dimensions implied by the state and input
partitions. Compared with the DAE~\eqref{eq:dae_model}, the ODE
\eqref{eq:state_space_form} eliminates the algebraic variables while
preserving the important physical parameters (topology, inertia, damping, droop gains, etc.) that determine the primary frequency response, thereby maintaining a direct
connection to the underlying generator, inverter, and network models. The ODE structure permits application of standard state-space theory for modeling, analysis, and design. 

\begin{notebox}[Modeling note: stability of primary control]
Internal stability of the dynamics of the system with primary control is a prerequisite for the implementation
of traditional secondary control methods. Due to the particularly simple structure of \eqref{eq:state_space_form}, internal stability is guaranteed unconditionally if all primary control droop parameters are positive. In more general modelling situations, internal stability will be parameter-dependent, and will require restrictions on the primary control gains. 

A broader open problem concerns stability certificates for heterogeneous
systems described by higher-fidelity models. Such models retain GFM and GFL
inner controls, PLL dynamics, voltage and reactive-power dynamics, delays,
saturation, deadband, and dynamic network effects. Determining scalable conditions under
which an arbitrary mixture of these resources is stable remains an active
research direction
\cite{gross2022compensating,haberle2026decentralized,pates2019robust,smith2021transient,iraniparast2025networked}.
\end{notebox}

\begin{derivationbox}[Derivation: ODE via frequency divider]
To see how~\eqref{eq:representative_control_model} can be derived, define the net active-power input at the algebraic buses as
\[
\bar p_\cL
:=
\begin{bmatrix}
\bar p^{\rm gfl}\\
-p^{\rm load}
\end{bmatrix},
\qquad
p_\cL^{\rm inj}
:=
\bar p_\cL-D_\cL\Omega_\cL.
\]
The $\cL$-bus block of~\eqref{eq:dae_model} is
\begin{equation}
\label{eq:algebraic_constraint}
0
=
p_\cL^{\rm inj}
-B_\cLG\theta_\cG
-B_\cLL\theta_\cL.
\end{equation}
Since $\Omega_\cL=\dot\theta_\cL$, this relation can also be written as
\begin{equation}
\label{eq:L_bus_descriptor_relation}
D_\cL\dot\theta_\cL+B_\cLL\theta_\cL
=
\bar p_\cL-B_\cLG\theta_\cG.
\end{equation}
When $D_\cL=0$, this is an algebraic constraint. Nonzero GFL droop gains
or load-frequency sensitivities introduce  frequency dependence.

Differentiating~\eqref{eq:L_bus_descriptor_relation} gives the exact
frequency relation
\begin{equation}
\label{eq:dc_frequency_divider}
D_\cL\dot\Omega_\cL+B_\cLL\Omega_\cL
=
-B_\cLG\Omega_\cG+\dot{\bar p}_\cL.
\end{equation}
Using $\Phi=-B_\cLL^{-1}B_\cLG$, this becomes
\begin{equation}
\label{eq:dc_frequency_divider_stacked}
D_\cL\dot\Omega_\cL
+
B_\cLL\bigl(\Omega_\cL-\Phi\Omega_\cG\bigr)
=
\dot{\bar p}_\cL.
\end{equation}
When these dynamics are fast relative to the electromechanical time scale and
$\bar p_\cL$ is constant or slowly varying, the terms
$D_\cL\dot\Omega_\cL$ and $\dot{\bar p}_\cL$ can be neglected, yielding
\begin{equation}
\label{eq:dc_frequency_divider_derivation}
\Omega_\cL
\approx
\Phi\Omega_\cG.
\end{equation}
Because $\Phi$ is entry-wise nonnegative and
$\Phi\mathbf 1_{n_G}=\mathbf 1_{n_L}$, each reconstructed frequency is a
convex combination of the frequencies at buses in $\cG$.

The algebraic constraint~\eqref{eq:algebraic_constraint} also gives
\begin{equation}
\label{eq:delta_L_elimination}
\theta_\cL
=
B_\cLL^{-1}p_\cL^{\rm inj}
-
B_\cLL^{-1}B_\cLG\theta_\cG.
\end{equation}
Substitution into the $\cG$-bus network injection yields
\begin{align}
p_\cG^{\rm inj}
&=
B_\cGG\theta_\cG+B_\cGL\theta_\cL
\nonumber\\
&=
B_{\rm red}\theta_\cG
+
B_\cGL B_\cLL^{-1}p_\cL^{\rm inj}.
\label{eq:kron_reduction}
\end{align}
Since the network susceptance matrix is symmetric,
\[
B_\cGL B_\cLL^{-1}
=
-\Phi^\top,
\]
and therefore
\begin{equation}
\label{eq:kron_compact}
p_\cG^{\rm inj}
=
B_{\rm red}\theta_\cG
-
\Phi^\top p_\cL^{\rm inj}.
\end{equation}
The frequency-divider approximation gives
\begin{equation}
\label{eq:pL_compact}
p_\cL^{\rm inj}
\approx
\bar p_\cL-D_\cL\Phi\Omega_\cG.
\end{equation}
Substituting~\eqref{eq:pL_compact} into~\eqref{eq:kron_compact} yields
\begin{align}
p_\cG^{\rm inj}
&\approx
B_{\rm red}\theta_\cG
+\Phi^\top D_\cL\Phi\Omega_\cG
-\Phi^\top\bar p_\cL.
\label{eq:pG_substituted}
\end{align}
Inserting this expression into the swing equations produces
\eqref{eq:representative_frequency}, with
$D=D_\cG+\Phi^\top D_\cL\Phi$.
\end{derivationbox}

\subsubsection{Understanding the Reduced Model}

The physics of the reduced model \eqref{eq:representative_control_model} (or equivalently, \eqref{eq:state_space_form}) admits a straightforward and purely mechanical interpretation. The dynamics \eqref{eq:representative_angle}--\eqref{eq:representative_frequency} are effectively a coupled mass-spring system with damping $-D\Omega_{\cG}$ and ``spring'' forces $-B_{\rm red}\theta_{\cG}$, and ``external forces'' arising from the power injections $(p^{\rm sg},\bar{p}^{\rm gfm},\bar{p}^{\rm gfl},p^{\rm load})$. The turbine-governor equation \eqref{eq:representative_governor} provides additional dynamics which are superficially similar to damping, but with lags induced by the turbine-governor time constants $T$. 

Consider incremental changes $(\Delta\bar{p}^{\rm sg},\Delta\bar{p}^{\rm gfm},\Delta\bar{p}^{\rm gfl},\Delta p^{\rm load})$ in the setpoints entering~\eqref{eq:representative_control_model}. Depending on the magnitudes and spatial distribution of these changes, some generator speeds will initially increase, while others decrease. In either case, the injected power from ``damping'' $-D\Omega_{\cG}$ and the mechanical power $p^{\rm sg}$ from the turbine will counter-act the changes until a new equilibrium is reached; this is the primary frequency response of these units. At this equilibrium, an important phenomenon that occurs is \emph{frequency synchronization}, meaning that all generator frequencies asymptotically converge to a common value $\Omega_{\cG} \to \Omega^{\rm eq}\mathbf{1}_{n_G}$. The common equilibrium frequency $\Omega^{\rm eq}$ is, in fact, easily determined by setting the left-hand sides of \eqref{eq:representative_frequency}, \eqref{eq:representative_governor} to zero:
\[
\begin{aligned}
0 &= \!-\!D\Omega^{\rm eq}\mathbf{1}_{n_G}
\!+\!
\begin{bmatrix}
\Delta \bar{p}^{\rm sg}-K^{\rm sg}\Omega^{\rm eq}\mathbf{1}\\
\Delta \bar{p}^{\rm gfm}
\end{bmatrix}\\
&\qquad \!-\!B_{\rm red}\theta_\cG
+\Phi^\top \!
\begin{bmatrix}
\Delta \bar{p}^{\rm gfl}\\
-\Delta p^{\rm load}
\end{bmatrix}.
\end{aligned}
\]
Left multiplying this by $\mathbf{1}_{n_G}^{\top}$, using $\mathbf{1}_{n_G}^{\top}B_{\rm red} = 0$ and $\mathbf{1}_{n_{G}}^{\top}\Phi^{\top} = \mathbf{1}_{n_L}^{\top}$, one obtains 
\begin{equation}
\label{eq:primary_frequency_equilibrium}
\Omega^{\rm eq}
=
\frac{1}{\beta_{\rm tot}}\underbrace{(\mathbf{1}^{\top}\Delta\bar{p}^{\rm sg} + \mathbf{1}^{\top}\Delta\bar{p}^{\rm gfm} + \mathbf{1}^{\top}\Delta\bar{p}^{\rm gfl} - \mathbf{1}^{\top}\Delta p^{\rm load})}_{\text{Total set-point change}},
\end{equation}
where
\begin{align}
\beta_{\rm tot}
:={}&
\sum_{i\in\cG^{\rm sg}}
\left(D_i^{\rm sg}+K_i^{\rm sg}\right)
+
\sum_{i\in\cG^{\rm gfm}}D_i^{\rm gfm}
\nonumber\\
&+
\sum_{i\in\cL^{\rm gfl}}D_i^{\rm gfl}
+
\sum_{i\in\cL^{\rm l}}D_i^{\rm load},
\label{eq:total_primary_response}
\end{align}
is known as the \emph{frequency characteristic} of the system. Notice that the equilibrium frequency is determined only by the net generation-load mismatch and the damping/primary control gains. Increasing the primary control gains decreases the steady-state frequency error; however, this error cannot be eliminated entirely with finite gains (as expected from proportional control action). Restoring the frequency back to its nominal value is the role of secondary control.

\paragraph{\textbf{Comparison of EMT and reduced model trajectories}}

\begin{figure}[h!]
    \centering
{
        \includegraphics[width=.8\linewidth]
        {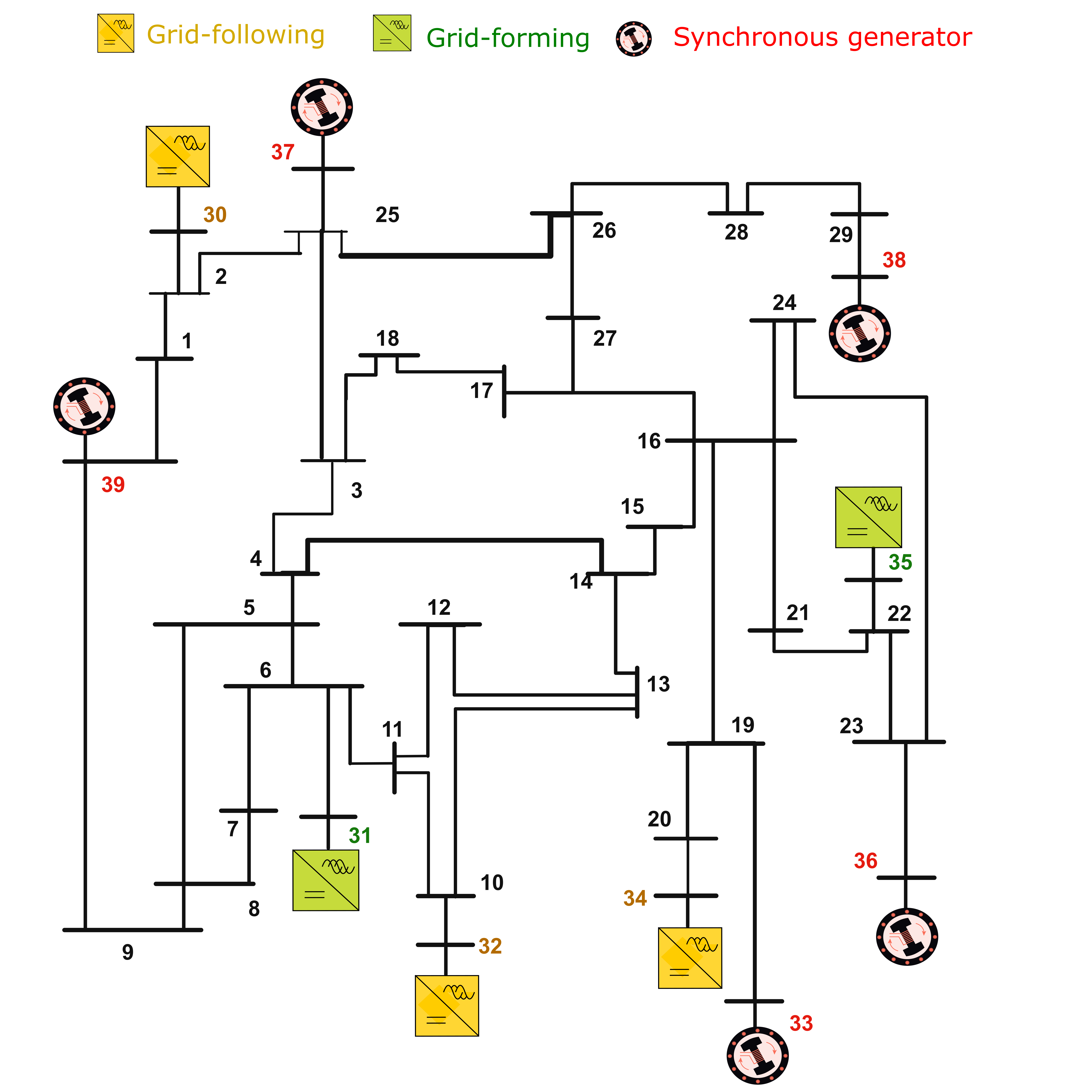}
    }
    \vspace{0.5em}
{
        \includegraphics[width=\linewidth]
        {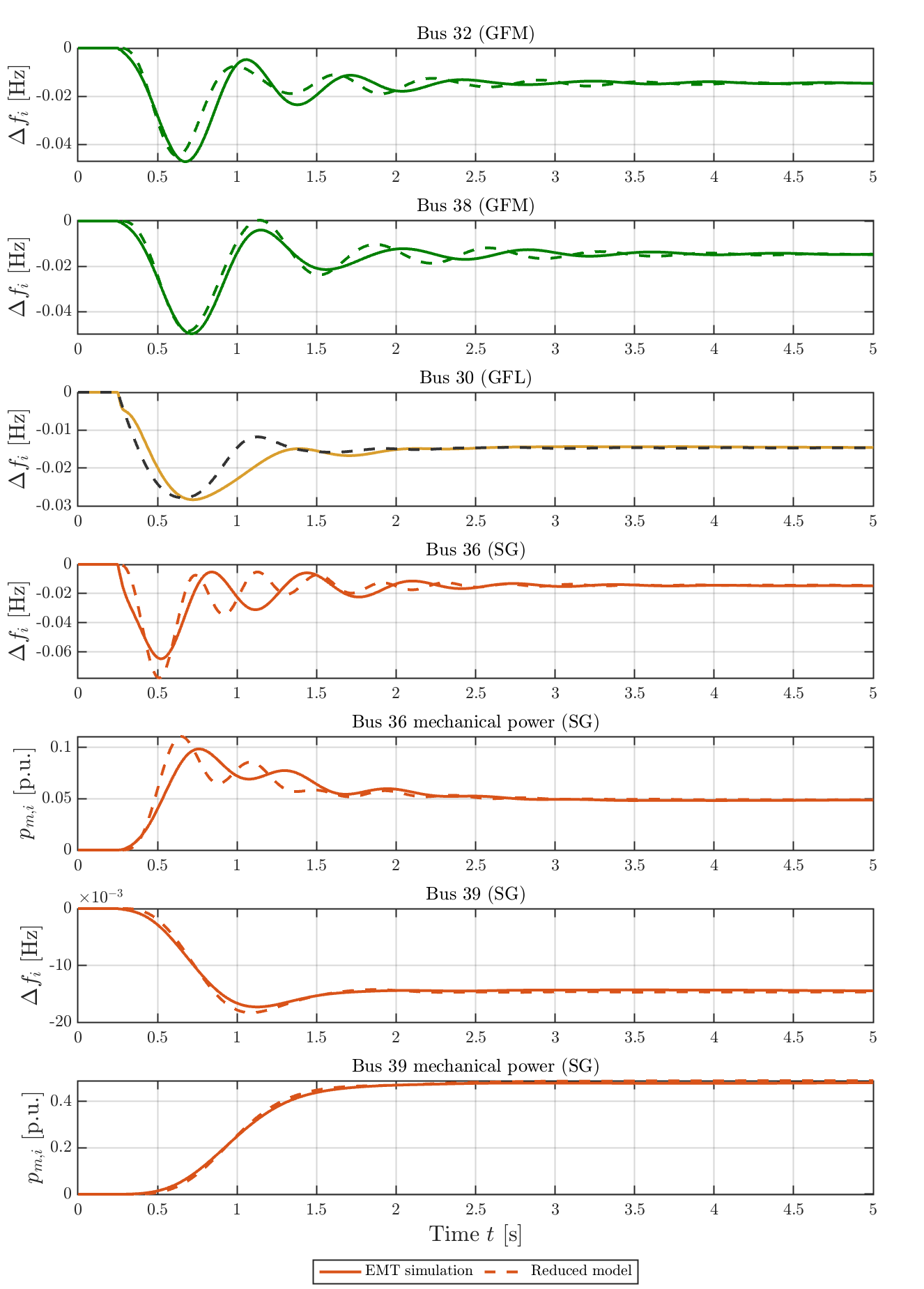}
    }
    \caption{Comparison between dq EMT simulations and the analytical
    model~\eqref{eq:state_space_form} on a modified IEEE 39-bus system
    representing ISO-NE. The test system features synchronous generators,
    GFM inverters, and GFL inverters.}
    \label{fig:F_model}
\end{figure}

To illustrate how the model~\eqref{eq:state_space_form} compares to high-fidelity EMT simulations, Fig.~\ref{fig:F_model} provides an illustrative numerical simulation for a modified IEEE 39-bus test system. The EMT simulations were
performed using \texttt{PowerSimulationsDynamics.jl},
an open-source Julia package for time-domain simulation of power system dynamics,
including detailed models of synchronous generators, inverter-based resources,
controllers, and network dynamics. The reader is referred to~\cite{lara2023powersimulationsdynamics} for a detailed description of the models (alternative platforms for EMT simulations include PSCAD, TyphoonSim, etc). Within the modeling scope of this study, the EMT simulation retains the complete network, synchronous generator, governor, converter, and converter-control dynamics. The converters are represented by their standard averaged models; consequently, individual semiconductor switching events, switching-frequency ripple, and the associated harmonics are not represented. These switching phenomena are outside the bandwidth and control objectives considered here. 

A $250$ MW active-power load increase is applied at bus~35 at $t=0.25$ s on a $100$ MVA system base. The figure compares the frequency deviations obtained from the EMT simulation and from the analytical model~\eqref{eq:state_space_form} at representative GFM, GFL, and synchronous-generator buses, together with the mechanical power deviations of selected synchronous generators. Both data sets are sampled with a time step of $0.01$ seconds over a $10$ seconds simulation horizon.

Figure~\ref{fig:F_model} shows that the analytical model preserves the dominant response observed in the EMT simulations, where dominant behavior refers to the principal frequency excursion, the subsequent recovery, the associated mechanical-power adjustment, and convergence toward a common steady-state frequency. To quantify the agreement, the error for each signal is normalized by the peak EMT excursion from its pre-disturbance value. For the five frequency trajectories shown, the normalized root-mean-square error over the complete post-disturbance interval ranges from 1.8\% to 8.0\%; over the recovery interval $t\geq 2$~s, it decreases to 0.5\%--2.4\%. For the two mechanical-power trajectories, the corresponding errors range from 1.3\% to 5.9\% over the complete post-disturbance interval and from 0.7\% to 1.2\% over the recovery interval. The larger discrepancies during the initial response arise primarily from differences in transient amplitude, timing, and damping, rather than from an error in the eventual power balance or steady-state response. The transient separation among the plotted frequency signals should also not be interpreted as loss of synchronism: the EMT quantities correspond to generator rotor speeds, GFM voltage frequencies, and the GFL PLL frequency, which are distinct local frequency measures and need not coincide immediately following the localized disturbance. Both models reproduce comparable transient spatial variation and subsequent convergence toward a common steady-state frequency. The close agreement during the recovery and steady-state portions of the response supports the use of the analytical model for the AGC and feedback optimization developments that will be considered in the ensuing sections.

\section{Two-time-scale models and methods for secondary frequency control}

Primary frequency control is essentially decentralized proportional control at each generation unit, and acts on a time scale of seconds to stabilize the frequency. The dynamics of an entire interconnected system with primary control and the resulting equilibrium frequency deviation were described in~\eqref{eq:representative_control_model}--\eqref{eq:primary_frequency_equilibrium}. The  frequency deviation is subsequently eliminated by AGC, which is a slower centralized controller operating \emph{within} each balancing area of the system. Describing the AGC therefore requires introducing a fine-grained description of~\eqref{eq:dae_model} (or the reduced ODE model~\eqref{eq:representative_control_model}) as an interconnected set of areas. This section describes AGC, formulates the required multi-area model, and uses two time-scale analysis to study stability of the closed-loop model obtained by interconnecting the reduced plant~\eqref{eq:state_space_form} with AGC.

\subsection{Area control errors and AGC}
While the implementation of primary control is decentralized at each unit, the situation in equilibrium is that \emph{all} generation units in the interconnected system provide compensating control action, irrespective of the spatial location of the generation-load mismatch. Thus, not only does the interconnected system settle to an off-nominal frequency, but any agreed-upon power exchanges between balancing areas will be disrupted. AGC addresses both these issues simultaneously through regulation of a specially constructed error signal---the \emph{area control error (ACE)}---which is based on both the so-called net-interchange error and the frequency error for a given balancing area.

\paragraph{\textbf{Net interchange}}

The interconnected transmission system is partitioned into $N_a$ balancing areas,  indexed by $a_1,\ldots,a_{N_a}$, each managed by a balancing authority (e.g., ISO, RTO, or utility); see
Fig.~\ref{fig:spatiotemporal_organization}. Transmission lines interconnecting balancing areas are known as tie lines. With the DC power flow model~\eqref{eq:dc_pf}, the active power flow on the line from bus $i$ to bus $j$ is
\[
p_{ij}^{\rm line}
=
b_{ij}(\theta_i-\theta_j).
\]
Let $\cT_{k\ell}\subset\mathcal E$ denote the set of tie lines oriented from area $a_k$ to area $a_\ell$. The total tie line flow from $a_k$ to $a_{\ell}$ is then
\begin{equation}
\label{eq:area_pair_interchange}
p_{k\ell}^{\rm tie}
:=
\sum_{(i,j)\in\cT_{k\ell}}
b_{ij}(\theta_i-\theta_j),
\end{equation}
where $p_{k\ell}^{\rm tie}>0$ denotes an export from area $k$ to area
$\ell$. Reversing the orientation gives
$p_{\ell k}^{\rm tie}=-p_{k\ell}^{\rm tie}$. The \emph{net interchange} $p_{k}^{\rm NI}$ of area $a_k$ is the total power exported to all neighboring areas, given by
\begin{equation}
\label{eq:area_net_actual_interchange}
p_k^{\rm NI}
:=
\sum_{\ell\in\cN_k^{\rm A}}p_{k\ell}^{\rm tie},
\end{equation}
where $\cN_k^{\rm A}$ is the set of areas connected to area $a_k$. Because
every tie-line flow enters the net interchange of its two incident areas with
opposite signs, it always holds that
\[
\sum_{k=1}^{N_a}p_k^{\rm NI}=0.
\]
Let $p_k^{\rm SNI}$ denote the \emph{scheduled net interchange} of area $a_k$; these set-points would be determined through interchange agreements, bilateral transactions, and, where wholesale markets exist, day-ahead and real-time market clearing. Consistent interchange schedules satisfy
\[
\sum_{k=1}^{N_a}p_k^{\rm SNI}=0.
\]
The deviation of the net interchange from its scheduled value is then
\begin{equation}
\label{eq:area_interchange_error}
\Delta p_k^{\rm NI}
:=
p_k^{\rm NI}-p_k^{\rm SNI}.
\end{equation}

\paragraph{\textbf{Area control error}}

AGC has two objectives: restoring nominal frequency
and restoring the net interchange of each balancing area to its scheduled
value. Let $\widetilde\Omega_k$ denote a representative frequency deviation
for area $a_k$, obtained, for example, from the frequency
measurement from a selected generator  or from a weighted average of measurements within the area. The \emph{area control error}
$\mathrm{ACE}_k$ of area $a_k$ combines the frequency and interchange errors as
\begin{equation}
\label{eq:ace_definition}
{\rm ACE}_k
=
\Delta p_k^{\rm NI}
+
b_k\widetilde\Omega_k,
\end{equation}
where $b_k>0$ is the \emph{frequency bias coefficient} selected by the $k$th
balancing authority~\cite{wood2013power,kundur1994power}. Both terms in
\eqref{eq:ace_definition} are available from measurements: tie-line meters
provide $p_k^{\rm NI}$, while local frequency measurements provide
$\widetilde\Omega_k$.

The main objective of AGC is to regulate ${\rm ACE}_k$ to zero in every
area. Although each ACE is the sum of two errors, simultaneous regulation of \emph{all}
ACEs eliminates both errors in equilibrium:
$$
{\rm ACE}_k=0\ \text{for all }k
\quad\Leftrightarrow\quad
\widetilde\Omega_k=0 
\text{~and~}
\Delta p_k^{\rm NI}=0\ \text{for all }k.
$$
Thus, zero ACE restores both nominal frequency and the scheduled
interchanges~\cite[Lemma 3.4]{JWSP-NM:20m}.

\begin{notebox}[Modeling note: why zero ACE restores both objectives]
At equilibrium, all representative area frequencies synchronize to a common value $\Omega^{\rm eq}$. Consequently, ${\rm ACE}_k
=
\Delta p_k^{\rm NI}
+
b_k\Omega^{\rm eq}$. If ${\rm ACE}_k=0$ for every area, summing these equations gives
\[
0
=
\sum_{k=1}^{N_a}\Delta p_k^{\rm NI}
+
\left(\sum_{k=1}^{N_a}b_k\right)\Omega^{\rm eq}.
\]
The first term is zero because
\[
\sum_{k=1}^{N_a}\Delta p_k^{\rm NI}
=
\sum_{k=1}^{N_a}p_k^{\rm NI}
-
\sum_{k=1}^{N_a}p_k^{\rm SNI}
=
0.
\]
Since every $b_k$ is positive, it follows that
$\Omega^{\rm eq}=0$. Each ACE equation then yields
$\Delta p_k^{\rm NI}=0$.
\end{notebox}

\paragraph{\textbf{Automatic generation control}}

With the area control errors defined in \eqref{eq:ace_definition}, each balancing authority must implement AGC controller to regulate its respective ACE to zero. In practice, AGC systems involve
ACE measurement, filtering, sampling, participation-factor allocation, reserve
logic, and setpoint dispatch to participating generators~\cite{apostolopoulou2014automatic}.
For analysis however, its action may be idealized as integral feedback on ACE~\cite{JWSP-NM:20m,jung2024inferences}. Let $\eta_k$ denote the AGC state of balancing area~$k$. A 
continuous-time model can be written as 
\begin{equation}
\label{eq:agc_integrator_area}
    \tau_k^{\rm agc} \dot \eta_k
    =
    -
    {\rm ACE}_k,
    \qquad k=1,\ldots,N_a,
\end{equation}
where $\tau_k^{\rm agc}>0$ is the AGC integral time constant. 

\begin{figure}[t!]
    \centering
\includegraphics[width=1.0\linewidth]
        {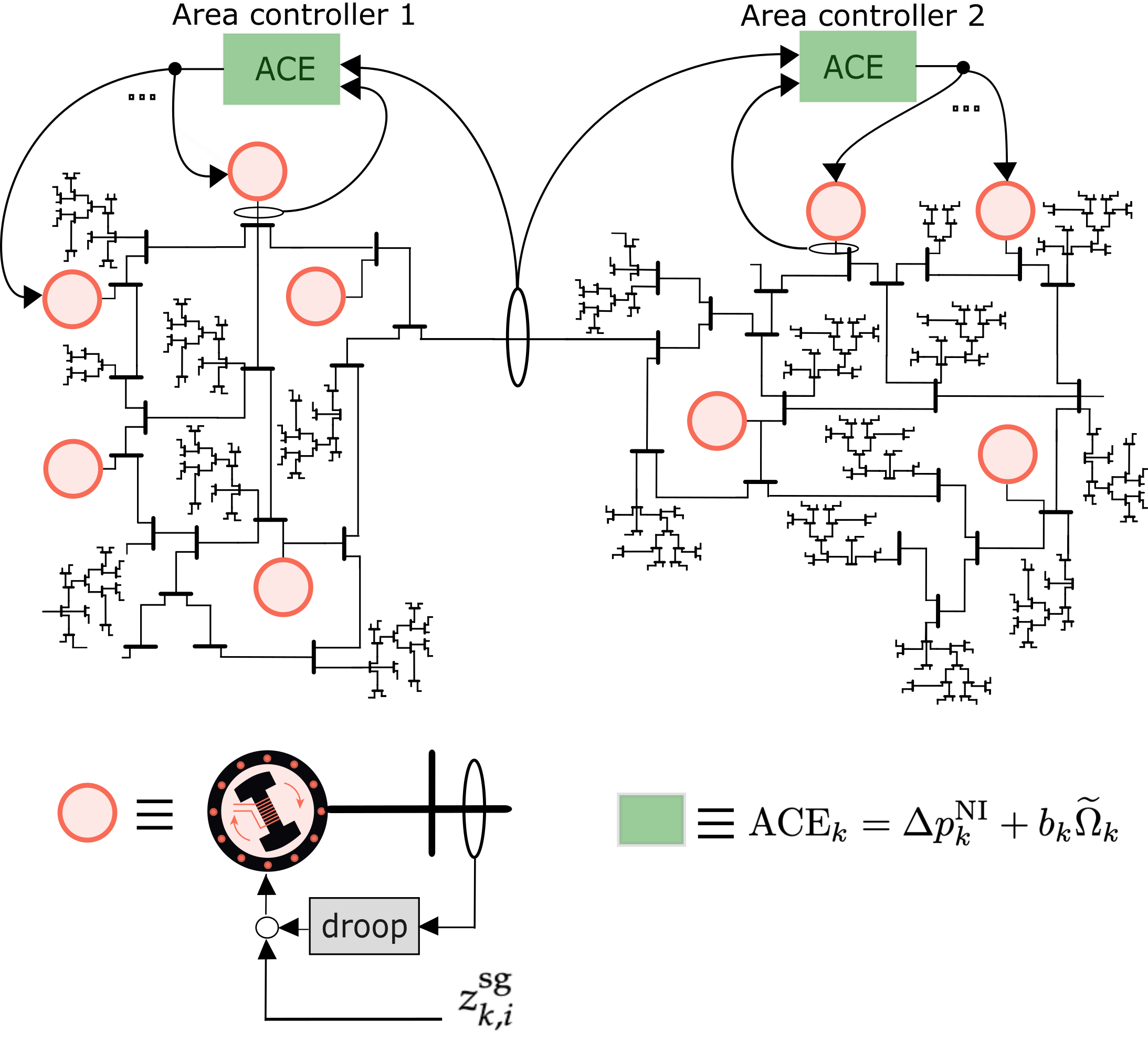}
    \caption{AGC architecture for a two-area system.}
    \label{fig:F_two_areas_ace}
\end{figure}

To close the loop, the scalar power set-point signal $\eta_k$ is then allocated among the generation units participating in AGC. Let $\cG_k^{\rm sg}$ denote the set of participating synchronous generators in area~$k$. References $\bar p^{\rm sg}$ for the mechanical powers are decomposed as
\begin{equation}
\label{eq:pm_agc_decomposition}
\bar p^{\rm sg}
=
p^{\rm sg, \star}
+
z^{\rm sg},
\end{equation}
where $p^{\rm sg, \star}$ is the scheduled reference obtained from economic dispatch
or an OPF problem, and $z^{\rm sg}$ is the power correction
generated by AGC. If $\alpha_{k,i}\ge 0$ is a nonnegative \emph{ participation factor} for
generator~$i$ in area~$k$, with
$$
    \sum_{i\in\cG_k^{\rm sg}}\alpha_{k,i}=1,
$$
then, in the unsaturated case,
\begin{equation}
\label{eq:agc_participation_area}
    z_{k,i}^{\rm sg}
    =
    \alpha_{k,i}\eta_k,
    \qquad i\in\cG_k^{\rm sg}.
\end{equation}
Economically efficient participation factors can be directly calculated using the marginal generation costs at optimality under economic dispatch or OPF~\cite{schweppe2013spot,SB-YCC-SVD:21}. The complete set of AGC controllers is now described by~\eqref{eq:ace_definition}--\eqref{eq:agc_participation_area}, and a high-level architecture of AGC is shown in  Fig.~\ref{fig:F_two_areas_ace} for a two-area system.

\begin{notebox}[Modeling note: practical implementations of AGC]
In practical  implementations, ACE signals are
filtered and sampled, power commands are updated at discrete times, generator
participation factors are combined with ramp-rate and reserve logic, and the
resulting commands are held constant between updates~\cite{apostolopoulou2014automatic,kundur1994power}. The continuous-time
integrator model \eqref{eq:agc_integrator_area} is reasonable
when AGC evolves slowly relative to the primary frequency dynamics, and when sampling periods are sufficiently fast. Specifically, if $T_s$ is the AGC sampling period, then~\eqref{eq:agc_integrator_area} can be
interpreted as the continuous-time approximation of the sampled-data update
\begin{equation}
\label{eq:sampled_agc_update}
    \eta_k((\ell+1)T_s)
    =
    \eta_k(\ell T_s)
    -
    \frac{T_s}{\tau_k^{\rm agc}} {\rm ACE}_k(\ell T_s) , \,\, \ell \in \mathbb{N} 
\end{equation}
with the resulting generator setpoint corrections held constant between sampling instants. Concretely, if the effective time constant of traditional AGC is on the order of $100$ seconds, then a sampling period of a few seconds represents roughly an order-of-magnitude separation. In this regime,
zero-order-hold and sampling effects may have limited influence on the dominant
secondary-control behavior, and singular
perturbation arguments can be used to analyze the closed-loop dynamics (see, for example,~\cite{kokotovic1976singular,peponides1982singular,khalil96nonlinear}, and contraction theoretic approaches to two-time-scale systems~\cite{del2012contraction,cothren2026online}).
\end{notebox}

\subsubsection{Stability analysis of AGC}
The objective is now to analyze closed-loop stability of the primary-controlled dynamics \eqref{eq:state_space_form} with the AGC controllers \eqref{eq:ace_definition}--\eqref{eq:agc_participation_area}. Since only synchronous generators traditionally participate in AGC, set-points for GFM and GFL inverter units are held constant. Substituting~\eqref{eq:pm_agc_decomposition} into~\eqref{eq:state_space_form}
gives the dynamic model
\begin{equation}
\label{eq:plant_with_agc}
\dot x
=
A x
+
B^{\rm sg} z^{\rm sg}
+
d^{\rm agc},
\end{equation}
where
\begin{equation}
\label{eq:dm_agc_definition}
d^{\rm agc}
:=
B^{\rm sg} p^{\rm sg, \star}
+
B^{\rm gfm}\bar p^{\rm gfm}
+
B^{\rm gfl}\bar p^{\rm gfl}
+
F^{\rm load}p^{\rm load}
\end{equation}
for brevity. Stacking all area-level AGC states into
$\eta:=\col(\eta_1,\ldots,\eta_{N_a})$, the allocation rules \eqref{eq:agc_participation_area} can be written as
\begin{equation}
\label{eq:agc_participation_matrix}
    z^{\rm sg}
    =
    \Pi^{\rm agc} \eta,
\end{equation}
where $\Pi^{\rm agc}$ is the generator participation matrix.

To write~\eqref{eq:ace_definition} in vector form, let ${\rm ACE}
    :=
    \col({\rm ACE}_1,\ldots,{\rm ACE}_{N_a})$ and 
    $p^{\rm SNI}
    :=
    \col(p_1^{\rm SNI},\ldots,p_{N_a}^{\rm SNI})$. Let
$\widetilde\Omega^{\rm area}=C_\Omega x$ collect the representative
area-frequency deviations, and define
$B^{\rm bias}:=\operatorname{diag}(b_1,\ldots,b_{N_a})$. By examining \eqref{eq:area_pair_interchange}--\eqref{eq:area_net_actual_interchange} and recalling the angle elimination procedure in \eqref{eq:delta_L_elimination},\eqref{eq:kron_reduction}, it follows that the vector of net interchanges
$p^{\rm NI}:=\col(p_1^{\rm NI},\ldots,p_{N_a}^{\rm NI})$ can be expressed as
\begin{equation}
\label{eq:nai_output_model}
p^{\rm NI}
=
C_{\rm ni}x
+
D_{\rm ni}^{\rm gfl}\bar p^{\rm gfl}
+
D_{\rm ni}^{\rm load}p^{\rm load},
\end{equation}
where the matrices in~\eqref{eq:nai_output_model} are obtained by substituting the
reconstructed bus angles into
\eqref{eq:area_pair_interchange}--\eqref{eq:area_net_actual_interchange}. Using the
net-interchange output model~\eqref{eq:nai_output_model}, the vector of ACEs
can be expressed as an affine function of $x$ as
\begin{align}
\label{eq:ace_output_explicit}
    {\rm ACE}(x)
    &=
    C^{\rm ace}x
    +
    r^{\rm ace},
\end{align}
where
\begin{equation*}
\begin{aligned}
C^{\rm ace} &:=C_{\rm ni}+B^{\rm bias}C_\Omega,\\
r^{\rm ace} &:=
    D_{\rm ni}^{\rm gfl}\bar p^{\rm gfl}
    +
    D_{\rm ni}^{\rm load} p^{\rm load}
    -
    p^{\rm SNI}.
\end{aligned}
\end{equation*}
Combining \eqref{eq:agc_integrator_area},
\eqref{eq:plant_with_agc}, \eqref{eq:agc_participation_matrix}, \eqref{eq:ace_output_explicit} gives the model
\begin{subequations}
\label{eq:closed_loop_agc_ct}
\begin{align}\label{eq:closed_loop_agc_ct_1}
    \dot x
    &=
    A x
    +
    B^{\rm sg}z^{\rm sg}
    +
    d^{\rm agc},
    \\
    \label{eq:closed_loop_agc_ct_2}
    T^{\rm agc}\dot\eta
    &= -\mathrm{ACE}(x) = 
    -
    C^{\rm ace}x
    -
    r^{\rm ace},\\
    z^{\rm sg} &= \Pi^{\rm agc}\eta,
\end{align}
\end{subequations}
where $T^{\rm agc}:=\diag(\tau_1^{\rm agc},\ldots,\tau_{N_a}^{\rm agc})$. For a given network and given controller parameters, closed-loop stability can now be checked numerically by computing the eigenvalues of
\[
A_{\rm cl}^{\rm agc} := \begin{bmatrix}
    A & B^{\rm sg}\Pi^{\rm agc}\\
    -(T^{\rm agc})^{-1}C^{\rm ace} & 0
\end{bmatrix}
\]
and verifying that they have negative real parts. 

However, because the time constants $\tau_{k}^{\rm agc}$ are always large compared to the time-scale of primary control, the model \eqref{eq:closed_loop_agc_ct} exhibits two time-scales \cite{khalil96nonlinear,kokotovic1976singular}. The state $x$ contains the fast and stable dynamics of the transmission system with primary frequency control, while $\eta$ (and, accordingly, $z^{\rm sg}$) evolves on the slower
secondary-control time scale. The state $x$ will converge on a fast timescale towards the quasi-equilibrium value 
\begin{equation}
\label{eq:agc_fixed_eta_equilibrium}
    x^{\star}(z^{\rm sg})
    \coloneqq
    -
    A^{-1}
    \left(
        B^{\rm sg}z^{\rm sg}
        +
        d^{\rm agc}
    \right),
\end{equation}
and the dynamics \eqref{eq:closed_loop_agc_ct_2} will then evolve on a slower time-scale according to
\begin{equation}\label{eq:reduced_agc}
\begin{aligned}
T^{\rm agc}\dot{\eta} &= -\mathrm{ACE}(x^{\star}(z^{\rm sg}))\\
&= -C^{\rm ace}x^{\star}(z^{\rm sg}) - r^{\rm ace}.\\ 
\end{aligned}
\end{equation}
In the language of singular perturbation theory, \eqref{eq:closed_loop_agc_ct_1} is the \emph{boundary layer} dynamics, \eqref{eq:agc_fixed_eta_equilibrium} describes the \emph{slow manifold}, \eqref{eq:reduced_agc} is the \emph{reduced dynamics}, and $(T^{\rm agc})^{-1}$ plays the role of the classical time-scale separation parameter ``$\epsilon$''. The reduced dynamics \eqref{eq:reduced_agc} were analyzed in \cite{JWSP-NM:20m}, where it was shown that they simplify to
\begin{equation}\label{Eq:AGC_reduced}
T^{\rm agc}\dot{\eta} = -H^{\rm agc}\begin{bmatrix}
    \mathrm{1}^{\top}z^{\rm sg}_1 \\ \vdots \\ \mathrm{1}^{\top}z^{\rm sg}_{N_a}
    \end{bmatrix} - q^{\rm agc},
\end{equation}
where $q^{\rm agc}
    :=
    r^{\rm ace}
    -
    C^{\rm ace}A^{-1}d^{\rm agc}$ is constant and where the $N_a \times N_a$ matrix $H^{\rm agc}$ has the simple form
\begin{equation}\label{Eq:H_agc_explicit}
H^{\rm agc}
= 
\frac{1}{\beta}
\begin{bmatrix}
\beta+d_1 & d_1 & \cdots & d_1 \\
d_2 & \beta+d_2 & \cdots & d_2 \\
\vdots & \vdots & \ddots & \vdots \\
d_{N_a} & d_{N_a} & \cdots & \beta+d_{N_a}
\end{bmatrix},
\end{equation}
where $d_k = b_k - \beta_k$ and $\beta = \sum_{k=1}^{N_{a}}\beta_k$, with
\begin{align}
\beta_k
:= & 
\sum_{i\in\cG_k^{\rm sg}}\!\left(D_i^{\rm sg}+K_i^{\rm sg}\right)
+\sum_{i\in\cG_k^{\rm gfm}}\!D_i^{\rm gfm} \nonumber \\
& 
~~~~~~~~~ +\sum_{i\in\cL_k^{\rm gfl}}\!D_i^{\rm gfl}
+\sum_{i\in\cL_k^{\rm l}}\!D_i^{\rm load} 
\label{Eq:Betak}
\end{align}
the total effective primary frequency response
coefficient of area $k$. The matrix $H^{\rm agc}$ is the DC gain matrix from total commanded AGC power $\mathbf{1}^{\top}z^{\rm sg}_k$ in each area to steady-state ACE values. Since $\mathbf{1}^{\top}z^{\rm sg}_k = \sum_{i\in\mathcal{G}^{\rm agc}_{k}}\alpha_{k,i}\eta_k = \eta_k$, \eqref{Eq:AGC_reduced} is simply the linear system
\begin{equation}\label{Eq:AGC_reduced2}
T^{\rm agc}\dot{\eta} = -H^{\rm agc}\eta - q^{\rm agc}.
\end{equation}

It follows from standard results that \eqref{eq:closed_loop_agc_ct} will be stable for all sufficiently large tunings of $T^{\rm agc}$ if the reduced dynamics \eqref{Eq:AGC_reduced2} are stable, i.e., if $-(T^{\rm agc})^{-1}H^{\rm agc}$ is Hurwitz. It is known that $-H^{\rm agc}$ is \emph{diagonally stable} for all bias selections $b_k > 0$, and thus $-(T^{\rm agc})^{-1}H^{\rm agc}$ is, in particular, Hurwitz. Thus, for all sufficiently slow AGC tunings, the closed-loop system is stable and all area control errors are regulated asymptotically to zero. See \cite{JWSP-NM:20m} for details of this analysis, along with further discussion of the reduced model \eqref{eq:reduced_agc}, and \cite{jung2024inferences,SB-YCC-SVD:21} for related work on AGC.

\begin{notebox}[Modelling note: AGC stability analysis with command limits]
When generator command limits are included, the allocation rule
\eqref{eq:agc_participation_area} is replaced by a saturated map. Specifically, the generation limits
$\underline \ell_{k,i}^{\rm sg}\le \bar p_{k,i}^{\rm sg}\le \overline \ell_{k,i}^{\rm sg}$
induce correction limits
$\underline z_{k,i}^{\rm sg}:=\underline \ell_{k,i}^{\rm sg}-p_{k,i}^{{\rm sg},\star}$
and
$\overline z_{k,i}^{\rm sg}:=\overline \ell_{k,i}^{\rm sg}-p_{k,i}^{{\rm sg},\star}$, and \eqref{eq:agc_participation_area} is replaced by
\begin{equation}
\label{eq:agc_saturated_allocation}
    z_{k,i}^{\rm sg}
    =
    \sat_{[\underline z_{k,i}^{\rm sg},\overline z_{k,i}^{\rm sg}]}
    \left(\alpha_{k,i}\eta_k\right), \quad i\in\cG_k^{\rm sg},
\end{equation}
where the scalar saturation operator is defined as $\sat_{[a,b]}(z)
    :=
    \min\{b,\max\{a,z\}\}$ for $b > a$. With this change, the closed-loop dynamics \eqref{eq:closed_loop_agc_ct_1} become
\begin{subequations}
\label{eq:sidebar_agc_interconnection_saturated}
\begin{align}
    \dot x
    &=
    A x+B^{\rm sg}\phi(\eta)+d^{\rm agc},
    \\
    T^{\rm agc}\dot\eta
    &=
    -C^{\rm ace}x-r^{\rm ace},
\end{align}
\end{subequations}
where
$\phi(\eta):=\operatorname{col}_{k,i}
\big(\sat_{[\underline z_{k,i}^{\rm sg},\overline z_{k,i}^{\rm sg}]}(\alpha_{k,i}\eta_k)\big)$ denotes the
saturated AGC correction vector. Two time-scale analysis can again be used to study stability of \eqref{eq:sidebar_agc_interconnection_saturated}. Following the same steps as in the main body, the reduced dynamics of \eqref{eq:sidebar_agc_interconnection_saturated} are
\begin{equation}\label{Eq:ReducedNonlinear}
T^{\rm agc}\dot{\eta} = -H^{\rm agc}\phi(\eta) - q^{\rm agc}.
\end{equation}
Suppose that the equilibrium $\eta^{\rm eq}$ of~\eqref{Eq:ReducedNonlinear}satisfies
$\underline z_{k,i}^{\rm sg}<\alpha_{k,i}\eta_k^{\rm eq}<\overline z_{k,i}^{\rm sg}$ for all participating
generators. From the results in \cite{JWSP-NM:20m} the matrix $-H^{\rm agc}$ is diagonally stable, i.e., there exists a diagonal $D_\eta = \mathrm{diag}(d_1,\ldots,d_{N_a}) \succ 0$ such that $D_{\eta}H^{\rm agc} + (D_{\eta}H^{\rm agc})^{\top} \prec 0$. The Popov-type Lyapunov function
\[
V_{\eta}(\eta) = \sum_{k=1}^{N_a}d_{k}\tau_{k}^{\rm agc} \int_{0}^{\eta^{\rm eq}_{k}}\phi_k(\xi_k) - \phi_k(\eta^{\rm eq}_k) \,\mathrm{d}\xi_k
\]
can then be used to establish both local exponential and global asymptotic stability of the equilibrium $\eta^{\rm eq}$ for \eqref{Eq:ReducedNonlinear}. Singular perturbation arguments can then once again be used to establish local exponential stability for the closed-loop system \eqref{eq:sidebar_agc_interconnection_saturated}. 
\end{notebox}

\paragraph{\textbf{Comparison of AGC model trajectories}}

To illustrate the typical dynamics of AGC, consider a two-area network consisting of two identical copies of the IEEE 39-bus network of Fig.~\ref{fig:F_model}. The copies are  interconnected with two tie lines, one from bus $17$ of area $a_1$ to bus 17 of area $a_2$, the other from bus $26$ of $a_1$ to bus 26 of $a_2$, with a scheduled net interchange for $a_1$ of $100$MW. The frequency characteristic of each area is $\beta_1 = \beta_2 = 17,016$ MW/Hz. The AGC time constants are set as $\tau_{1}^{\rm agc} = \tau_{2}^{\rm agc} = 64$s, and the bias factors are set as $b_1 = 1.2\beta_1$ and $b_2 = 0.9\beta_2$; the tuning of area $a_1$ is said to be \emph{overbiased}, while the tuning of area $a_2$ is \emph{underbiased}. From \eqref{Eq:H_agc_explicit}, observe that the effect of overbiased or underbiased tuning is to introduce cross-coupling between the dynamics of AGC states on a long time-scale. For each AGC controller, the participation factors are set such that the synchronous generators at buses 33, 36, 37, and 39 participate equally in AGC response. 

\begin{figure}[h!]
    \centering
\includegraphics[width=1.0\linewidth]
        {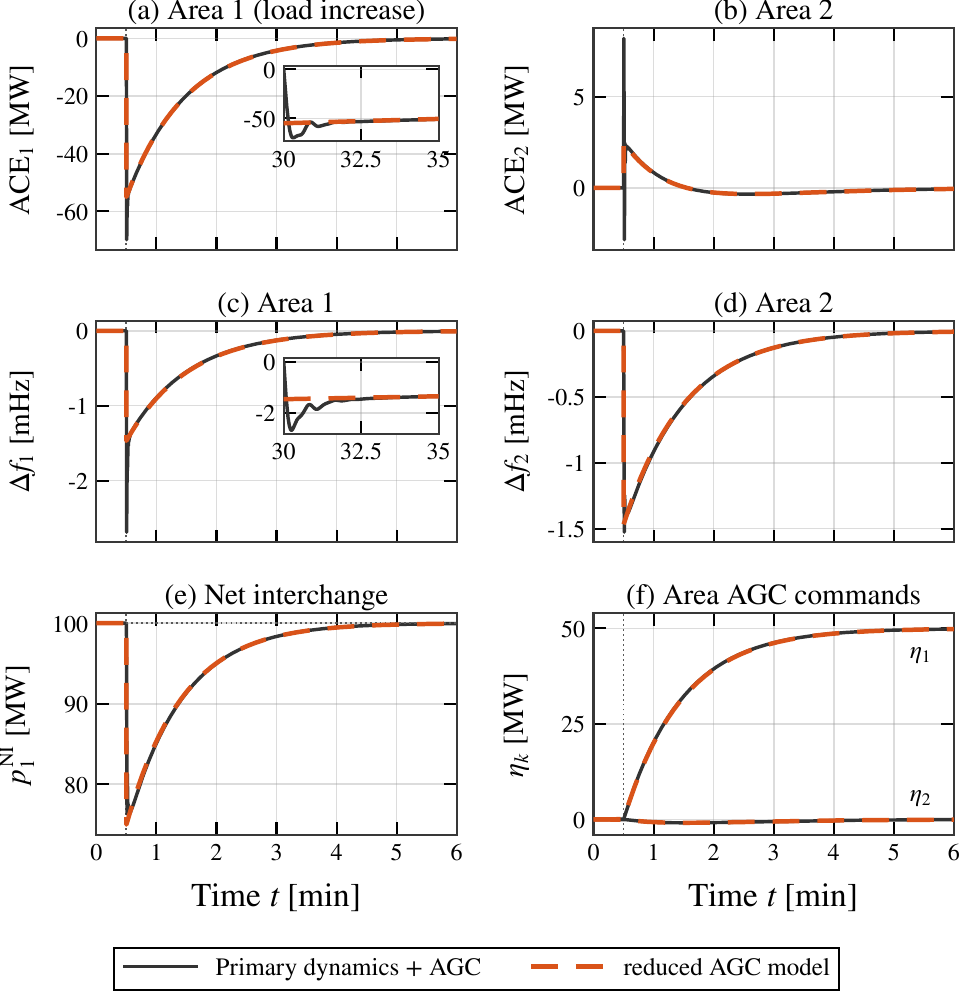}
    \caption{Comparison between closed-loop analytical model \eqref{eq:closed_loop_agc_ct} and  reduced AGC model \eqref{eq:reduced_agc} for two interconnected IEEE 39-bus systems.}
    \label{fig:agc_response}
\end{figure}

A 50MW load increase is applied in area $a_1$ at $t = 30$s. Fig.~\ref{fig:agc_response} plots the response for both the closed-loop model \eqref{eq:closed_loop_agc_ct} and the reduced AGC model \eqref{eq:reduced_agc}. 

After the initial transient due to primary frequency response, the models agree closely. Area control errors, frequency deviations, and net interchange errors are slowly regulated to zero by the AGC scheme. From Fig.~\ref{fig:agc_response} (a) and (b), and as expected, the disturbance in area $a_1$ is primarily detected through $\mathrm{ACE}_1$, with $\mathrm{ACE}_2$ having a more muted response, mostly arising from transient frequency changes and tie line flows. The slight underbiasing of the tuning for $b_2$ causes a slight undershoot in $\mathrm{ACE}_2$ and in the AGC response $\eta_2$.

\section{Research Opportunities Within the Existing Layered Architecture}

This section discusses opportunities for the development of new optimization and control methods that fit \emph{within} the layered architecture of Fig.~\ref{fig:spatiotemporal_organization}(a) as a means to address emerging challenges in power systems. By preserving the existing hierarchies, new methods can be analyzed, validated, and introduced incrementally,  offering a risk-mitigated path toward adoption by utilities and power system operators. In the context of the evolving resource mix illustrated in Fig.~\ref{fig:F_future},  this section focuses on four  problems: 
\begin{itemize}
    \item  Integrating  IBRs and aggregations of IBRs into primary and secondary control;
    \item Integrating AI data centers into secondary frequency control;
    \item Redesign of primary frequency controllers to incorporate stability and economic specifications;  
    \item Design of auxiliary controllers for IBRs to improve the dynamic response after a disturbance. 
\end{itemize}

\begin{figure}[t!]
    \centering
\includegraphics[width=1.0\linewidth]
        {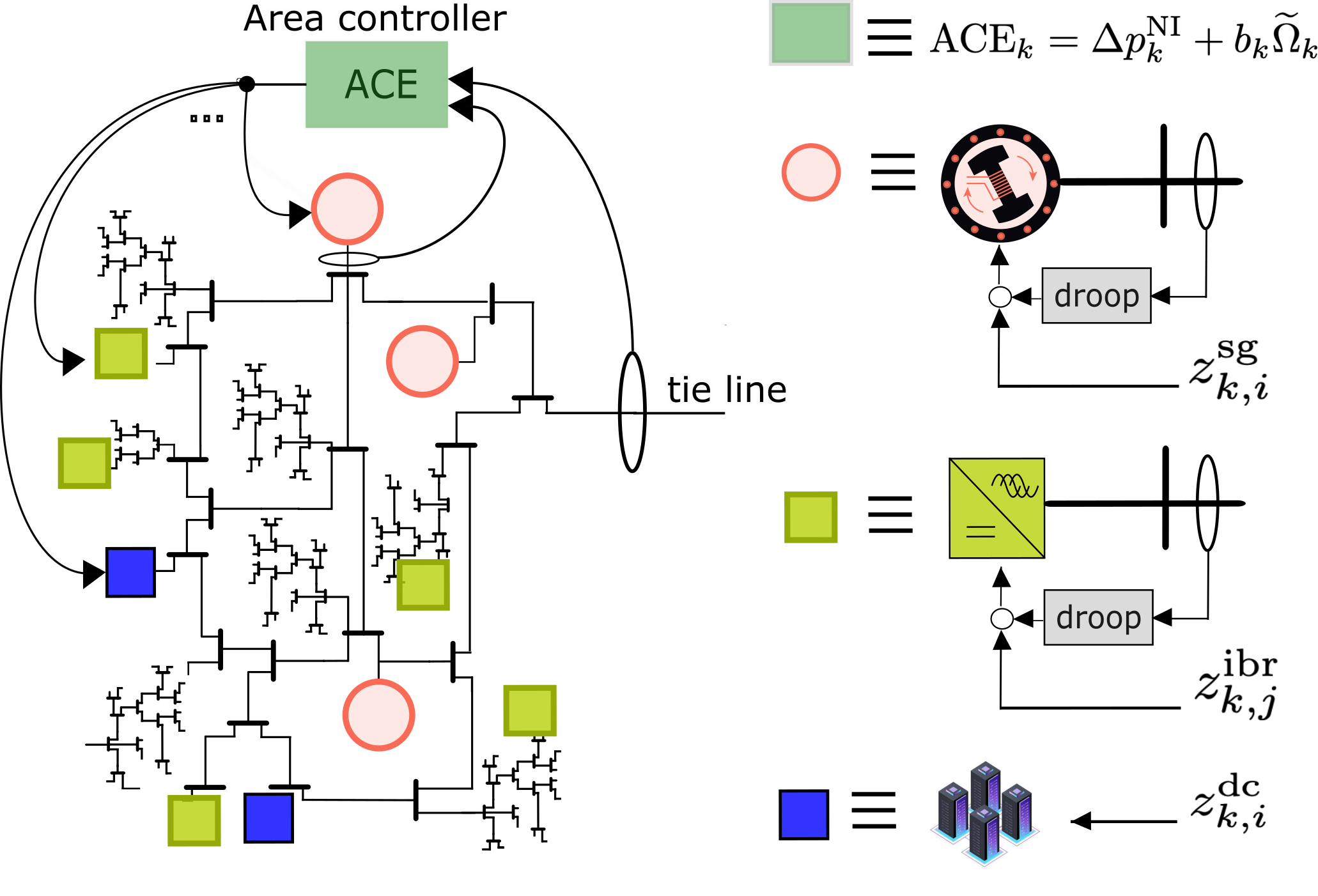}
    \caption{Example of AGC architecture for synchronous generators, IBRs, and data centers. Here,  data centers  participates in secondary frequency control. }
    \label{fig:F_one_area_ace_others}
\end{figure}

\begin{figure*}[t]
\centering
\resizebox{0.95\textwidth}{!}{%
\includegraphics[width=503.5271pt,height=243.30276pt]{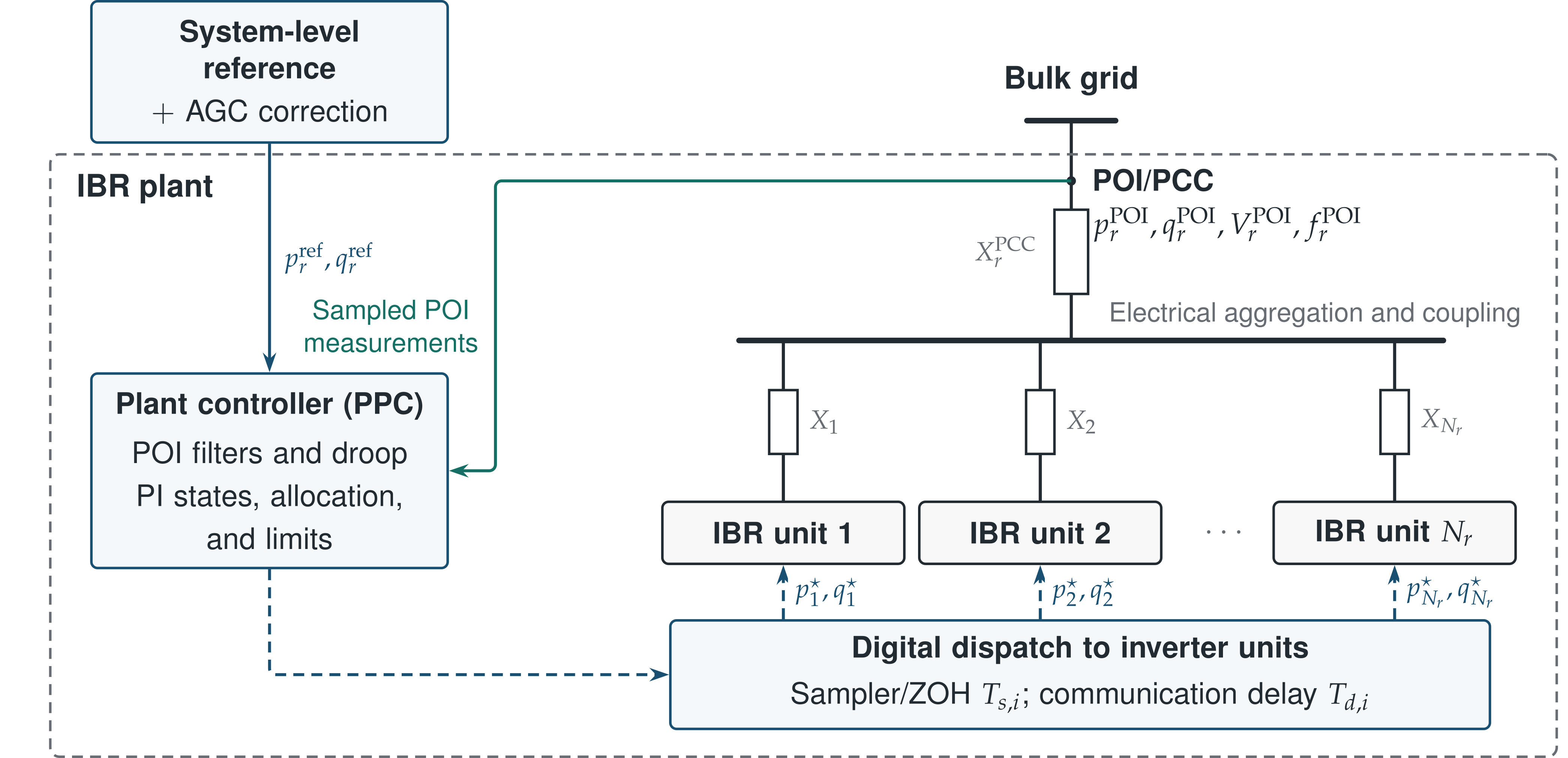}%
}
\caption{Electrical and control architecture of an IBR plant
(conceptual diagram based on
\cite{epri2023modelguide,ghimire2026delaysampling}). The plant controller distributes
commands to inverter units connected through the collector network.
Sampling, communication delays, and inverter-control dynamics
therefore affect the primary-frequency response and AGC tracking.}
\label{fig:ibr_plant_schematic}
\end{figure*}

\subsection{Integrating  IBRs into primary and secondary control}

As discussed in the previous section, IBRs provide primary
frequency response via droop laws; see the models in~\eqref{eq:gfm_model} and~\eqref{eq:gfl_static_droop} for GFM and GFL inverters, respectively. IBRs could also participate in secondary control by accepting
AGC corrections as shown in Fig.~\ref{fig:F_one_area_ace_others}. To model IBR participation in secondary control, let $\cR_k^{\rm agc}$ denote the AGC-participating resources in
area $k$, including IBRs. The nominal  allocation extends to
IBRs as $z_{r,k}=\alpha_{r,k}\eta_k$, $\sum_{r\in\cR_k^{\rm agc}}\alpha_{r,k}=1$,
where $\alpha_{r,k}\geq 0$ and $z_{r,k}$ is the requested correction
in net power injection. Accordingly, the reference point $\bar p_r^{\rm gfm}$ of the IBR can be decoupled as $\bar p_r^{\rm gfm} = \alpha_{r,k}\eta_k +  p_r^{\rm gfm, \star}$, where $p_r^{\rm gfm, \star}$ is the scheduled reference obtained from economic dispatch or an OPF problem; similarly, for GFL one has $\bar p_r^{\rm gfl} = \alpha_{r,k}\eta_k +  p_r^{\rm gfl, \star}$, with $p_r^{\rm gfl, \star}$ determined at the tertiary layer.

However, delivering these services requires sufficient power and
energy capability at the point of interconnection~(POI) over the
required service interval~\cite{lin2020research}. For wind and
photovoltaic plants without storage,  upward regulation
requires maintaining headroom by operating below the maximum available
renewable power; downward regulation is provided by reducing
generation, subject to plant operating limits. For energy storage systems,
regulation capability depends on charging and discharging power
limits, state of charge, and the cumulative
energy required by the scheduled operation and regulation response.
The AGC allocation $\alpha_{r,k}\eta_k$ must respect the resource's
committed regulation range~\cite{pjm2026manual12}. A  research opportunity, beyond the scope of this article,  is to develop resource-level optimization
methods for determining how much regulation capacity can be
committed over a specified service interval, accounting for
renewable forecast uncertainty and power and energy constraints. Alternatively, resources can be pooled in a (dynamic) VPP to collectively match the requirements of the frequency response service~\cite{marinescu2022dynamic,haberle2021control}.

Once the regulation capacity is accounted for, one may argue that this AGC extension would work well when modeling bulk-level IBR plants as an
\emph{individual} unit (i.e., as in the models in~\eqref{eq:gfm_model} and~\eqref{eq:gfl_static_droop}).  For a plant containing \emph{multiple} inverters, the response to primary and secondary commands is typically specified at the \emph{plant level}. As illustrated in Fig.~\ref{fig:ibr_plant_schematic}, the plant controller regulates the aggregate output and distributes power commands among individual inverters. Their local power-control and current-control loops then track these commands~\cite{epri2023genericibrmodels,fan2023operational,fan2024ibrfrequency}. Consequently, plant controller dynamics,  power allocation strategy, and communication enter the  primary-frequency feedback loop. 

The WECC/EPRI generic models represent this hierarchy through converter (\texttt{REGC}), electrical-control (\texttt{REEC}), and plant-controller (\texttt{REPC}) modules~\cite{epri2023genericibrmodels}. Depending on the selected modules and settings, these models include measurement filters, plant-level frequency droop, voltage or reactive-power regulation, deadbands, PI control, and ramp and output limits. Measurements and commands are typically updated digitally, and commands reach individual units through the plant communication network. Representing the updates and transmission delays explicitly results in a sampled-data control system with delayed actuation. For the plants tabulated in~\cite{ghimire2026delaysampling}, sampling periods range from $60$ to $1000$~ms and communication delays from $9$ to $750$~ms. Figure~\ref{fig:ibr_plant_schematic} illustrates the architecture.

For illustration, consider a GFL plant with a PI controller and fixed allocation coefficients $\beta_{r,i}\geq 0$, where $i\in\mathcal I_r$ indexes its inverter units and $\sum_{i\in\mathcal I_r}\beta_{r,i}=1$. Recall that the plant has to emulate droop control at the POI and follow AGC signals. Assume a common sampling period $T_{s,r}>0$ and a constant command delay $T_{d,r}\geq 0$. Measurements are acquired at $nT_{s,r}$ before any command update, and the resulting commands arrive at $t_{r,n}=nT_{s,r}+T_{d,r}$. The above attributes can be captured via the model
\begin{subequations}
\label{eq:IBRplants}
\begin{align}
\bar p_r[n]
&=
p_r^{\star}[n]
+z_{r,k}(nT_{s,r})
-K_{f,r}
\widehat{\Omega}_r[n],
\\
e_r[n]
&=
\bar p_r[n]-p_r^{\rm POI}(nT_{s,r}), \\
\xi_r[n+1]
&=
\xi_r[n]+T_{s,r}e_r[n],
\\
u_r[n]
&=
\operatorname{sat}_{\mathcal U_r}
\bigl(K_{p,r}e_r[n]+K_{i,r}\xi_r[n]\bigr),
\\
\nu_{r,i}(t)
&=
\beta_{r,i}u_r[n],
\quad t\in[t_{r,n},t_{r,n+1}),
\end{align}
\end{subequations}
where $\bar p_r[n]$ combines the scheduled injection
$p_r^{\star}[n]$ (from the tertiary level), the AGC correction
$z_{r,k}(nT_{s,r})$, and the primary control contribution
computed by the plant-level frequency-power droop controller.  The PI state is $\xi_r$, its gains are
$K_{p,r}$ and $K_{i,r}$, and $u_r$ is the aggregate command.
The fixed interval $\mathcal U_r$ is chosen to respect the unit
power limits. Each $\nu_{r,i}$ is
the power reference delivered to the inverter $i$. Equation~\eqref{eq:IBRplants} highlights the droop, PI, allocation,
and communication mechanisms at the IBR plant level. Practical implementations
also require treatment of integrator windup, time-dependent limits,
and coordination with voltage and reactive power control (which is not covered in this article). 

Sampling and communication delay may destabilize the plant control loop (even if the corresponding continuous, delay-free model is stable~\cite{fan2024ibrfrequency,ghimire2026delaysampling}).  An open research opportunity is to develop scalable stability conditions that account for the interactions among IBR plant controllers,  individual controller of inverters, and the transmission network.  Time-delay methods~\cite{gu2003stability}, sampled-data analyses of load-frequency control~\cite{luo2020stability}, and recent IBR-plant studies~\cite{ghimire2026delaysampling} provide relevant tools and initial results.

\subsection{Integrating  data centers into secondary control}

Data centers are envisioned to participate in AGC by adjusting their power
demand~\cite{colangelo2026ai,wierman14ppportunities}. In the model considered in this article, they can
therefore be included in the set of participating resources $\cR_k^{\rm agc}$, as illustrated in
Fig.~\ref{fig:F_one_area_ace_others}. The data center operator determines the regulation capability it
can offer and implements the internal controls needed to track AGC signals~\cite{acun2026investigating}.

Available control actions for power modulation in a data center include workload scheduling  and server
provisioning~\cite{lin2011online}, as well as processor power
capping. Assessing the resulting flexibility requires models
relating power consumption to workload arrivals, processing
rates, and queue evolution. Reducing power can delay job execution,
whereas increasing power through computation requires sufficient
executable work and available computing resources. The admissible
response therefore depends on workload composition and requirements
on waiting times, execution times, and service quality
\cite{zhang2024data}. If cooling systems or on-site storage
participate, their thermal or energy dynamics and operating limits
must also be incorporated.

\begin{figure}[t!]
\centering
\resizebox{\columnwidth}{!}{%
    \includegraphics[width=267.26454pt,height=266.3682pt]{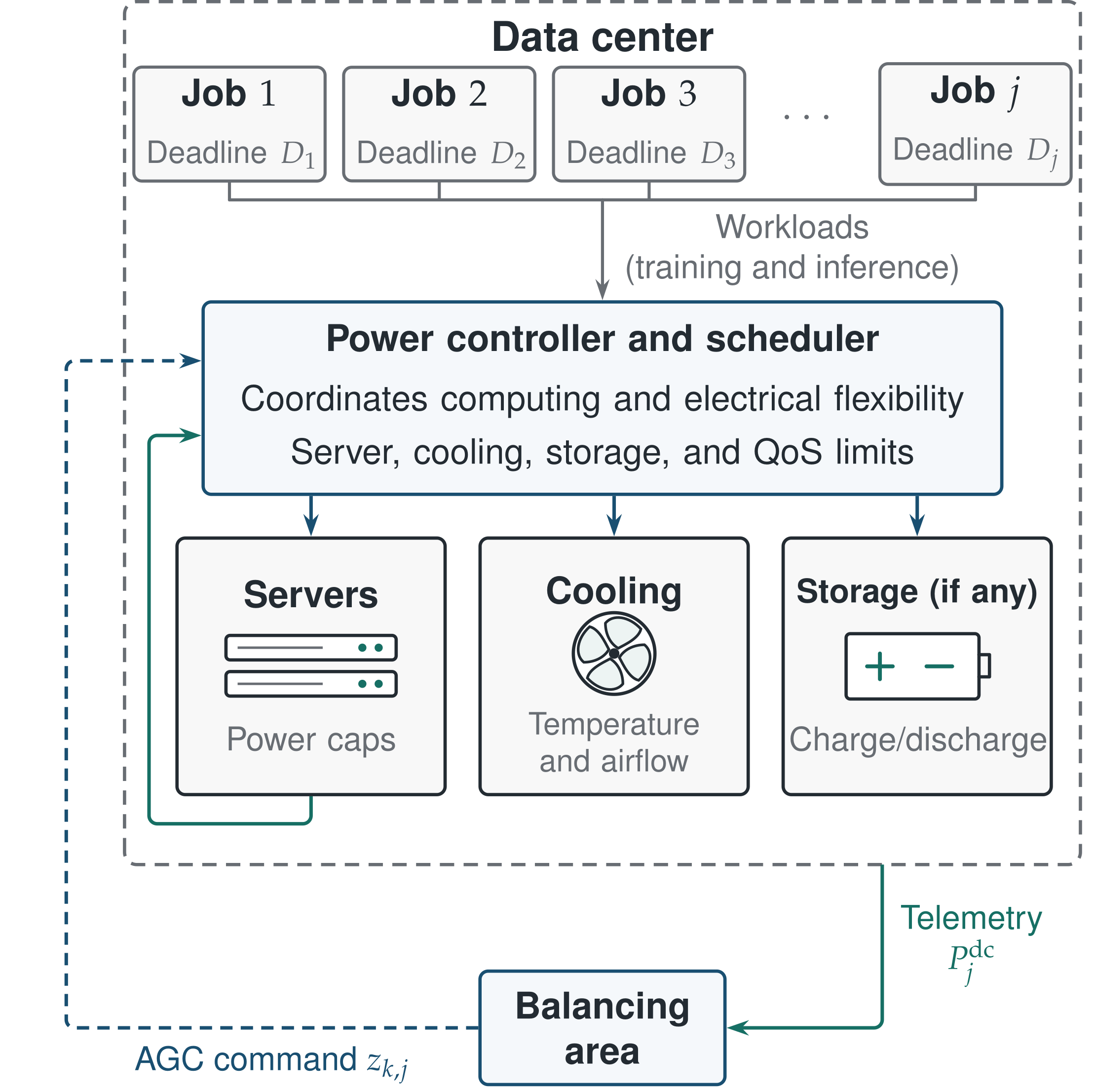}%
}
\caption{Data center participation in AGC. The requested
net-injection correction $z_{r,k}$ corresponds to a consumption
reference $\bar p_r-z_{r,k}$. A power controller and workload
scheduler coordinate computing resources and, where available,
cooling and storage to track this reference subject to computational
and operating constraints~\cite{colangelo2026ai,zhang2024data}.}
\label{fig:F_data_center}
\end{figure}

These constraints enter both the selection of a regulation
commitment and the design of the runtime policy that delivers it. Although these are not problems related to transmission system operation, they are nevertheless of interest to the control community: the design of the runtime policy for a data center is a dynamic power-allocation problem where  the data center controller
must distribute power adjustments among computing resources and,
where available, cooling and storage, while accounting for how
power modulation affects  operating constraints such as the quality of service for workloads~\cite{colangelo2026ai,zhang2022hpc}. Figure~\ref{fig:F_data_center} illustrates this
internal coordination. The inability to track AGC signals will affect the restoration  of the nominal frequency; therefore, the tracking performance of a data center is directly related to robustness,  practical stability, and tracking of transmission system dynamics. Analysis of the effects of tracking errors into the AGC performance is then an additional research line. 

Several studies express the requested consumption of a data center using a baseline
and a regulation capacity:~\cite{zhang2022hpc}
\begin{equation}
p_{{\rm dc},r}^{\rm ref}(t)= p_r^{\rm baseline}+y_r(t)R_r,
\label{eq:dc_agc_flexibility}
\end{equation}
where $p_r^{\rm baseline}$ is the baseline consumption for the service
interval, $R_r\geq 0$ is the committed regulation capacity, and
$y_r(t)\in[-1,1]$ is the normalized regulation signal. If the
baseline equals scheduled consumption, the corresponding
net-injection correction can be written as $z_{r,k}(t)=-R_r y_r(t)$.
This representation specifies the amplitude of the requested
adjustment. This model suggests a research direction: how to use $R_r$
to determine the AGC participation factors $\alpha_{r,k}$. The allocation must account for both the magnitude of each power adjustment
and the duration for which it can be offered. For example, a data center
may initially accommodate a demand reduction but subsequently violate job
deadlines if that reduction persists. This motivates the joint design of flexibility estimation methods, AGC
allocation rules, and local tracking policies, especially when in the AGC mechanism explained in the section ``Area control errors and automatic generation
control'' the balancing area needs to compute participation factors $\alpha_{r,k}$ for various assets including generators, IBRs, and data centers. Such methods should update
the allocation as resource conditions change, redistribute regulation
requests when a resource approaches its limits.

\subsection{Reshaping frequency dynamics}

In the design of the secondary
control layer, the parameters of the primary frequency controllers  are commonly
treated as fixed. An alternative is
to treat the primary control parameters as \emph{design
variables}, with objectives that include shaping the frequency response (e.g., nadir, time constant, etc), 
allocation of the power imbalance, and robustness.

Several contributions illustrate this design perspective. For example, a reduced
aggregate-frequency model was used in~\cite{guggilam2018optimizing} to select synthetic-inertia and droop coefficients of IBRs according to specifications on frequency nadir and
steady-state frequency deviation; the optimization framework
allocates the required response among IBRs in proportion to their
power ratings. Local load-control
laws from a convex optimal load-control problem were derived in ~\cite{zhao2014loadsideprimary}, and 
convergence of the associated network dynamics to an optimal allocation problem was established.
Both generation and controllable loads were considered in~\cite{zhao2014optimalprimary}, designing local frequency
feedback so that closed-loop equilibria solve an economic allocation
problem. 

Other contributions address certificates for stability  and transient
performance, with primary control parameters entering the certificates. For example,~\cite{pates2019robust} developed
decentralized sufficient conditions for small-signal stability based
on local bus dynamics and bounds on incident network coupling. The work in~\cite{paganini2020global} analyzed 
frequency and synchronization performance for heterogeneous
generators. Under proportionality assumptions on parameters,
~\cite{paganini2020global}  separates aggregate frequency deviation from deviations among buses,
obtaining analytical expressions for performance measures that
include frequency nadir, rate of change of frequency, and a
synchronization cost. A dynamic-droop
compensation was developed in~\cite{jiang2021dynamicdroop} that permits transient-response tuning while preserving
the steady-state droop characteristic. Their analysis characterizes
tradeoffs among disturbance response, measurement-noise amplification,
and synchronization performance, with tuning results established
under stated assumptions on the resource dynamics and parameters.

\begin{figure}[t]
    \centering
    \resizebox{0.9\columnwidth}{!}{%
        \includegraphics[width=230.15086pt,height=256.82468pt]{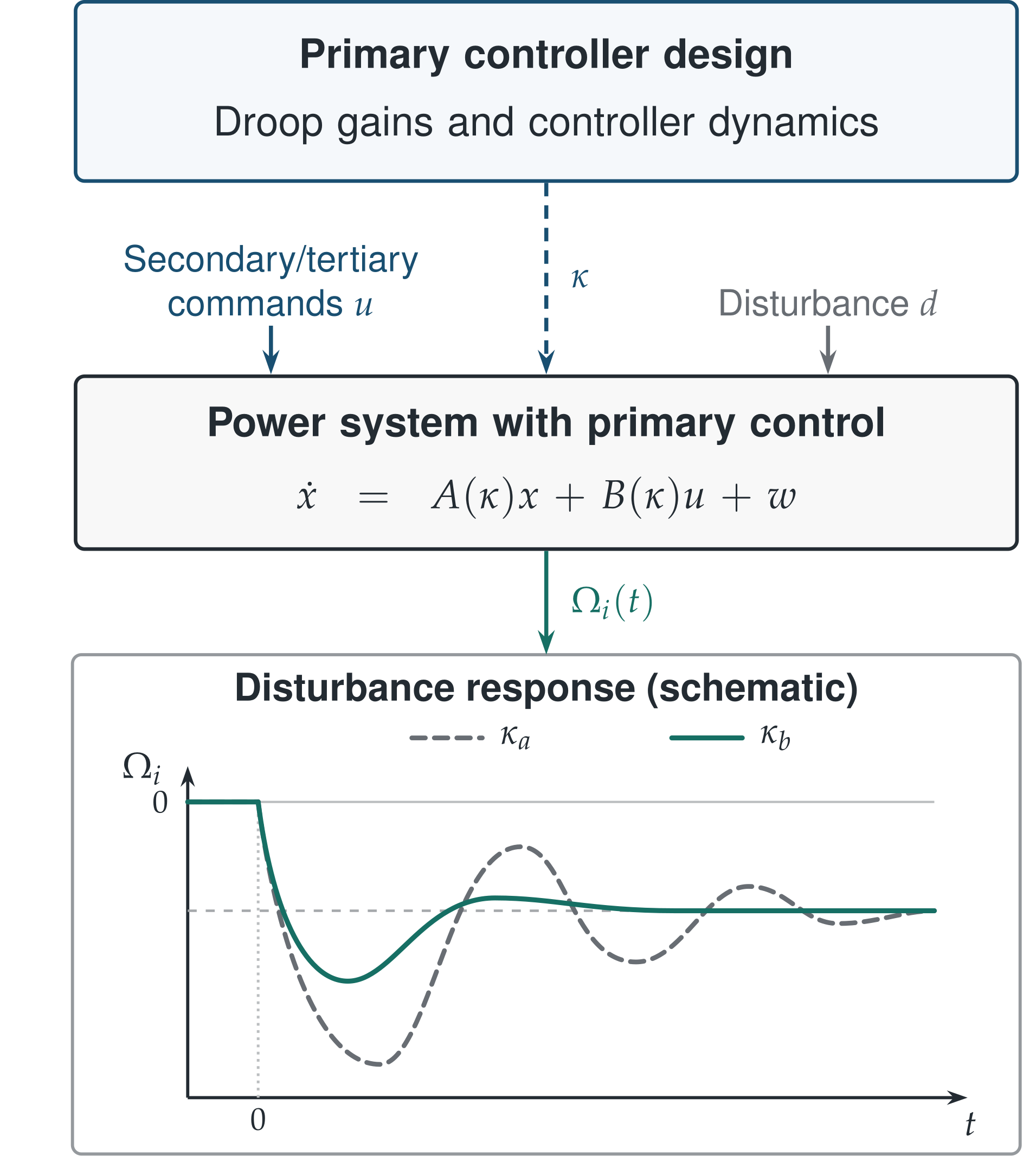}%
    }
    \caption{Reshaping primary frequency dynamics through the
    controller parameters $\kappa$. The schematic responses
    illustrate different nadirs and damping for two tunings
    with the same steady-state droop characteristic. Nominal-frequency
    restoration remains a secondary-control objective.}
    \label{fig:F_reshaping_frequency}
\end{figure}

Using the notation of this article, one example of 
design of primary controllers  can be expressed through a parameterized version of the model~\eqref{eq:state_space_form} as:
\begin{equation}
    \dot x
    =
    A(\kappa)x
    +
    B(\kappa)u
    +
    w,
    \label{eq:primary_control_codesign}
\end{equation}
where $x$ has the same meanings of the previous section, $u$ collects the powers for controllable resources, and $w$ represents the contributions of non-controllable loads and non-controllable resources. The vector $\kappa$ in~\eqref{eq:primary_control_codesign} collects the primary control parameters
available for design, such as droop coefficients.  A  model such as~\eqref{eq:primary_control_codesign} can be derived using the model approximation approach adopted in the article, or via linearization around operating points; a stability certificate for~\eqref{eq:primary_control_codesign} can then be derived accordingly. 

An open research opportunity is to develop scalable synthesis
methods to design $\kappa$ based on well-defined performance and robustness criteria, and that certify stability across a family of detailed network models. The objective is to characterize admissible controller parameters
$\kappa$ together with the operating conditions and network topology covered by the analytical guarantees. When synthesis
relies on a reduced model, the associated error bounds and
time-scale assumptions must remain valid throughout the certified
parameter range. Specific to IBRs, a further challenge is
to translate these parameters $\kappa$ of the reduced models into parameters of the higher-fidelity  converter models. EMT simulations and hardware
experiments can inform the design and analysis.

\subsection{Constraining the frequency dynamics}

A related potential research direction is to augment primary and secondary  controllers with auxiliary modules that intervene only when operational limits (such as frequency limits or line flow limits) are at risk.
 Control Barrier Function (CBF) based safety filters \cite{ames2019cbf} can minimally modify a nominal
control action to render a prescribed safe set forward invariant; its application to frequency control has been recently explored in \cite{dall2026local} via a two-time-scale dynamic
implementations that approximate centralized safety filters using only local
state and derivative estimates. While \cite{dall2026local} represents a preliminary framework applicable to a network with GFL inverters and generators at every bus, extending these control design and analysis  approaches  to transmission systems with GFM inverters and  dynamic loads remains an open
problem of possible interest.  Key questions include whether these CBF-based filters can be implemented in GFM inverters and loads, and how to construct compatible
local safety conditions, guarantee feasibility under actuator and energy
limits, and preserve closed-loop stability when the filters interact with
different device dynamics \cite{zhang2019transient}.

\begin{figure}[t]
    \centering
    \resizebox{0.9\columnwidth}{!}{%
        \includegraphics[width=247.86569pt,height=212.61948pt]{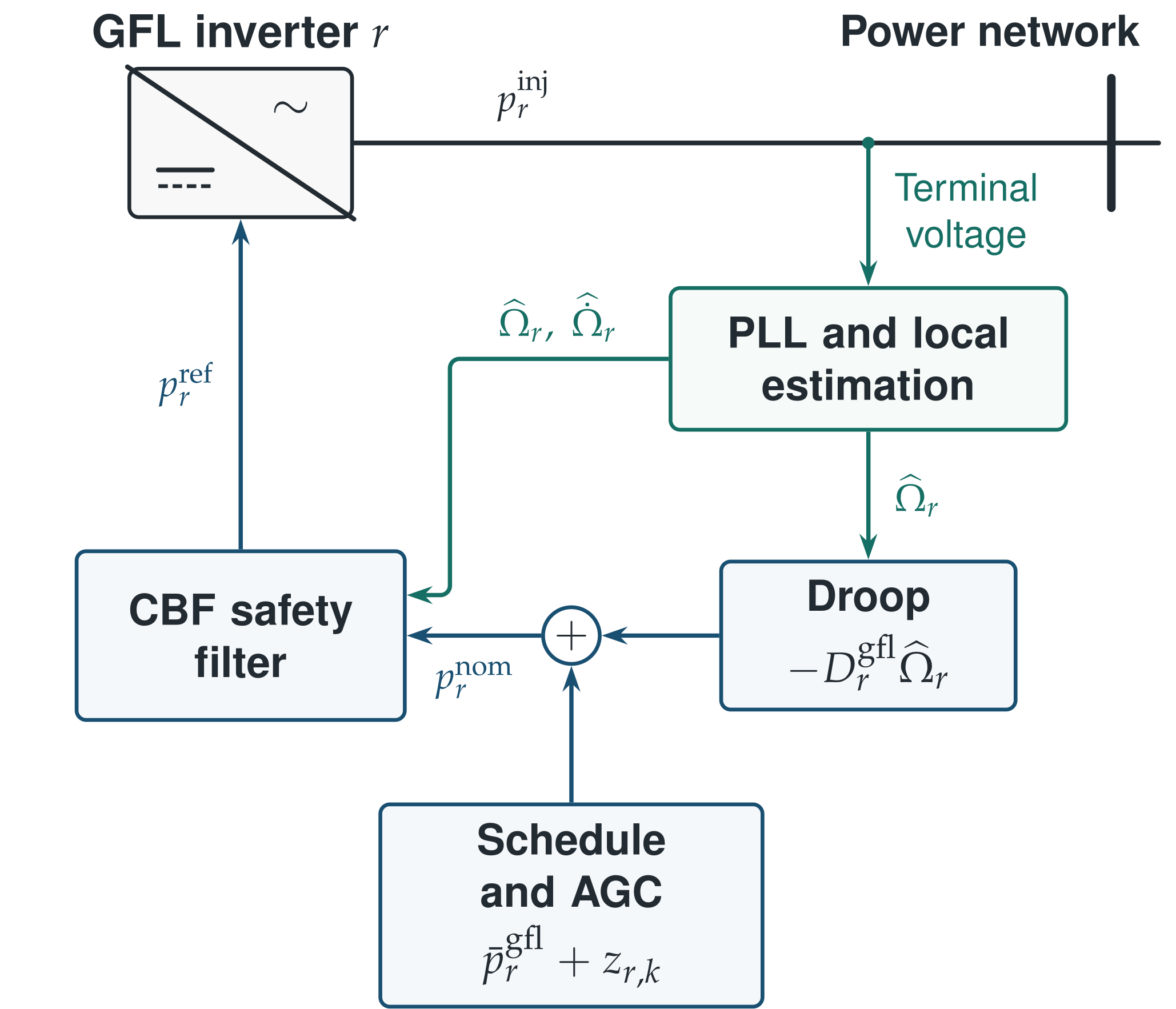}%
    }
    \caption{Local CBF safety filter for a GFL inverter.
    The nominal active power reference combines the scheduled
    setpoint, AGC correction, and droop response. The filter
    uses local frequency and derivative estimates.}
    \label{fig:F_cbf}
\end{figure}

Embedding a dynamic safety filter into a GFL inverter also raises measurement and
implementation questions. The phase-locked loop (PLL)
already produces a local estimate of frequency, and its internal dynamics may be used together with an observer or filtered differentiator to estimate
$\dot{\omega}_i$ \cite{golestan2017pll}. Direct numerical differentiation of the
PLL frequency estimate, however, can amplify measurement noise. The PLL,
derivative estimator, and dynamic safety filter must therefore be co-designed
to account for bandwidth, phase lag, sampling, and estimation errors.

Related concepts can be introduced at the secondary control layer. A CBF-based
filter could supervise the commands generated by AGC and modify them when necessary to prevent violations of
transmission line limits during events \cite{zhao2014loadsideprimary}. Within the linearized network model considered in this article, transmission line flows (and, in particular, flows in the tie lines)
are affine functions of the secondary control variables; a CBF-based filter acting
on the updates of these variables can therefore enforce line-flow
limits. Open research questions
include robustness to linearization errors and unmodeled dynamics,
sampled-data implementation, feasibility under simultaneous network and
resource constraints, and distributed implementations across balancing areas.


\section{Research Opportunities Beyond the Existing Layered Architecture}

The separation of primary frequency control, secondary control,
and economic dispatch evolved gradually as power systems developed, as explained in the sidebar ``History of Bulk Power System Control and Operations''.  This section examines
opportunities to reconsider the temporal and spatial structure of control frameworks. A first motivation for such reconsideration is to improve economic efficiency, robustness, and equilibrium tracking by coordinating decisions across layers; the article refers  to this as \emph{time-scale compression}. For example, there may be economic benefits to coordinating frequency regulation and optimization of the grid operating point as operating conditions change, compressing the secondary and tertiary layers~\cite{sawant2026integrated}.  Such architectures may enable more rapid response to changes in generation and demand, reduce the balancing burden assigned to regulation reserves, and coordinate economic objectives with network and resource constraints.

The second motivation is to reconsider how
control responsibilities are distributed among
transmission operators, distribution operators, and resource
aggregators. For example, the resources available at the distribution level can be aggregated in the form of a virtual power plant (VPP) so that they can collectively respond to a transmission-level request~\cite{marinescu2022dynamic,haberle2021control} while satisfying their limits and the distribution network's
constraints.
Revising these coordination interfaces provides an opportunity to incorporate such dependencies into system-level decisions.

It goes without saying that a wide variety of technical approaches can be brought to bear to enable compression of timescales and to develop optimized control architectures with different spatial structures. The next subsection outlines one such framework for analyzing and addressing these design problems.

\subsection{From Existing Controllers to New Architectures}

\begin{figure*}[t]
    \centering
        \resizebox{0.8\textwidth}{!}{%
\includegraphics[width=465.22939pt,height=142.47185pt]{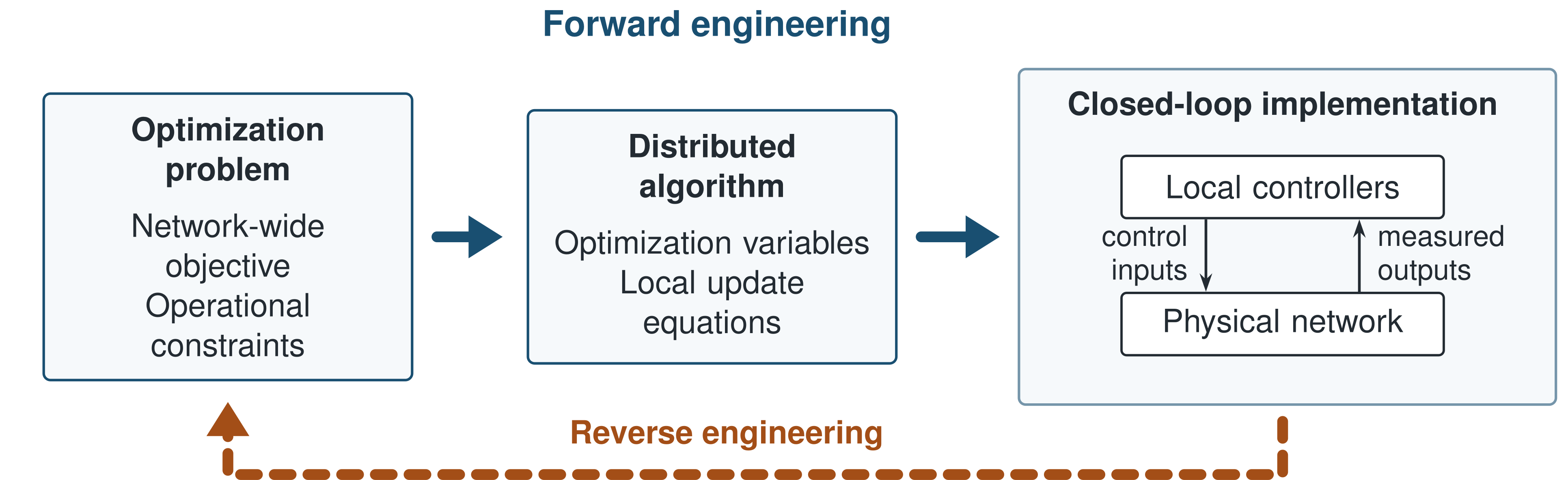}%
    }
    \caption{Optimization as a framework for interpreting and
    designing feedback systems. Forward engineering proceeds
    from an optimization problem to an algorithm and its
    closed-loop realization. Reverse engineering identifies
    an optimization problem associated with the equilibria
    of an existing feedback system.}  \label{fig:F_controller_optimization_philosophy}
\end{figure*}

Feedback laws which bring a dynamical system to equilibrium can sometimes be fruitfully reinterpreted as solvers for an optimization problem whose optimality conditions are met at the closed-loop equilibrium. Low, Paganini, and Doyle developed
this approach for Internet congestion control~\cite{low2002internet}:
source rates are allocation variables, link capacities impose
shared constraints, and congestion feedback coordinates the
sources. Their analysis connects the equilibrium allocation to an
optimization objective, and subsequently uses this connection to guide protocol
design and analysis. Interpreting an existing controller in this way is referred to as
\emph{reverse engineering}. Constructing a controller from a
specified optimization problem through decomposition methods is instead termed \emph{forward engineering}.
Figure~\ref{fig:F_controller_optimization_philosophy} illustrates
these two directions.

\begin{sidebar}{AGC as a Dual-ascent Method for a Generalized Nash Equilibrium}
\sdbarinitial{T}his sidebar gives an (apparently, novel) interpretation of classical AGC as a
dual-ascent algorithm associated with a generalized Nash equilibrium (GNE) problem \cite{facchinei2007generalized}. This interpretation is an instance of the reverse engineering methodology described in the main body.

The notation follows that introduced in the section ``Two-time-scale models and methods for secondary frequency control''. For
area $k$, let
$z_k:=\operatorname{col}(z_{k,i})_{i\in\cG_k^{\rm sg}}$ denote candidate AGC
corrections, where $z_{k,i}$ is the correction for generator $i$ in area $k$. For a fixed vector $z:=\operatorname{col}(z_1,\ldots,z_{N_a})$ of AGC
corrections, based on \eqref{eq:reduced_agc}--\eqref{Eq:H_agc_explicit} let
\[
\mathrm{ACE}^{\star}(z) \coloneqq 
H^{\rm agc}\begin{bmatrix}
\mathbf{1}^{\top}z_1 \\ \vdots \\ \mathbf{1}^{\top}z_{N_a}
\end{bmatrix}+q^{\rm agc}.
\]
denote the vector of equilibrium ACEs. 

For area $k$, consider the equilibrium selection problem
\begin{equation}
\label{eq:sidebar_area_sharing_problem}
\begin{aligned}
    \minimize_{z_k}
    \quad &
    J_k(z_k):=
    \sum_{i\in\cG_k^{\rm sg}}\frac{1}{2\alpha_{k,i}}z_{k,i}^2
    \\
    \subto \quad &
    \mathrm{ACE}^{\star}_k(z_k,z_{-k})=0,
\end{aligned}
\end{equation}
where $z_{-k}$ represents the vector of AGC corrections in all \emph{other} areas and where $\alpha_{k,i} \geq 0$ are participation factors satisfying $\sum_{i\in\mathcal{G}_{k}^{\rm sg}} \alpha_{k,i} = 1$. In \eqref{eq:sidebar_area_sharing_problem}, each area seeks to minimize cost of provided power, subject to regulation of its local ACE. However, the allocation problems \eqref{eq:sidebar_area_sharing_problem} are coupled, because each ACE
depends on the AGC corrections selected by all other areas due to interconnected nature of the network. The relevant optimal equilibrium concept is a GNE: a selection $\{z_k\}_{k=1}^{N_a}$ which simultaneously solves \eqref{eq:sidebar_area_sharing_problem} for every individual area.

To connect the allocation problem \eqref{eq:sidebar_area_sharing_problem} to the AGC controller, first write the $k$th steady-state ACE equation as
$$
    {\rm ACE}^{\star}_k(z_k,z_{-k})
    =
    \sigma_k\mathbf{1}^\top z_k
    +
    \psi_k(z_{-k}),
$$
where $\sigma_k:=[H^{\rm agc}]_{kk}$ and
\[
\psi_k(z_{-k})
:=
\sum_{j\neq k}[H^{\rm agc}]_{kj}\mathbf 1^\top z_j+[q^{\rm agc}]_k.
\]
The Lagrangian of \eqref{eq:sidebar_area_sharing_problem} is
$$
    \mathcal{L}_k(z_k,\lambda_k)
    =
    J_k(z_k)+\lambda_k{\rm ACE}^{\star}_k(z_k,z_{-k}).
$$
For fixed $\lambda_k$, the Lagrangian can be minimized over the primal variable, and the minimization separates element-wise into
\begin{equation}\label{Eq:PrimalMin}
    z_{k,i}(\lambda_k)
    =
    \arg\min_{\zeta}
    \left\{
        \frac{1}{2\alpha_{k,i}}\zeta^2
        +
        \lambda_k\sigma_k \zeta
    \right\}
\end{equation}
with unique minimizer $z_{k,i}(\lambda_k) = -\alpha_{k,i}\sigma_{k}\lambda_k$. A dual-ascent algorithm for the dual problem can now be written as
$$
    \kappa_k\dot\lambda_k={\rm ACE}^{\star}_k(z_k(\lambda_k),z_{-k}), \quad \kappa_k > 0,
$$
or equivalently, with the the change of variables $\eta_k:=-\sigma_k\lambda_k$,
$$
    \tau_k\dot\eta_k=-{\rm ACE}^{\star}_k(z_k(\eta_k),z_{-k}),
    \quad
    \tau_k:=\kappa_k/\sigma_k,
$$
where $z_{k,i}(\eta_k) = \alpha_{k,i}\eta_k$. Replacing the steady-state expression ${\rm ACE}^{\star}_k(z)$ by the \emph{measured
real-time ACE} now recovers the real-time AGC controller
$$
    \tau_k\dot\eta_k=-{\rm ACE}_k.
$$
This interpretation of AGC as a GNE-seeking method extends easily to the case of saturated operation, wherein the minimization in \eqref{eq:sidebar_area_sharing_problem} is subject to the constraints $\underline z_{k,i}^{\rm sg} \leq z_{k,i} \leq \overline z_{k,i}^{\rm sg}$; in this situation, the primal minimization \eqref{Eq:PrimalMin} simply recovers the saturated control law $z_{k,i}
    =
    \sat_{[\underline z_{k,i}^{\rm sg},\overline z_{k,i}^{\rm sg}]}
    \left(\alpha_{k,i}\eta_k\right)$. An open research problem is to exploit the forward engineering methodology and develop new AGC architectures based on this generalized Nash equilibrium interpretation; one such design is presented in the subsection ``Example \#1: Economic secondary control via feedback optimization''.
\end{sidebar}

Earlier work has interpreted frequency control through a single
system-wide optimization problem and used this interpretation
to develop new control architectures. For example,
\cite{li2015connecting} interprets the dynamics of a reduced
power-system model with AGC as a partial primal-dual algorithm.
This interpretation motivates modifications to the control law
that incorporate economic dispatch into AGC. The framework
in~\cite{chen2017reverse} similarly reverse engineers primary
frequency dynamics and uses the resulting optimization
interpretation to design distributed controllers that combine
frequency restoration with economic optimization.

Related work incorporates economic objectives directly into
primary frequency control. In~\cite{zhao2014loadsideprimary},
frequency-responsive load controllers are designed so that their
interconnection with a reduced network model implements a
partial primal-dual algorithm for an optimal load-control
problem. The formulation in~\cite{zhao2014optimalprimary}
coordinates generation and controllable loads through local
frequency feedback, with synchronous equilibria that minimize
an aggregate cost subject to power balance and resource limits.

The sidebar ``AGC as a Dual-ascent Method for a Generalized
Nash Equilibrium'' provides a concrete example of reverse engineering. For the AGC mechanism described in the section
``Two-time-scale Models
and Methods for Secondary Frequency Control'', each area is associated with a quadratic
power-allocation objective whose weights are determined by
the participation factors $\alpha_{r,k}$. Its constraint
requires zero steady-state ACE and depends on the decisions
of other areas through the interconnected network. This
coupling gives rise to a generalized Nash game.
Under the assumptions described in the sidebar,
the AGC integral dynamics admit an interpretation as coupled
dual updates, and their equilibria satisfy the corresponding
game's optimality conditions. This interpretation makes
explicit both the allocation policy within each area and
the coupling among areas.

Reverse engineering provides a starting point for architectural
design by identifying which optimization updates the existing
controllers implement. Modifying and redistributing these updates
can then produce new control architectures. For time-scale compression, one can replace AGC's fixed allocation based on participation factors with updates of resource power references that incorporate marginal costs and network constraint
multipliers, while keeping ACE dual updates for
frequency regulation. Making the scheduled operating points
variables of these coupled dynamics brings economic dispatch
and secondary regulation into a common feedback
layer~\cite{sawant2026integrated}. The architectural change
therefore concerns which decisions are updated within the
feedback loop, as well as their update rates. Bringing communications and coordination at the primary layer, if feasible, can also enable a time-scale compression between primary and secondary layers. 

For spatial coordination, reverse engineering could start from the
existing control laws and information exchanges among transmission
operators, distribution operators, and resource
owners~\cite{ilic2007hierarchical}. One seeks an optimization
or game formulation consistent with these interconnected
controllers. For example, in a transmission-distribution
architecture, this interpretation can reveal whether the
upstream dispatch accounts for the costs and network constraints
governing the downstream allocation. Comparing the resulting
equilibrium conditions with those of a specified joint
optimization problem can identify missing coordination terms
and explain why separate operator decisions yield a different
allocation~\cite{lecadre2019game}. The research opportunity is
to determine which changes to local objectives, controller
updates, or exchanged signals would eliminate this discrepancy.
Designing these changes, potentially through optimization
decomposition, is the subsequent
forward engineering step.

\subsection{Feedback Optimization as a  Design Framework}
Feedback optimization implements optimization updates in feedback
with a physical system, with the objective of regulating its operation
toward solutions of a prescribed optimization problem
\cite{Jokic2009controller,brunner2012feedback,colombino2019online,hauswirth2020timescale,lawrence2021linear,tang2017real,bianchin2021time}. Measurements of physical output enter the updates of decision
and coordination variables, replacing the corresponding model of the system.  The sidebar ``What Is Feedback Optimization?''
presents a representative formulation and explains how the measurements
enter the controller.

\begin{sidebar}{What is Feedback Optimization?}
\label{sidebar:what_is_fo}

\setcounter{sfigure}{3}
\renewcommand{\thesfigure}{S\arabic{sfigure}}

\sdbarinitial{F}eedback optimization refers to controllers that drive a physical system towards an optimal equilibrium point in closed loop; the reader is referred to the early works in~\cite{Jokic2009controller,brunner2012feedback}, the more recent works in~\cite{colombino2019online,hauswirth2020timescale,lawrence2021linear,tang2017real,bianchin2021time}, and the survey~\cite{hauswirth2024optimization} for some representative reading material. Consider
a system
\begin{equation}
\dot x=f(x,u,w),
~~~~
y=g(x,u,w),
\label{eq:sidebar_fo_plant}
\end{equation}
where $u$ is a controllable reference, $y$ is a measured output, and $w$
collects unknown disturbances. Suppose that, for fixed $(u,w)$, the system has
an exponentially stable equilibrium $x^{\star} = h_x(u,w)$ with steady-state output $y^\star=h(u,w)$. An equilibrium selection  objective can be expressed as
\begin{equation}
\label{eq:sidebar_fo_problem}
\begin{aligned}
\underset{u\in\mathcal U}{\operatorname{minimize}}
~~&
\phi(u)+\psi\bigl(h(u,w)\bigr)
\\
\operatorname{subject~to}
~~&
\ell\bigl(h(u,w)\bigr)\leq0
\end{aligned}
\end{equation}
where the functions $\phi$ and $\psi$ may represent equilibrium operating costs or performance
objectives, while $\mathcal U$ and $\ell$ describe actuator and equilibrium
output constraints.

In a feedforward solution approach (Fig.~\ref{fig:F_sidebar_feedback_optimization_figure}(a)), 
\eqref{eq:sidebar_fo_problem} is solved using a model and a
forecast of $w$, and the resulting optimizer is fed as a reference to a conventional set-point tracking controller. Feedback optimization (Fig.~\ref{fig:F_sidebar_feedback_optimization_figure}(b)) instead integrates the optimization and tracking modules, often by adapting an optimization algorithm as a dynamic controller. For example, a
measurement-based projected primal--dual flow applied to solving~\eqref{eq:sidebar_fo_problem} has the schematic form
\begin{subequations}
\label{eq:sidebar_fo_flow}
\begin{align}
\tau\dot u
&=
\proj_{\mathcal U}
\left(
u+\eta_u d_u(u,y,\mu)
\right)-u,
\\
\tau \dot\mu
&=
\proj_{\mathbb R_+}
\left(
\mu+\eta_\mu\ell(y)
\right)-\mu,
\end{align}
\end{subequations}
where $\tau,\eta_u,\eta_{\mu} > 0$ are gains, $d_u$ is a descent direction for the Lagrangian of \eqref{eq:sidebar_fo_problem}, and $\mu$ contains the
multipliers associated with the output constraints. The measured output
$y(t)$ replaces the steady-state map $h(u,w)$ in the controller
updates. Consequently, changes in $w$ are reflected through feedback rather than utilizing
forecasts or disturbance observers.

\sdbarfig{%
  \resizebox{1.0\linewidth}{!}{%
    \includegraphics[width=315.32793pt,height=225.05676pt]{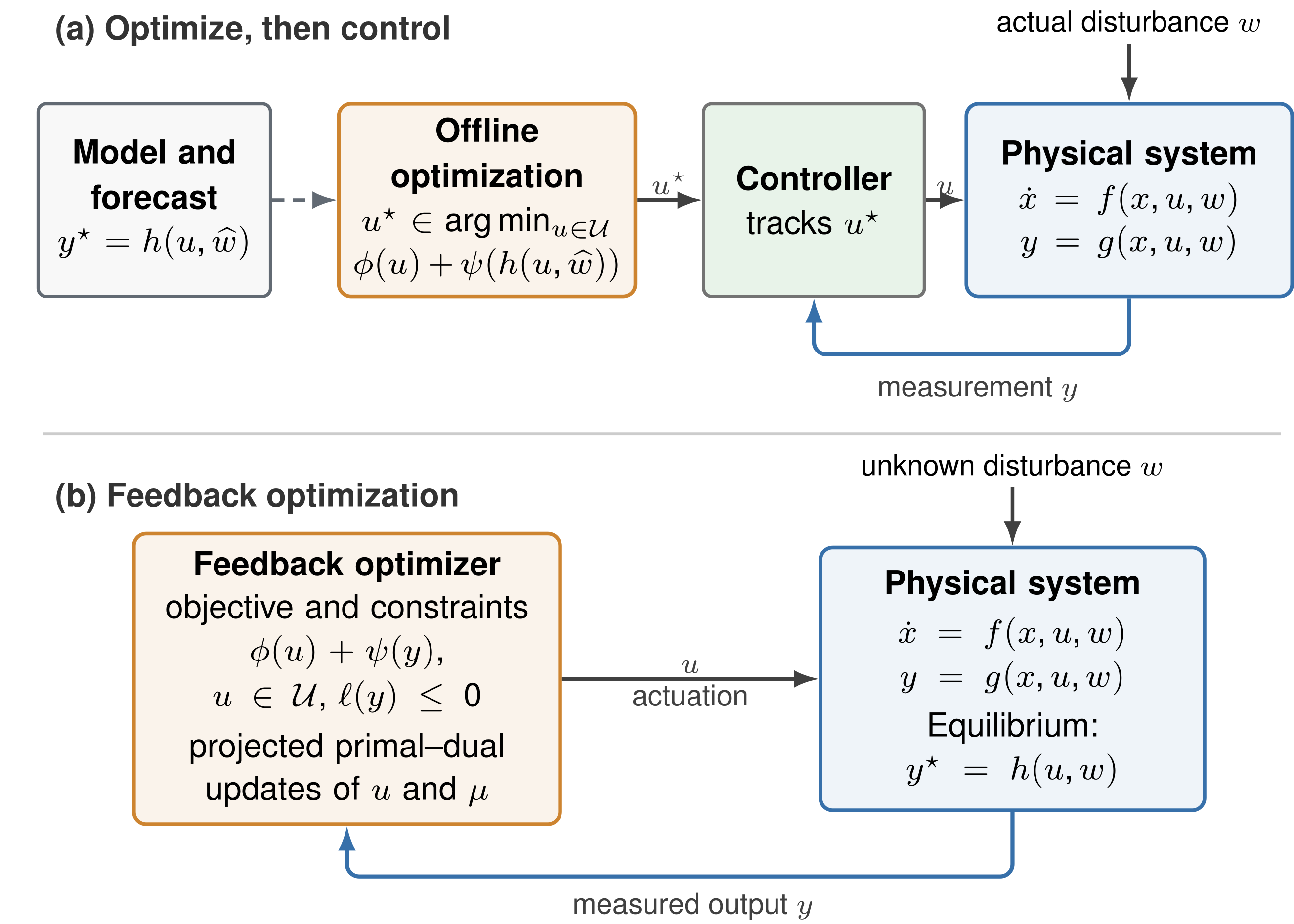}%
  }%
}{%
  Comparison of conventional optimize--then--control and feedback-optimization
  architectures; the latter is customized for the algorithm~\eqref{eq:sidebar_fo_flow} as an example. \label{fig:F_sidebar_feedback_optimization_figure}%
}

Closed-loop stability of~\eqref{eq:sidebar_fo_plant},\eqref{eq:sidebar_fo_flow} can be studied using singular perturbation arguments; see, for example, Lyapunov-based and contraction theory based approaches in~\cite{hauswirth2020timescale} and~\cite{cothren2026online}, respectively, as examples. Under suitable convexity, regularity, and stability conditions,
equilibria of~\eqref{eq:sidebar_fo_flow} satisfy the optimality conditions of
\eqref{eq:sidebar_fo_problem}. When disturbances vary with time, the
controller instead tracks a moving optimizer~\cite{bianchin2021time}.

Feedback optimization differs from model predictive control (MPC) in that FO targets only steady-state optimization of a pre-stabilized system, while MPC pursues stabilization and optimization of the transient response. As a result, FO designs require less model information, typically  only the sensitivity $\frac{\partial h}{\partial u}$ of system equilibrium map $h$, which is embedded in the descent direction $d_u$ in \eqref{eq:sidebar_fo_flow}. Feedback optimization is also distinct from other online optimization methods such as extremum seeking, as no dithering is used in FO; see \cite{hauswirth2024optimization} for more information.

\end{sidebar}

As a forward engineering framework, feedback optimization provides
a way to design control architectures from specified objectives
and constraints formalized in an optimization problem. An optimization problem does not determine a unique
controller: primal-dual dynamics~\cite{bianchin2021time}, projected gradient
methods~\cite{hauswirth2020timescale}, proximal augmented Lagrangian
methods~\cite{ding2020global},  alternating direction method of multipliers~\cite{boyd2011distributed}, and safe gradient
flows~\cite{allibhoy2023control} provide different ways to design a control architecture. In fact, the algorithm and its
decomposition specify the variables updated by each controller
and the information exchanged among controllers. Their feedback-based
implementation must account for the available measurements and
the dynamics of the power system.

The potential of feedback optimization for real-time power grid operations has been demonstrated by employing the proposed forward engineering procedure on standard OPF problems.
This design approach yields fast re-dispatch actions for congestion management and economic rescheduling, that can be employed both at the transmission and distribution level \cite{dall2016optimal,Gan2016,tang2017real,hauswirth_online_2017}.
Since those seminal papers, feedback optimization for economic redispatch has been validated in experiments \cite{Ortmann2020}, deployed on distribution feeders \cite{Ortmann2023}, and demonstrated via numerical experiments on real subtransmission grid models \cite{ortmann_subtransmission_2026}.
However, all these works live on the quasi-static grid operation, and do not consider the primary and secondary frequency dynamics that we reviewed.

For a continuous-time implementation,
a representative 
interconnection with the power system model considered
in this article is
\begin{subequations}
\label{eq:feedback_architecture}
\begin{align}
    \dot{\zeta} &= \varepsilon\mathcal{C}(x,\zeta), \\
    \dot{x} &= Ax + Bu + w, \qquad u=S\zeta,
\end{align}
\end{subequations}
where $x$ includes angles, frequency deviations, and mechanical powers, $w$ collects the effect of non-controllable loads,
and $\zeta$ collects the optimizer states, including power references for the controllable resources (generators, IBRs, data centers, etc)
and any dual or coordination variables. The matrix $S$ selects
the references applied to the resources. The map $\mathcal{C}$ specifies
the optimization updates and depends on $x$ through the available
measurements. When power references are the only optimizer states,
$\zeta=u$ and the controller reduces to
$\dot{u}=\varepsilon\mathcal{C}(x,u)$.

The partition of $\mathcal{C}$ into local updates (at, e.g., generators, IBRs, data centers, etc), centralized updates (at, e.g., the controller for the balancing area) and the information
required by each update describe the spatial organization of the 
control  architecture. For example, resource controllers may update power
references locally, while coordinating controllers update
multipliers for shared network constraints (e.g., line flow limits, tie-line exchanges, etc). Singular perturbation theory~\cite{kokotovic1976singular} provides
a framework for establishing sufficient conditions on $\epsilon$ for stability of the interconnection with the power systems with resources implementing primary control. Results based on Lyapunov
analysis~\cite{hauswirth2020timescale} and contraction
theory~\cite{cothren2026online} quantify admissible gains $\epsilon$ under
given stability and regularity assumptions.

This approach offers opportunities to combine secondary frequency control and
economic dispatch within a common feedback controller~\cite{colombino2019online}, reducing the separation between secondary and tertiary decisions.  The associated control problem is to establish stability and
tracking guarantees for the interconnection of the form~\eqref{eq:feedback_architecture}. While existing results \cite{colombino2019online,bianchin2021time,hauswirth2020timescale,lawrence2021linear} provide starting points
for convex problems coupled to reduced models such as the one in~\eqref{eq:plant_with_agc}, open research problems arise when integrating resources such as IBR plants and data centers: sampling, communication delays, and changes in resource capability involve time-varying, delayed and sample-data reformulations of the system~\eqref{eq:feedback_architecture}. The section ``Examples of New Control Methods and Architectures'' will provide detailed examples on how feedback optimization can be used to modify the existing AGC-based secondary control layer.

A second research opportunity is to combine bi-level optimization and
feedback optimization to: (i) enable 
transmission-distribution coordination, (ii) allow aggregations of resources forming a VPP to dispatch individual resources, and (iii) account for internal constraints of resources. A bi-level formulation is appropriate when a power system operator selects aggregate power requests or
incentives (upper-level), while anticipating the optimal responses of independently
operated resources or networks (lower level). 
The feedback optimization design framework proves to be very versatile, as it allows to model a system response that goes beyond the physical transient of the devices: the lower-level problems can describe
the automated allocation of power within a VPP or distribution system,
accounting for local economic objectives, resource capabilities, and
 network constraints. Measurements of power exchanges and network
conditions can enter the feedback updates at both levels.
Related work has combined bilevel games and feedback optimization
for incentive-based transmission-distribution voltage
support~\cite{jiang2025voltage} and load curtailment~\cite{jiang_incentive-based_2026}. The section ``Examples of New Control Methods and Architectures'' will provide an example  on how to enable coordination across power systems operators and VPPs.

\section{Examples of new control methods and architectures}

Building on the authors' prior work and expertise, this section presents concrete examples of new architectures. The first example  involves the re-design of a secondary control method via feedback
optimization; this method is compatible with the present layered structure. The second and third examples go beyond the existing layered structure and seek to achieve time-scale compression: feedback optimization that combines secondary regulation with economic dispatch, and disturbance feedback control that brings selected secondary actions onto the time scale of the primary dynamics. The fourth example shows how to integrate VPPs ny leveraging bilevel optimization. 

\subsection{Example \#1: Congestion-aware AGC via feedback optimization}

This section uses feedback optimization to derive a line constraint-aware version of AGC. The formulation will preserve the existing time-scale separation between market dispatch and real-time
control, and can be implemented using  measurements already available to
the operator. While transmission line constraints are typically respected through appropriate dispatch of generation, this design transfers some of this responsibility into the secondary-control layer, enabling dynamic reallocation of reserves to manage transmission line congestion.  

Following the same logic as in the sidebar ``AGC as a dual-ascent method for a generalized Nash equilibrium,'' the method will be developed for a single balancing area, and the decisions of other balancing areas are suppressed for notational simplicity. Simultaneous implementation of the method by all balancing areas will lead to a GNE interpretation of the resulting equilibrium.

Consider a balancing area~$k$, and assume for generality that synchronous generators, GFL inverters, and GFM inverters are all eligible to participate in the secondary control. The market layer (i.e., economic-dispatch or OPF) provides the scheduled active-power references
\begin{equation}
\label{eq:ofo_market_dispatch}
u_k^{\rm mkt}
:=
\col\left(
p_k^{{\rm sg},\star},
p_k^{{\rm gfm},\star},
p_k^{{\rm gfl},\star}
\right)
\end{equation}
for the resources in the area. The feedback optimization layer does not recompute this dispatch, but determines a
real-time correction
\begin{equation}
\label{eq:ofo_reference_decomposition}
z_k:=u_k-u_k^{\rm mkt}
=
\col\left(
z_k^{\rm sg},z_k^{\rm gfm},z_k^{\rm gfl}
\right),
\end{equation}
where
$u_k:=\col(\bar p_k^{\rm sg},\bar p_k^{\rm gfm},\bar p_k^{\rm gfl})$
collects the references sent to the participating resources in area~$k$. The system with primary control can therefore be written as
$$
\dot x =
Ax+\sum_{a=1}^{N_a}B_{z,a}z_a+w^{\rm ext}(t),
$$
where $B_{z,k}$ contains the columns of the input matrix associated with the
resources controlled within area~$k$, and the term $w^{\rm ext}(t)$ collects non-dispatchable loads,
fixed injections (from non-participating IBRs), and market setpoints; see~\eqref{eq:state_space_form}. For brevity, define  $w_k = \sum_{a=1, a \neq k}^{N_a}B_{z,a}z_a+w^{\rm ext}(t)$, so that the system can be represented as
\begin{equation}
\label{eq:ofo_plant}
\dot x=Ax + B_{z,k}z_k + w_k.
\end{equation}

\begin{figure*}[t!]
    \centering
    \resizebox{1.0\textwidth}{!}{%
        \includegraphics[width=700.84163pt,height=258.09705pt]{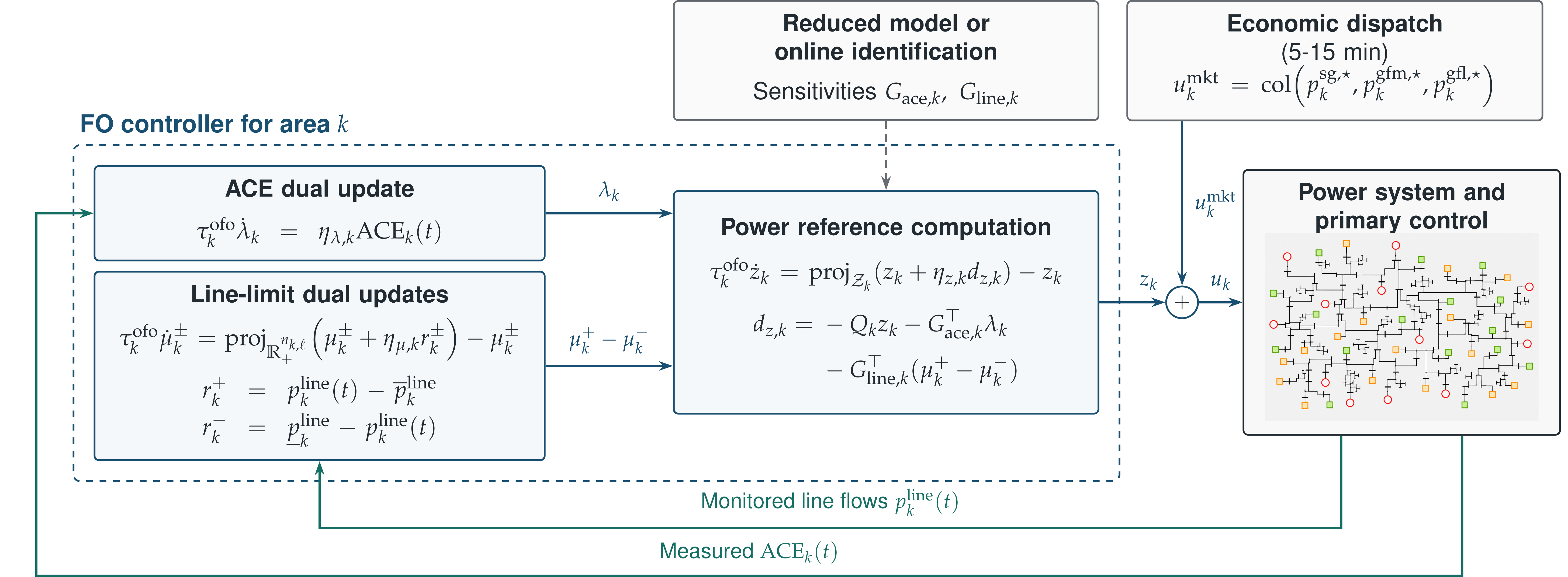}%
    }
    \caption{Closed-loop implementation of the
    feedback optimization (FO) controller~\eqref{eq:ofo_primal_dual_flow} in area~$k$. The optimization state $z_k$ provides real-time
    corrections to the participating-resource references, while the physical
    system returns the measured ACE and measured or estimated monitored line
    flows. The residual values are therefore obtained from feedback, whereas
    the sensitivities $G_{{\rm ace},k}$ and $G_{{\rm line},k}$ are supplied by
    a reduced model, online linearization, or data-driven identification. }
    \label{fig:ofo_closed_loop}
\end{figure*}

For constant $z_k$ and $w_k$, internal stability of the transmission system dynamics with primary frequency control defines
the equilibrium map
\begin{equation}
\label{eq:ofo_equilibrium_map}
x^\star(z_k,w_k)
\coloneqq
-A^{-1}\left(B_{z,k}z_k+w_k\right).
\end{equation}
Let $e_k$ denote the $k$th canonical vector. Using \eqref{eq:ace_output_explicit}, the steady-state ACE of area~$k$
is then
\begin{align}\label{eq:ofo_ss_ace_map}
h_k(z_k,w_k)
&:= e_k^{\top}\mathrm{ACE}(x^{\star}(z_k,w_k))\\
&= e_k^{\top}(C^{\rm ace}x^{\star}(z_k,w_k) + r^{\rm ace}).
\end{align}
Let $p_k^{\rm line}$ collect the active-power flows monitored by
area~$k$, including selected internal lines and boundary lines for which the
area is responsible. Following the same algebraic-angle reconstruction used
for~\eqref{eq:nai_output_model}, these flows can be written as
\begin{equation}
\label{eq:ofo_line_output}
p_k^{\rm line}
=
C_k^{\rm line}x+D_{z,k}^{\rm line}z_k+r_k^{\rm line},
\end{equation}
where $D_{z,k}^{\rm line}$ includes  contribution from GFL
setpoints; effects of the market schedule, uncontrolled injections/consumptions, and
neighboring-area actions are included in $r_k^{\rm line}$ and $w_k$. The
steady-state values of the monitored flows are then
\begin{align}
g_k(z_k,w_k)
& :=
p_k^{{\rm line},\star}(z_k,w_k) \nonumber  \\
& =
C_k^{\rm line}x^\star(z_k,w_k)
+D_{z,k}^{\rm line}z_k+r_k^{\rm line}. \label{eq:ofo_ss_line_map}
\end{align}
A constraint-aware generalization of the AGC equilibrium selection problem \eqref{eq:sidebar_area_sharing_problem} can now be defined by the steady-state
optimization problem
\begin{subequations}
\label{eq:ofo_secondary_problem}
\begin{align}
\min_{z_k\in\mathcal Z_k}
~~&
f_k(z_k):=\frac{1}{2}z_k^\top Q_kz_k
\label{eq:ofo_secondary_objective}
\\
\text{subject to}
~~&
h_k(z_k,w_k)=0,
\label{eq:ofo_secondary_ace_constraint}
\\
&
\underline p_k^{\rm line}
\leq g_k(z_k,w_k)\leq
\overline p_k^{\rm line},
\label{eq:ofo_secondary_line_constraints}
\end{align}
\end{subequations}
where $Q_k\succ0$ assigns different weights to deviations from the market
schedule to different generators and IBR plants. The set
\begin{equation}
\label{eq:ofo_resource_set}
\mathcal Z_k
:=
\left\{
z_k:
\underline u_k-u_k^{\rm mkt}
\leq z_k\leq
\overline u_k-u_k^{\rm mkt}
\right\}
\end{equation}
represents the upward and downward headroom currently available from the
participating resources in area~$k$. If the market setpoints remain
feasible after a disturbance, the optimal correction is $z_k^\star=0$.
Otherwise,
\eqref{eq:ofo_secondary_problem} selects the smallest weighted correction that
regulates the ACE, respects the available headroom, and avoids steady-state line
overloads. If the line constraints are removed, the problem
reduces to \eqref{eq:sidebar_area_sharing_problem}. More sophisticated variants of the problem formulation \eqref{eq:ofo_secondary_problem} could include
ramp-rate limits, energy or state-of-charge constraints for storage,
time-varying flexibility limits for renewable-based IBR plants and controllable loads,
reserve requirements, and contingency-dependent line-flow limits. See \cite{JL-DEO-JWSP-MK-DZ:18d} for a variation on \eqref{eq:ofo_secondary_problem} applied to microgrids.

As an example of the feedback optimization methodology, consider a continuous-time projected primal--dual flow; see, for example, the general frameworks in \cite{wang2011control,cherukuri2017saddle, bianchin2021time}, and the projected dynamics in~\cite{xia2000stability}. Introduce a dual variable
$\lambda_k$ for the zero-ACE constraint and nonnegative dual variables
$\mu_k^+$ and $\mu_k^-$ for the upper and lower local line-flow limits,
respectively; let $n_{k,\ell}:=\dim(p_k^{\rm line})$. Define the primal descent direction
\begin{equation}
\label{eq:ofo_primal_direction}
d_{z,k}
:=
-Q_kz_k
-G_{{\rm ace},k}^{\top}\lambda_k
-G_{{\rm line},k}^{\top}(\mu_k^+-\mu_k^-),
\end{equation}
where
\begin{align}
\label{eq:ofo_sensitivities}
G_{{\rm ace},k}
&:=
\nabla_{z_k} h_k(z_k;w_k)
=
e_k^\top
\left(
D_{z,k}^{\rm ace}-C^{\rm ace}A^{-1}B_{z,k}
\right)\\
\nonumber
G_{{\rm line},k}
&:=
\nabla_{z_k} g_k(z_k;w_k)
=
D_{z,k}^{\rm line}-C_k^{\rm line}A^{-1}B_{z,k}.
\end{align}
are the (constant) sensitivities of the affine maps defined in \eqref{eq:ofo_ss_ace_map} and
\eqref{eq:ofo_ss_line_map}. These matrices can be computed from the reduced model, updated from an online
linearization, or estimated from data~\cite{ZT-EE-JWSP-EF-MP-AH-HH:22b,picallo2022adaptive}.

With this, the closed-loop system with a projected primal-dual controller is \cite{bianchin2021time}
\begin{subequations}
\label{eq:ofo_primal_dual_flow}
\begin{align}
\dot x
&=
Ax+B_{z,k}z_k+w_k,
\label{eq:ofo_closed_loop_plant}
\\
\tau_k^{\rm ofo}\dot z_k
&=
\proj_{\mathcal Z_k}\left(
z_k+\eta_{z,k}d_{z,k}
\right)-z_k,
\label{eq:ofo_primal_flow}
\\
\tau_k^{\rm ofo}\dot\lambda_k
&=
\eta_{\lambda,k}{\rm ACE}_k(t),
\label{eq:ofo_ace_dual_flow}
\\
\tau_k^{\rm ofo}\dot\mu_k^+
&=
\proj_{\mathbb R_+^{n_{k,\ell}}}\left(
\mu_k^+
+\eta_{\mu,k}
\left[p_k^{\rm line}(t)-\overline p_k^{\rm line}\right]
\right)-\mu_k^+,
\label{eq:ofo_upper_line_dual_flow}
\\
\tau_k^{\rm ofo}\dot\mu_k^-
&=
\proj_{\mathbb R_+^{n_{k,\ell}}}\left(
\mu_k^-
+\eta_{\mu,k}
\left[\underline p_k^{\rm line}-p_k^{\rm line}(t)\right]
\right)-\mu_k^-,
\label{eq:ofo_lower_line_dual_flow}
\end{align}
\end{subequations}
where $\eta_{z,k},\eta_{\lambda,k},\eta_{\mu,k}>0$ are internal controller gains and the scalar $\tau_k^{\rm ofo}>0$ sets the time scale of the
optimization dynamics in area~$k$. For a closed
convex set $\mathcal S$, $\proj_{\mathcal S}(v)$ denotes the Euclidean projection
of the point $v$ onto $\mathcal S$, $\proj_{\mathcal S}(v)
:=
\arg\min_{y\in\mathcal S}\frac{1}{2}\|y-v\|^2$.

The controller \eqref{eq:ofo_primal_flow}--\eqref{eq:ofo_lower_line_dual_flow} uses the real-time ACE measurement ${\rm ACE}_k(t)$ and the real-time line flow measurement $p_k^{\rm line}(t)$ to compute the control corrections $z_k$. If a line flow is not directly
measured, $p_k^{\rm line}(t)$ in \eqref{eq:ofo_primal_dual_flow} can be replaced
by an estimate $\widehat p_k^{\rm line}(t)$ obtained from a state estimator
using available telemetry and, possibly, pseudo-measurements for unobserved
injections. The overall architecture is illustrated in Fig.~\ref{fig:ofo_closed_loop}. For the closed-loop~\eqref{eq:ofo_primal_dual_flow}, under suitable time-scale separation, the
physical state tracks $x^\star(z_k;w_k)$, and stability will be governed by a set of reduced dynamics described by a collection of primal-dual gradient flows \cite{cherukuri2017saddle,bianchin2021time}, coupled by the equilibrium network equations, and with heterogeneous time constants $\tau_{k}^{\rm ofo}$. While singular perturbation arguments~\cite{kokotovic1976singular,hauswirth2024optimization,
menta2018stability} or contraction arguments~\cite{cothren2026online,del2012contraction} are likely applicable, deriving conditions for stability of~\eqref{eq:ofo_primal_dual_flow} is left as an open problem. 

While~\eqref{eq:ofo_primal_dual_flow}  illustrates an application of feedback optimization using primal-dual dynamics, 
alternative approaches to designing and analyzing feedback optimization approaches can be applied to yield different designs, including proximal augmented Lagrangian methods~\cite{ding2020global}, optimal steady-state control \cite{lawrence2021linear, simpsonporco2022lowgain,jakob2026generalized}, or the so-called safe gradient flow method~\cite{allibhoy2023control,colot2024optimal}. Additionally, leveraging the work in~\cite{hauswirth2024optimization},  
an alternative feedback--based primal--dual flow with the system dynamics in the loop can be derived using projected gradient on tangent cones. Finally, when the updates of the feedback optimizer need to be performed in a \emph{sampled-data} fashion, design approaches and conditions for stability can be derived using approaches similar to, e.g.,~\cite{belgioioso2021sampled,cothren2023perception}.

\begin{notebox}[Modeling note: time-varying loads] 
When $w_k$ varies with time, the closed-loop generally tracks a moving
optimizer rather than converging to a fixed equilibrium. Input-to-state
stability provides a useful framework for characterizing this behavior.
Under uniform stability of the frozen closed loop, Lipschitz continuity of
the optimizer map, suitable constraint regularity, and sufficient time-scale
separation, the tracking error can be bounded in terms of the rate of
variation of $w_k$, measurement errors, and inaccuracies in the
input--output sensitivities. In particular, one expects a bound of the form
\begin{align}
\left\|\xi_k(t)-\xi_k^\star(w_k(t))\right\|
\leq{}&
c e^{-\alpha t}
\left\|\xi_k(0)-\xi_k^\star(w_k(0))\right\|
\nonumber\\
&\hspace{-.2cm} +\gamma_w\left(\|\dot w_k\|_\infty\right)
+\gamma_e\left(\|e_k\|_\infty\right),
\label{eq:ofo_iss_bound}
\end{align}
where $\xi_k$ collects the physical, primal, and dual states, and $e_k$
collects measurement and sensitivity errors; see, for example,~\cite{bianchin2021time}. Thus, constant disturbances
lead to convergence to the corresponding equilibrium, whereas slowly varying
disturbances and small implementation errors lead to a correspondingly small
tracking neighborhood.
\end{notebox}

\subsection{Example \#2: Compressing the time scales between secondary and tertiary control}

The feedback optimization controller in the preceding subsection does not replace any function of economic dispatch at the tertiary control level. Indeed, the
dispatch $u^{\rm mkt}$ is computed first (e.g., every 5-15 minutes, or even every hour) and is held fixed while FO determines
the correction $z=u-u^{\rm mkt}$ around the prescribed economic operating point. The merging of some secondary and tertiary level functionality however offers opportunities for improved tracking of optimal operating points as system conditions change \cite{li2015connecting,sawant2026integrated}. This example illustrates how optimization of the dispatch point and the interchange schedules can be embedded within the secondary control level using feedback optimization.

\begin{figure}[t]
    \centering
    \resizebox{0.95\linewidth}{!}{%
        \includegraphics[width=218.56836pt,height=109.70311pt]{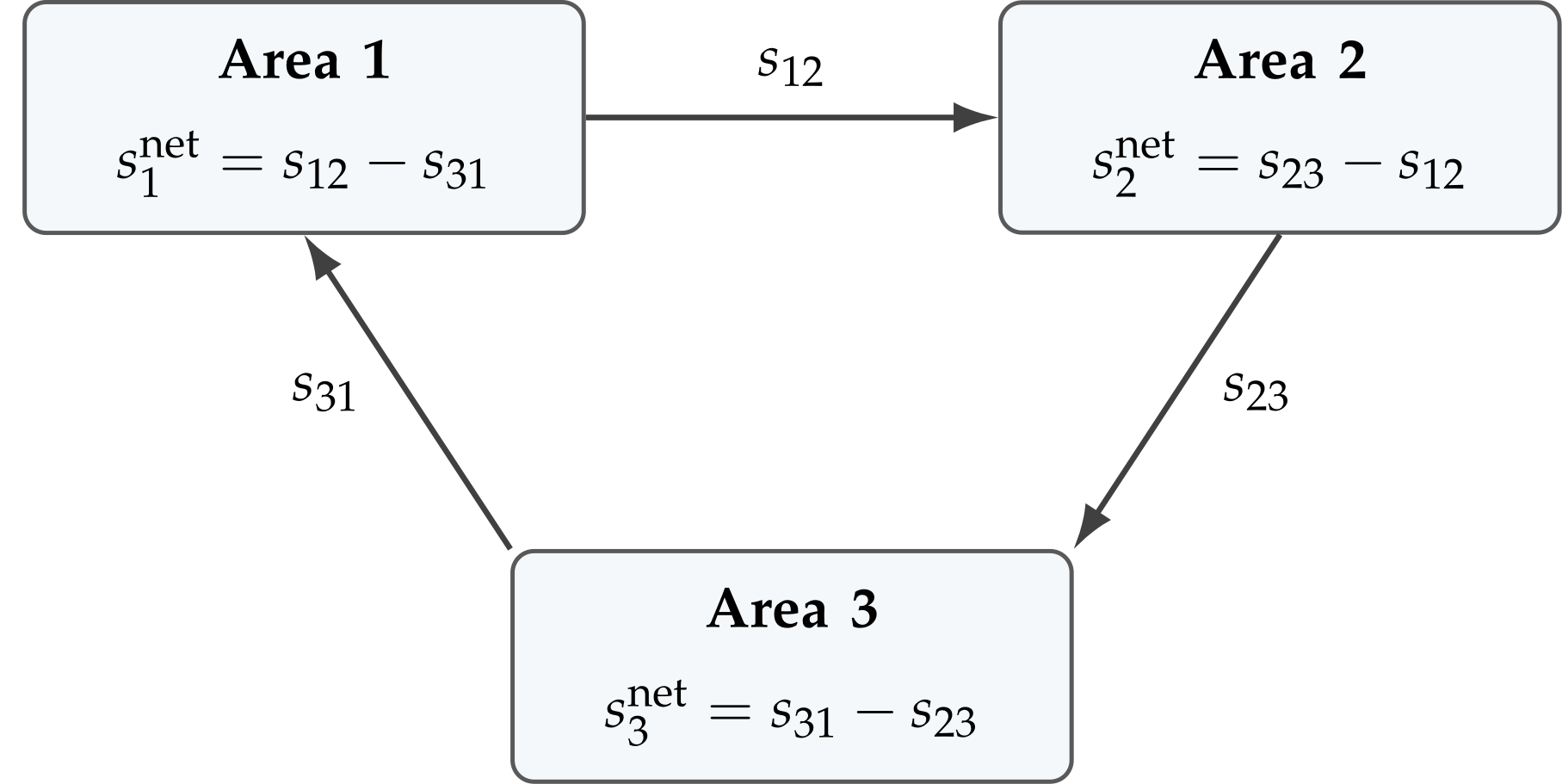}%
    }
    \caption{Scheduled transfers $s_{k\ell}$ and net interchange
    schedules $s_k^{\rm net}$ for three balancing areas.
    Arrows define the directions of the power transfers on tie lines.}
\label{fig:F_three_area_schedules}
\end{figure}

Consider the undirected graph
$\mathcal G_{\rm A}=(\{1,\ldots,N_a\},\mathcal E_{\rm A})$ whose nodes are
balancing areas and whose edges represent neighboring areas between which
tie line flows can be scheduled. Let $\mathcal N_k$ denote the neighbors of
area~$k$. Assign an arbitrary orientation to every edge
$(k,\ell)\in\mathcal E_{\rm A}$, and let $s_{k\ell}$ denote the scheduled
transfer from area~$k$ to area~$\ell$ under this orientation. The reverse transfer is defined by
\begin{equation}
s_{\ell k}:=-s_{k\ell}, 
\label{eq:interface_schedule_antisymmetry}
\end{equation}
as illustrated in Fig.~\ref{fig:F_three_area_schedules}. Thus, only one independent schedule variable is maintained for each
interface between two areas. Positive $s_{k\ell}$ represents a scheduled export from area~$k$
to area~$\ell$. The scheduled net interchange of area~$k$ is then
\begin{equation}
s_k^{\rm net}
:=
\sum_{\ell\in\mathcal N_k}s_{k\ell}.
\label{eq:area_net_schedule}
\end{equation}
The  condition~\eqref{eq:interface_schedule_antisymmetry}
automatically ensures $\sum_{k=1}^{N_a}s_k^{\rm net}=0$,
so scheduled exports and imports balance across the interconnected system.

The measured ACE of area~$k$ is now evaluated relative to the (possibly, time-varying) net
interchange schedule
\begin{equation}
\label{eq:compressed_ofo_ace}
{\rm ACE}_k(t;s)
=
b_k\widetilde\Omega_k(t)
+p_k^{\rm NI}(t)
-\sum_{\ell\in\mathcal N_k}s_{k\ell}(t),
\end{equation}
where $s$ stacks all the \emph{independent} oriented variables $s_{k\ell}$ for the edges $(k,\ell)\in\mathcal E_{\rm A}$. If limits are imposed on a scheduled
bilateral transfer, let
\begin{equation}
\mathcal S_{k\ell}
:=
[\underline s_{k\ell},\overline s_{k\ell}]
\label{eq:interface_schedule_set}
\end{equation}
denote its admissible set; otherwise,
$\mathcal S_{k\ell}=\mathbb R$. The feasible set of interface schedules is $\mathcal S_{\rm E}
:=
\prod_{(k,\ell)\in\mathcal E_{\rm A}}\mathcal S_{k\ell}$.

Rewrite the dynamics in~\eqref{eq:state_space_form}
as  $\dot x
=
Ax+B_u u+w$ where, for brevity, $w$ is a vector representing the contributions of non-controllable loads and non-controllable resources. For given constant inputs $(u,s)$, let $h(u,s)$ collect the steady-state values of
\eqref{eq:compressed_ofo_ace}, and let $g(u)$ collect the steady-state flows
of the monitored transmission lines. An integrated economic-dispatch and
secondary-regulation equilibrium selection problem can be formulated as
\begin{subequations}
\label{eq:compressed_ofo_problem}
\begin{align}
\underset{u\in\mathcal U,\,s\in\mathcal S_{\rm E}}
{\operatorname{minimize}}
~~&
C(u)
\label{eq:compressed_ofo_objective}
\\
\operatorname{subject~to}
~~&
h(u,s)=0,
\label{eq:compressed_ofo_ace_constraint}
\\
&
\underline p^{\rm line}
\leq
g(u)
\leq
\overline p^{\rm line},
\label{eq:compressed_ofo_line_constraints}
\end{align}
\end{subequations}
where $C(u)$ models a cost of generation and
\[
\mathcal U
:=
\left\{
u:
\underline u
\leq u\leq
\overline u
\right\}
\]
models generation constraints. Problem~\eqref{eq:compressed_ofo_problem} differs from
\eqref{eq:ofo_secondary_problem} in two ways: \emph{(i)} it optimizes the
actual operating points $u$, rather than corrections around a previously
computed dispatch; and \emph{(ii)} it optimizes the interchange schedules,
rather than treating them as fixed references. Hence, economic power
transfers across balancing-area interfaces and secondary frequency
regulation are determined by the same optimization problem. 

To express the schedule dependence compactly, let $E_{\rm A}$ be the
incidence matrix (e.g., \cite{dorfler2018electrical}) of the oriented balancing-area graph $\mathcal{G}_{A}$, with a $+1$ at the
sending area and a $-1$ at the receiving area of each oriented interface.
The vector of scheduled net interchanges is then $E_{\rm A}s$, and the
steady-state ACE map has the form
\begin{equation}
h(u,s)
=
h_0(u)-E_{\rm A}s.
\label{eq:edge_schedule_ace_map}
\end{equation}
Consequently, $G_{{\rm ace},u}
:=
\nabla_u h$, $G_{{\rm ace},s}
:=
\nabla_s h
=
-E_{\rm A}$, and 
$G_{\rm line}
:=
\nabla_u g$ are the (again, constant) sensitivity matrices of these maps. A measurement-based primal--dual method for solving
\eqref{eq:compressed_ofo_problem} is
\begin{subequations}
\label{eq:compressed_ofo_primal_dual_flow}
\begin{align}
\dot x
&=
Ax+B_u u+w,
\label{eq:compressed_ofo_closed_loop_plant}
\\
\tau^{\rm ofo}\dot u
&=
\proj_{\mathcal U}
\left(
u+\eta_u d_u
\right)-u,
\label{eq:compressed_ofo_primal_flow}
\\
\tau^{\rm ofo}\dot s_{k\ell}
&=
\proj_{\mathcal S_{k\ell}}
\left(
s_{k\ell}
+\eta_s(\lambda_k-\lambda_\ell)
\right)-s_{k\ell} 
,
\label{eq:compressed_ofo_schedule_flow}
\\
\tau^{\rm ofo}\dot\lambda
&=
\eta_\lambda{\rm ACE}(t;s),
\label{eq:compressed_ofo_ace_dual_flow}
\\
\tau^{\rm ofo}\dot\mu^+
&=
\proj_{\mathbb R_+^{n_\ell}}
\left(
\mu^+
+\eta_\mu
\left[
p^{\rm line}(t)-\overline p^{\rm line}
\right]
\right)-\mu^+,
\label{eq:compressed_ofo_upper_line_dual_flow}
\\
\tau^{\rm ofo}\dot\mu^-
&=
\proj_{\mathbb R_+^{n_\ell}}
\left(
\mu^-
+\eta_\mu
\left[
\underline p^{\rm line}-p^{\rm line}(t)
\right]
\right)-\mu^-,
\label{eq:compressed_ofo_lower_line_dual_flow}
\end{align}
\end{subequations}
for $(k,\ell)\in\mathcal E_{\rm A}$, where
\begin{equation}
d_u
:=
-\nabla C(u)
-G_{{\rm ace},u}^{\top}\lambda
-G_{\rm line}^{\top}(\mu^+-\mu^-),
\label{eq:compressed_ofo_resource_direction}
\end{equation}
and $\eta_u,\eta_s,\eta_\lambda,\eta_\mu>0$ are controller gains.

The  update
\eqref{eq:compressed_ofo_schedule_flow} follows directly from
\eqref{eq:edge_schedule_ace_map}. Moreover, from area~$k$ to
area~$\ell$, $\frac{\partial h_k}{\partial s_{k\ell}}=-1$,
$\frac{\partial h_\ell}{\partial s_{k\ell}}=1$.
 Therefore, the derivative of the Lagrangian with respect to $s_{k\ell}$
contains $-\lambda_k+\lambda_\ell$, and its projected descent direction is
$\lambda_k-\lambda_\ell$. Each interface schedule can consequently be updated
using only the ACE multipliers of its two incident areas.

The controller closes two loops simultaneously. The ACE feedback regulates both
frequency and interchange balance, while the cost gradient continuously
redispatches the participating resources. Unlike classical AGC, the
scheduled bilateral transfers are controller states and can now change on the time-scale of secondary control in
response to evolving operating conditions.

The FO algorithm~\eqref{eq:compressed_ofo_primal_dual_flow} can be implemented in a decentralized fashion. To see this, the message passing is explained for a two-area system as an example. Let $p_{12}$ denote the measured net
tie-line flow from area~1 to area~2, and let $s_{12}$ denote the corresponding
scheduled transfer. Since $s_{21}=-s_{12}$, the two ACE signals are
\begin{equation}
\begin{aligned}
{\rm ACE}_1(t;s_{12})
&=
b_1\widetilde\Omega_1(t)
+p_{12}(t)-s_{12},\\
{\rm ACE}_2(t;s_{12})
&=
b_2\widetilde\Omega_2(t)
-p_{12}(t)+s_{12}.
\end{aligned}
\label{eq:two_area_ace}
\end{equation}
Each area forms its ACE from its representative frequency, the measured
tie-line flow, and the shared interface schedule. It then updates its own ACE
multiplier according to
\begin{equation}
\tau^{\rm ofo}\dot\lambda_k
=
\eta_\lambda{\rm ACE}_k(t;s_{12}),
~~
k\in\{1,2\}.
\label{eq:two_area_ace_multiplier}
\end{equation}
The interface schedule is updated according to
\begin{equation}
\tau^{\rm ofo}\dot s_{12}
=
\proj_{\mathcal S_{12}}
\left(
s_{12}+\eta_s(\lambda_1-\lambda_2)
\right)-s_{12}.
\label{eq:two_area_interface_schedule_update}
\end{equation}
Thus, schedule coordination requires only the exchange of $\lambda_1$ and
$\lambda_2$ between the two neighboring balancing areas. 

The monitored transmission lines and their associated multipliers can also
be partitioned by balancing area. For the update of the power  references for generators and inverters, partition the references and costs per area as $u=\col(u_1,u_2)$ and $C(u)=C_1(u_1)+C_2(u_2)$, and let  $\mathcal U=\mathcal U_1\times\mathcal U_2$. Moreover,  let $g_j$ collect the monitored-line outputs assigned to area~$j$.
Define
\begin{equation}
G_{{\rm ace}_j,u_k}
:=
\nabla_{u_k}h_j,
~~
G_{{\rm line}_j,u_k}
:=
\nabla_{u_k}g_j,
~~
j,k\in\{1,2\}.
\label{eq:two_area_sensitivity_blocks}
\end{equation}
The contribution of the constraints monitored by area~$j$ to the resource
update in area~$k$ is
\begin{equation}
\xi_{j\rightarrow k}
:=
G_{{\rm ace}_j,u_k}^{\top}\lambda_j
+
G_{{\rm line}_j,u_k}^{\top}
\left(\mu_j^+-\mu_j^-\right).
\label{eq:two_area_marginal_signal}
\end{equation}
The  update for the power references for generators and inverters in area~$k$ can therefore be written as
\begin{subequations}
\label{eq:two_area_resource_update}
\begin{align}
d_{u,k}
&=
-\nabla C_k(u_k)
-\xi_{1\rightarrow k}
-\xi_{2\rightarrow k},
\\
\tau^{\rm ofo}\dot u_k
&=
\proj_{\mathcal U_k}
\left(
u_k+\eta_u d_{u,k}
\right)-u_k,
~~
k\in\{1,2\}.
\end{align}
\end{subequations}
Area~$j$ can compute $\xi_{j\rightarrow k}$ using its locally maintained
multipliers and the corresponding sensitivity blocks. While
$\xi_{k\rightarrow k}$ is computed locally,
$\xi_{j\rightarrow k}$, with $j\neq k$, must be communicated to the
neighboring area. This communication is required because a resource
adjustment in one area can affect the steady-state ACE and constrained line
flows monitored in the other.

This example clarifies the distributed structure of the controller. Each
balancing area measures its representative frequency, actual net
interchange, and monitored internal line flows; updates its own ACE and
line-limit multipliers; and computes its own resource references. Each
interface schedule is updated using only the ACE multipliers of the two
incident areas. Additional sensitivity-weighted signals may be required by
the resource updates because the physical transmission network remains
interconnected.

When the balancing-area graph contains cycles, the interface schedules need
not be uniquely determined by the area-level ACE constraints: adding a
circulating schedule around a cycle does not change any area's scheduled net
interchange. This non-uniqueness does not affect frequency restoration or
area-level interchange balance. 

At equilibrium, the states of the reduced
transmission-system model and of the controller satisfy the KKT conditions
of~\eqref{eq:compressed_ofo_problem}. The stationarity condition associated
with an interface schedule $s_{k\ell}$ is
\begin{equation}
0
\in
-\lambda_k^\star+\lambda_\ell^\star
+
N_{\mathcal S_{k\ell}}(s_{k\ell}^\star),
~~
(k,\ell)\in\mathcal E_{\rm A},
\label{eq:compressed_ofo_schedule_stationarity}
\end{equation}
where $N_{\mathcal S_{k\ell}}(s_{k\ell}^\star)$ is the normal cone to the
admissible schedule set. If $s_{k\ell}^\star$ is strictly inside its limits,
\eqref{eq:compressed_ofo_schedule_stationarity} gives
\begin{equation}
\lambda_k^\star=\lambda_\ell^\star.
\label{eq:neighboring_area_price_agreement}
\end{equation}
Hence, if the balancing-area graph is connected and no interface-schedule
constraint is active, then
\begin{equation}
\lambda^\star
=
\lambda_0^\star\mathds{1}.
\label{eq:area_balance_price_consensus}
\end{equation}
Optimizing the interface schedules therefore equalizes the area balance
prices across unconstrained interfaces. If a schedule reaches one of its
limits, a difference between the multipliers of the incident areas may
remain. Binding transmission-network constraints additionally produce the
location-dependent congestion components of the nodal prices described
below.

The interchange variables $\{s_{k\ell}\}$ determine the scheduled bilateral
power transfers, whereas the associated financial settlements are not
included in the objective of~\eqref{eq:compressed_ofo_problem}, since payments
between areas are transfers rather than physical operating costs. Under the
adopted sign convention, the area price component is
$\pi_k=-\lambda_k$. When an interface schedule is strictly inside its limits,
\eqref{eq:compressed_ofo_schedule_stationarity} implies equal prices in its
two incident areas. The importing area pays for the scheduled transfer and
the exporting area receives the corresponding payment; equal prices imply
zero congestion rent, not zero payment. If an interface-schedule or
transmission-network constraint is active, the effective marginal prices may
differ because of the corresponding constraint multipliers, producing a
congestion rent. Such settlements can be computed from the equilibrium dual
variables without modifying the feedback optimization dynamics.

\begin{figure}[h!]
    \centering
    \includegraphics[width=\linewidth]
        {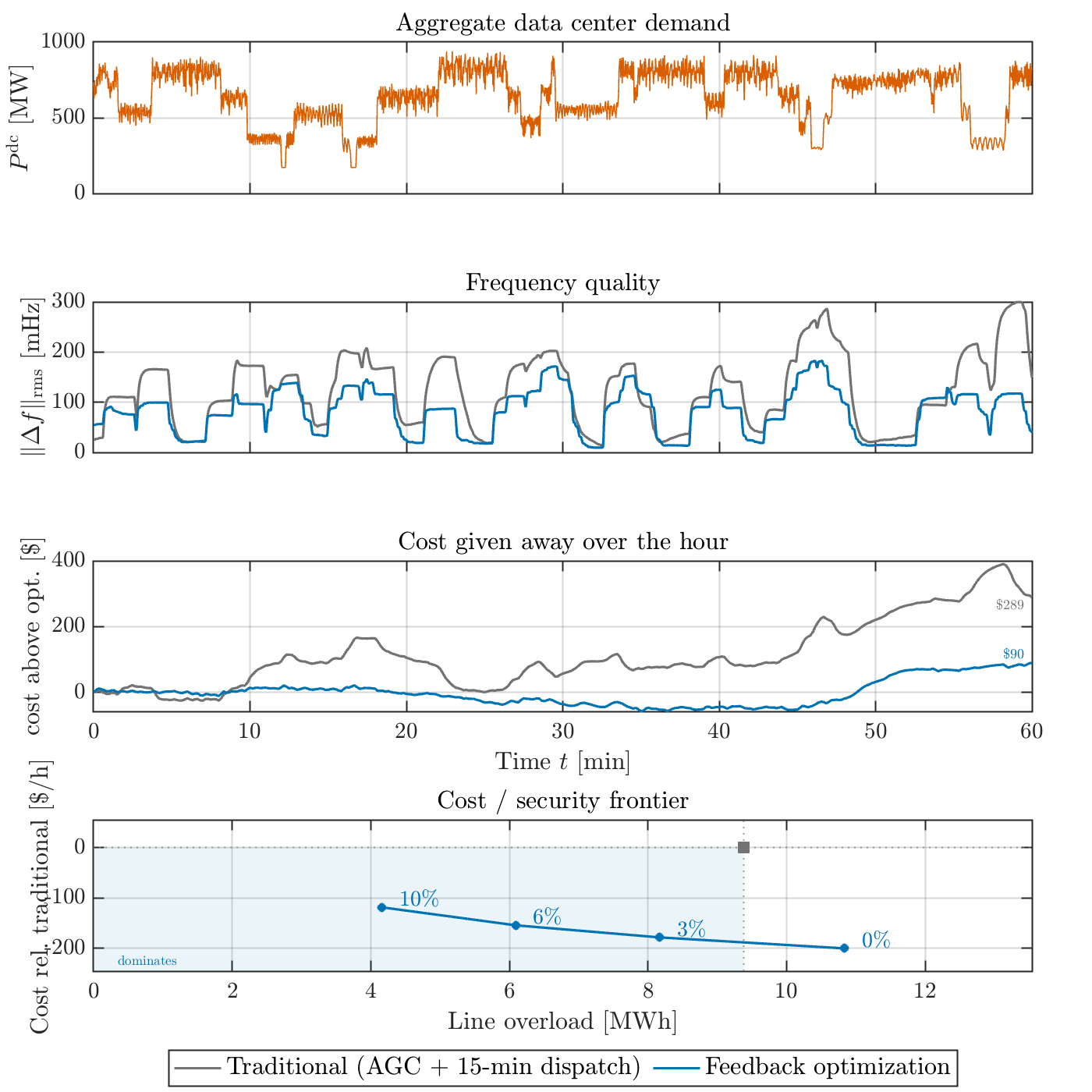}
    \caption{One hour under time-varying demand from data centers the traditional architecture
    (primary control, AGC, $15$-minute economic dispatch on a forecast) against
    the compressed controller~\eqref{eq:compressed_ofo_primal_dual_flow}, both
    tuned to a $21$~s closed-loop time constant. Panels show: 
    Aggregate data center demand, on top of a smooth feeder ramp; frequency deviation as a two-minute rolling r.m.s.\ value; cost accumulated above the clairvoyant optimum; cost/security frontier, cost relative to the traditional architecture
    and violation scored against the true limit.}
    \label{fig:F_newofo}
\end{figure}

As an example, consider testing~\eqref{eq:compressed_ofo_primal_dual_flow} on a modified IEEE 39-bus
system split into three balancing areas, with seven synchronous generators,
three grid-forming and five grid-following inverters, and three AI data centers
as large time-varying loads. Figure~\ref{fig:F_newofo} compares one hour of operation against the
traditional stack -- primary control, AGC, and an economic dispatch solved every
$15$~minutes on a causal forecast --under an identical load realization, with
both architectures tuned to the same $21$~s closed-loop time constant. The
frequency r.m.s.\ falls from $144$ to $93$~mHz and the ACE r.m.s.\ from $66$ to
$53$~MW, not because the secondary loop is faster but because it re-optimizes
the whole resource set continuously rather than correcting the synchronous units
around a set point that is $15$~minutes stale.
The economic advantage
accumulates over the dispatch intervals: the third panel shows the cost given away
relative to a clairvoyant benchmark, which drifts upward within each
$15$-minute interval and is only partially reset at the next re-dispatch, so the
traditional stack ends the hour \$$289$ above the optimum against \$$90$ for the
compressed controller (a factor of $3.2$).

The fourth panel is presented as a frontier for a reason. A controller that tracks the
true optimum operates \emph{on} the active constraint and has no margin, so its
tracking error appears directly as violation, whereas the economically slack
baseline buys margin by accident. Operating against a tightened limit trades
cost back for margin: at a $3\%$ margin the violation already falls below the
traditional $9.4$~MWh, to $8.2$~MWh, at a cost gap of \$$111$, and at $10\%$ it
drops to $4.2$~MWh at \$$171$. The compressed architecture dominates on both axes, but
only once the security margin is treated as a design variable.

\subsection{Example \#3: Compressing  primary and secondary control via disturbance feedback}

The analysis of AGC presented earlier highlighted a key characteristic: closed-loop stability is ensured only if AGC operates on a slower time-scale than the system dynamics with primary control. This should not be viewed as a defect, but as a natural consequence of the simplicity of AGC and of the slow frequency regulation needs of traditional power systems. Indeed, due to the simplifying nature of time-scale separation, the design of AGC requires no substantive dynamic model of the primary control, and closed-loop stability is robust to the particulars of those primary control dynamics. 

The price paid for these design and stability benefits is slow transient performance. Again, this is not a significant drawback when synchronous generators are the primary sources of power, as the slow time constants of turbine-governor systems preclude the possibility of fast regulation. Nonetheless, the requirement of time-scale separation means that the fast-ramping capabilities of inverter-interfaced generation cannot be effectively used for frequency regulation under the traditional AGC paradigm. New control approaches are needed if inverter-based resources are to provide fast regulation beyond their default primary control response --- so-called \emph{fast frequency control} --- and necessarily, this requires the inclusion of additional model information in the design process. This section describes one recent approach to fast frequency control based on active disturbance estimation and rejection \cite{EE-ZT-JWSP-EF-MP-HH:20k,EE-JWSP-EF-MP-AH-LZ:23f, JS-JWSP-AP:24e}. The approach is described assuming a single balancing area, but is easily extended to a decentralized design for multiple balancing areas.

The starting point is the (stable) reduced system dynamics \eqref{eq:state_space_form}, which are assumed to be available for design. The objective is to design an operations-level controller which can redispatch the GFM/GFL set-points $(\bar{p}^{\rm gfm}, \bar{p}^{\rm gfl})$ on a time-scale of seconds to reject step changes in load $\Delta p^{\rm load} \in \real^{|\cL^{\rm l}|}$, providing fast response and resilience to major grid events. The synchronous generator set-points $\bar{p}^{\rm sg}$ will be adjusted by traditional AGC on a time-scale of minutes, and thus for the purposes of this design are known and constant. The deviation dynamics of \eqref{eq:state_space_form} describing the evolution in the initial seconds after a load change are therefore
\begin{equation}\label{eq:state_space_form_measurement}
\begin{aligned}
\Delta \dot  x &=
A \Delta x + 
B\begin{bmatrix}\Delta \bar p^{\rm gfm} \\ 
\Delta\bar p^{\rm gfl}
\end{bmatrix}
+ F\Delta p^{\rm load},\\
\Delta y_{\rm m} &= C_{\rm m}\Delta x,\\
\Delta \Omega_{\rm coi} &= C_{\Omega}\Delta x,
\end{aligned}
\end{equation}
where for simplicity we label $B = \begin{bmatrix}B^{\rm gfm} & B^{\rm gfl}\end{bmatrix}$ and $F = F^{\rm load}$. The model \eqref{eq:state_space_form_measurement} has been augmented with measurements $\Delta y_{\rm m} \in \real^{n_{\rm m}}$, such as frequency, power generation, and line power flows, which are available in real-time to the system operator. An average ``center-of-inertia'' frequency deviation $\Delta\Omega_{\rm coi}$ defined as
\[
\Delta\Omega_{\rm coi} = \frac{\mathbf{1}^{\top}M_{\mathcal{G}}\Delta\Omega_{\mathcal{G}}}{\mathbf{1}^{\top}M_{\mathcal{G}}\mathbf{1}}
\]
has been selected as a typical performance output. 

Assuming the unknown load change $\Delta p^{\rm load}$ is constant, it is described by the simple differential equation $\Delta\dot{p}^{\rm load} = 0$. Extending the state of \eqref{eq:state_space_form_measurement} with this additional equation, one seeks to design an extended-state observer
\begin{equation}\label{eq:state_space_estimator}
\begin{aligned}
\begin{bmatrix}
\Delta\dot{\hat{x}}\\
\Delta\dot{\hat{p}}^{\rm load}
\end{bmatrix}
&=
\begin{bmatrix}
A & F\\
0 & 0
\end{bmatrix}
\begin{bmatrix}
\Delta\hat{x}\\
\Delta\hat{p}^{\rm load}
\end{bmatrix}
+
\begin{bmatrix}
B\\
0
\end{bmatrix}
\begin{bmatrix}
\Delta\bar p^{\rm gfm}\\
\Delta\bar p^{\rm gfl}
\end{bmatrix}\\
&\qquad +
\begin{bmatrix}
L_1\\
L_2
\end{bmatrix}
\left(
\Delta y_{\rm m}-\Delta\hat y_{\rm m}
\right),
\\
\Delta\hat y_{\rm m}
&=
\begin{bmatrix}
C_{\rm m} & 0
\end{bmatrix}
\begin{bmatrix}
\Delta\hat{x}\\
\Delta\hat{p}^{\rm load}
\end{bmatrix},
\end{aligned}
\end{equation}
with gains $(L_1, L_2)$ to be tuned. Since $A$ is Hurwitz, the pair $(C_{\rm m},A)$ is detectable, and by the PBH test the extended-state system is detectable if
\begin{equation}\label{Eq:NoZeros}
\mathrm{rank}\begin{bmatrix}
A & F\\
C_{\rm m} & 0
\end{bmatrix} = n_{x} + |\cL^{\rm l}|,
\end{equation}
which says that the constant load disturbances are observable through the given measurements. Under this condition, $(L_1, L_2)$ may be designed using, e.g., LQE methods. The rank condition \eqref{Eq:NoZeros} necessarily requires that the number of available measurements be greater than or equal to the number of load values to be estimated. In a tightly interconnected system or balancing area, it is reasonably accurate to consider all load as being concentrated at a single bus, in which case a single frequency measurement will typically suffice for satisfying \eqref{Eq:NoZeros}.

With the load change estimate $\Delta\hat{p}^{\rm load}$ produced by \eqref{eq:state_space_estimator}, one may redispatch the GFM and GFL using FO to reject the disturbance, subject to the constraint  that the aggregated change in GFM/GFL set-points should equal the total estimated load change $\mathbf{1}^{\top}\Delta\hat{p}^{\rm load}$. The simplest implementation would be
\begin{equation}\label{Eq:Allocator}
\begin{bmatrix}\Delta\bar p^{\rm gfm} \\ 
\Delta\bar p^{\rm gfl}
\end{bmatrix} = -\Lambda \Delta\hat{p}^{\rm load}
\end{equation}
where $\Lambda$ is a nonnegative column-stochastic matrix. More sophisticated control allocation procedures and constraint enforcement can be obtained by embedding this estimation procedure within a FO framework; see \cite{NY-JWSP:25b}.

\begin{figure}[h!]
    \centering
{      \includegraphics[width=\linewidth]
        {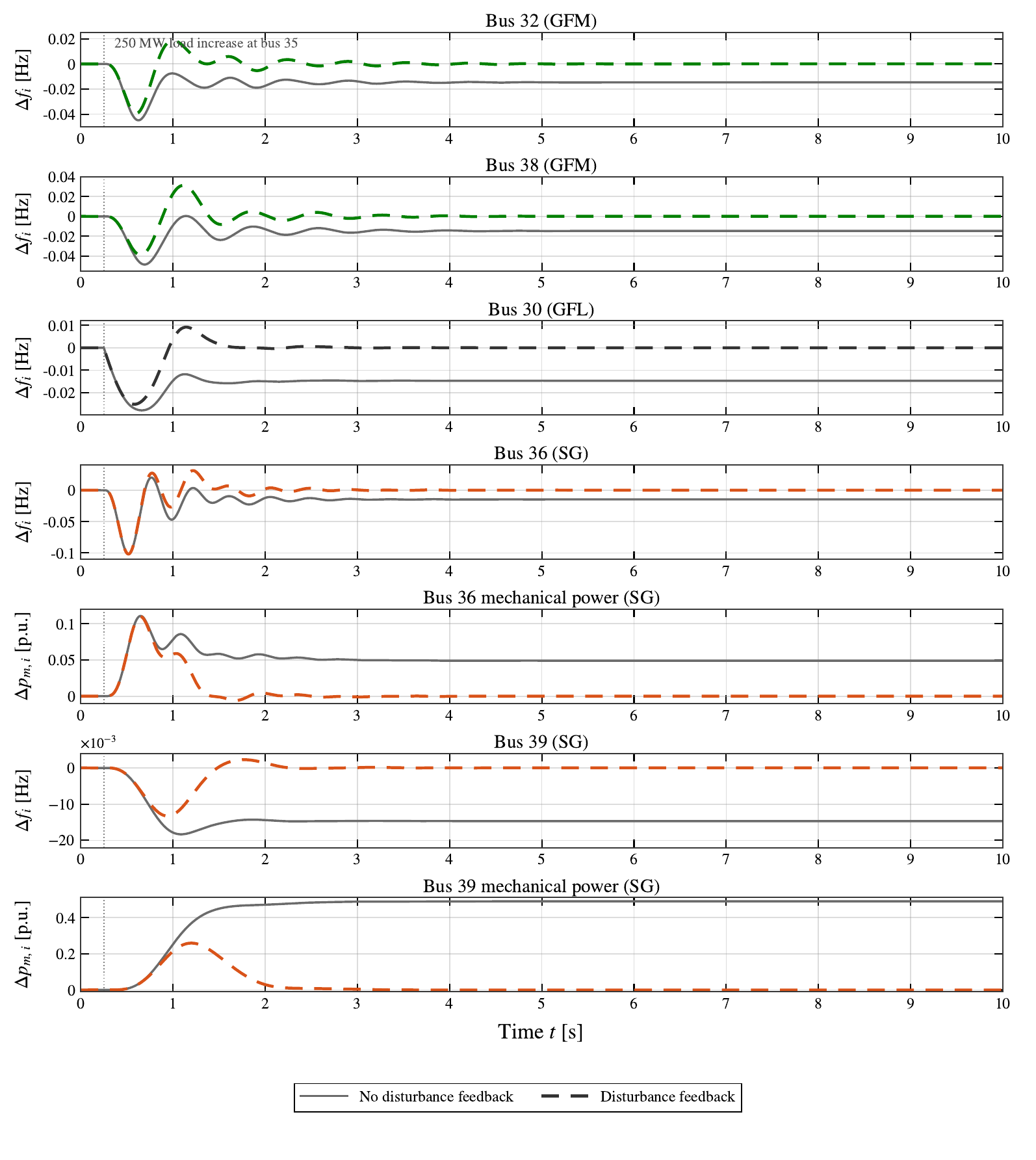}
    }
    \vspace{0.5em}
{
        \includegraphics[width=\linewidth]
        {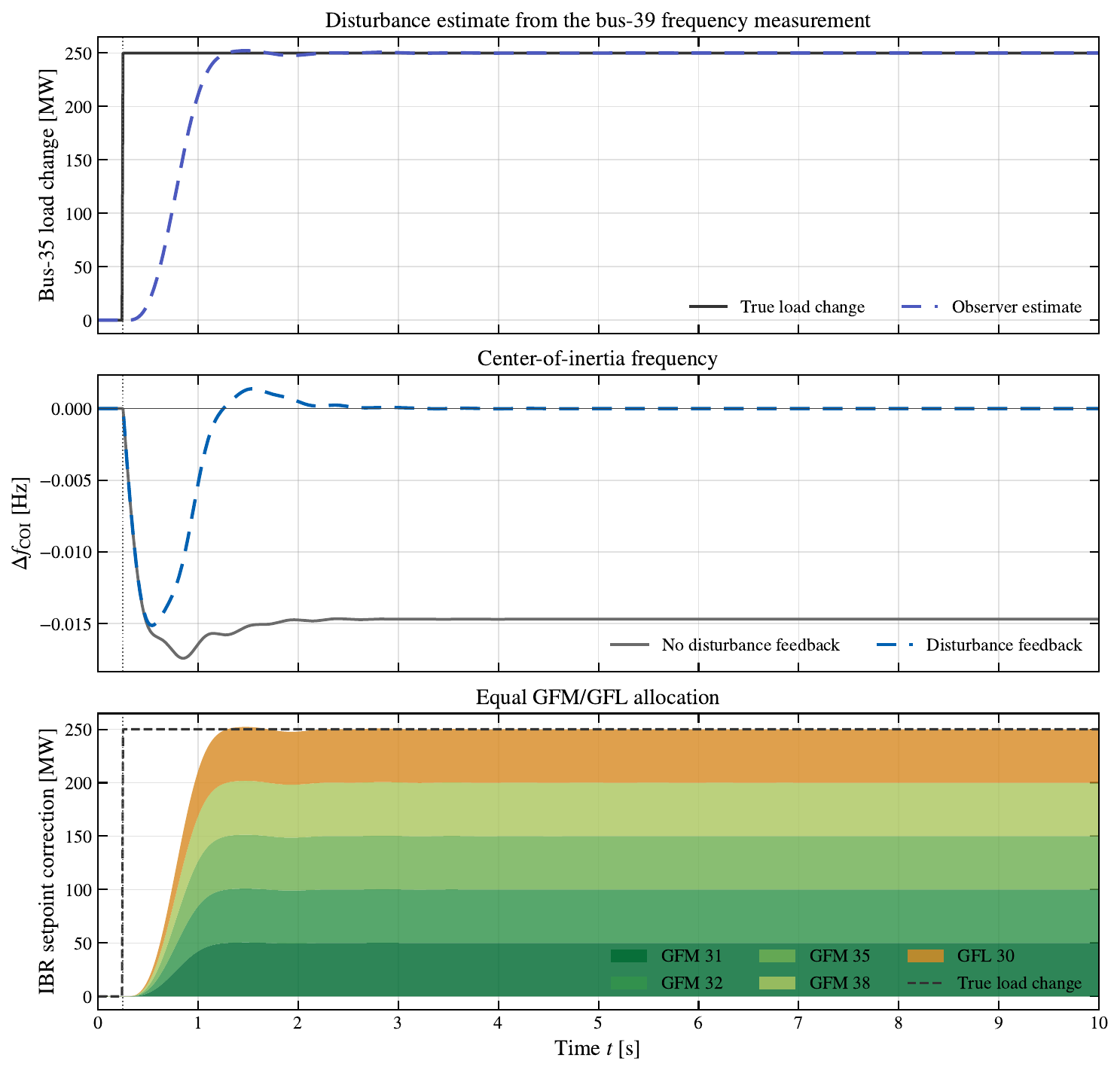}
    }
    \caption{Closed-loop response of disturbance feedback fast-frequency control design on the modified IEEE 39-bus system
    representing ISO-NE.}
    \label{Fig:Disturbance}
\end{figure}

The closed-loop fast frequency control dynamics are now described by the  dynamics with primary control \eqref{eq:state_space_form_measurement} with the controller \eqref{eq:state_space_estimator}, \eqref{Eq:Allocator}. Nominal stability and regulation performance can be assessed by replacing the estimator states with the error coordinates
\[
\Delta\tilde{x} = \Delta x - \Delta \hat{x}, \qquad \Delta \tilde{p}^{\rm load} = \Delta p^{\rm load} - \Delta \hat{p}^{\rm load},
\]
after which the closed-loop dynamics \eqref{eq:state_space_form_measurement}, \eqref{eq:state_space_estimator}, \eqref{Eq:Allocator} take the form
\begin{equation}\label{Eq:JohnErrorDynamics}
\begin{aligned}
\left[\begin{array}{c}
\Delta\dot{x}\\
\hline
\Delta\dot{\tilde{x}}\\
\Delta\dot{\tilde{p}}^{\rm load}
\end{array}\right] &= \left[\begin{array}{c|cc}
A & 0 & B\Lambda\\
\hline
0 & A-L_1C_{\rm m} & F\\
0 & -L_2C_{\rm m} & 0
\end{array}\right]\left[\begin{array}{c}
\Delta{x}\\
\hline
\Delta{\tilde{x}}\\
\Delta{\tilde{p}}^{\rm load}\end{array}\right]\\
&\qquad + \left[\begin{array}{c}
(F-B\Lambda)\\
\hline
0\\
0
\end{array}\right]\Delta p^{\rm load}\\
\Delta \Omega_{\rm coi} &= C_{\rm m}\Delta x.
\end{aligned}
\end{equation}
The error dynamics \eqref{Eq:JohnErrorDynamics} are block upper triangular, and separate into the stable eigenvalues of the dynamics with primary control and the stable eigenvalues of the designed estimator. It follows immediately that
\[
\lim_{t\to\infty}\Delta \Omega_{\rm coi}(t) = \Omega_{\rm coi, ss} \coloneqq -C_{\rm m}A^{-1}(F-B\Lambda)\Delta p^{\rm load}.
\]
However, an equilibrium analysis analogous to that in \eqref{eq:primary_frequency_equilibrium}--\eqref{eq:total_primary_response} shows that the DC gains have the simple form
\[
-C_{\rm m}A^{-1}F = \frac{1}{\beta}\mathbf{1}^{\top}, \qquad C_{\rm m}A^{-1}B\Lambda = -\frac{1}{\beta}\mathbf{1}^{\top}\Lambda = -\frac{1}{\beta}\mathbf{1}^{\top}
\]
by column stochasticity of $\Lambda$, and thus $\Delta \Omega_{\rm coi,ss} = 0$, i.e., nominal frequency regulation is achieved. 

The controller design \eqref{eq:state_space_estimator},\eqref{Eq:Allocator} requires no time-scale separation, as evidenced by  the error dynamics \eqref{Eq:JohnErrorDynamics}; one is free to place the estimation error eigenvalues on the same time-scale as the primary control dynamics. The effect will be a fast-acting secondary-like response from GFM/GFL units in the seconds immediately after a disturbance. On a longer time-scale of minutes, AGC will still act through synchronous generators to correct any regulation errors and transfer the burden of generation back to large capacity units. This appealing separation of time-scales and responsibilities has of course come at the cost of using more model information in the design of the estimator \eqref{eq:state_space_estimator}.

Figure \ref{Fig:Disturbance} shows the result of implementing this feedback design on the modified IEEE 39 bus system described previously, with the same configuration and disturbance. The measurement used to drive the estimator is the frequency measurement at bus $35$, the observer was designed using LQE, and the allocation matrix $\Lambda$ splits the disturbance estimate equally between all five GFM/GFL units. The controller quickly rejects the sudden disturbance, ramping up the GFM/GFL units and bringing the frequency back to its nominal value.

\subsection{Example \#4: VPP coordination via feedback optimization}

This section provides an illustrative example of how to combine a bi-level formulation for a dispatch problem with feedback
optimization to enable coordination with VPPs. Suppose that a balancing authority for area $k$ dispatches synchronous
generators and two VPPs; each VPP  receives its aggregate active power request, and it allocates it
among its resources. Let
$\bar p^{\rm sg}
=\operatorname{col}(\bar p_j^{\rm sg})_{j\in\mathcal G^{\rm sg}}$
collect the references for the  mechanical powers of the generators, and let
$\bar p_i^{\rm vpp}$ denote the aggregate net-injection request sent
to VPP $i$. Each VPP has $n_i$ resources, and has to compute the power references 
$u_i\in\mathcal U_i\subset\mathbb R^{n_i}$ for its resources, where $\mathcal U_i$ is
nonempty, compact, and convex.

As discussed previously in the article, the physical interconnection can be described by the approximate model
\begin{equation}
 \dot x=Ax+B^{\rm sg}\bar p^{\rm sg}+B_1u_1+B_2u_2+w,
 \label{eq:two_vpp_plant}
\end{equation}
with primary controllers included in the dynamics via the matrix $A$.
For this minimal example, assume that each VPP is connected at a single transmission
connection point. To track the aggregate active power references,  define the following VPP-level
allocation problem for computing the powers of individual resources: 
\begin{equation}
\begin{aligned}
 V_i(\bar p_i^{\rm vpp})
 &=\min_{v_i\in\mathcal U_i} f_i(v_i)\\
 &\quad\text{subject to }
 \mathbf1^\top v_i=\bar p_i^{\rm vpp},
\end{aligned}
\label{eq:two_vpp_lower}
\end{equation}
for each VPP $i = 1, 2$, where $f_i$ is a differentiable convex cost. Note the constraint $\mathcal P_i^{\rm vpp}
 =\{\mathbf1^\top v_i:v_i\in\mathcal U_i\}$, which is imposed to meet the aggregate reference $\bar p_i^{\rm vpp}$.  Define
\[
 \varrho=\begin{bmatrix}\bar p^{\rm sg}\\\bar p^{\rm vpp}\end{bmatrix},
 \qquad
 \bar p^{\rm vpp}
 =\begin{bmatrix}\bar p_1^{\rm vpp}\\\bar p_2^{\rm vpp}\end{bmatrix},
\]
and $\mathcal W
 =\mathcal U^{\rm sg}\times
  \mathcal P_1^{\rm vpp}\times\mathcal P_2^{\rm vpp}$,
where $\mathcal U^{\rm sg}$ is a nonempty closed convex set of
admissible generator references.

The balancing authority uses measurements of the  ACE
\begin{equation}
 {\rm ACE}_k(x)
 =p_k^{\rm NI}(x)-p_k^{\rm SNI}
  +b_k\widetilde\Omega_k(x),
 \label{eq:two_vpp_measured_ace}
\end{equation}
with net exports taken as positive, as in
\eqref{eq:ace_definition}. To formulate the dispatch problem at the level of the balancing area, suppose momentarily that $w$ is held constant. Let $h_k(\varrho)$ and $g(\varrho)$
denote the steady-state ACE and monitored transmission flows when
the VPPs deliver their requested aggregate injections.
Under the linear reduced model and the assumptions
above, these maps are affine:
\begin{subequations}
\label{eq:two_vpp_steady_outputs}
\begin{align}
 h_k(\varrho)&=G_{{\rm ace},k}\varrho+h_{k,0},\\
 g(\varrho)&=H\varrho+\ell_d,
 \qquad H=\begin{bmatrix}H^{\rm sg}&H^{\rm vpp}\end{bmatrix}, 
\end{align}
\end{subequations}
where  $G_{{\rm ace},k}$ is the row vector of steady-state ACE
sensitivities with respect to $\varrho$, as explained in the section ``Area control errors and
AGC''.  Then, an example of problem that the balancing authority can solve is the following: 
\begin{subequations}
\label{eq:two_vpp_upper}
\begin{align}
 \min_{\varrho\in\mathcal W}\quad&
 C^{\rm sg}(\bar p^{\rm sg})
 +V_1(\bar p_1^{\rm vpp})+V_2(\bar p_2^{\rm vpp}),\\
 \text{subject to}\quad&
 h_k(\varrho)=0,\\
 &-\bar\ell\leq g(\varrho)\leq\bar\ell,
\end{align}
\end{subequations}
where $C^{\rm sg}$ is differentiable and convex and
$\bar\ell\in\mathbb R_{>0}^{n_\ell}$ contains the transmission-line
limits. This formulation selects generator references and VPP
requests that minimize total cost subject to zero steady-state ACE
and the monitored line limits. The relation to an algebraic area power-balance constraint follows from the steady-state area balance. 

To construct the local feedback controllers, let $\nu_i$ be the
multiplier associated with
$\mathbf1^\top v_i-\bar p_i^{\rm vpp}=0$ in
\eqref{eq:two_vpp_lower}. Under a constraint qualification ensuring
strong duality and existence of a multiplier,
$-\nu_i^\star\in\partial V_i(\bar p_i^{\rm vpp})$.
Thus, each VPP can communicate its multiplier as marginal-cost
information, together with its aggregate capability
$\mathcal P_i^{\rm vpp}$. Define
\[
 \mathsf R_{\mathcal K}(v,h)
 =\mathrm{proj}_{\mathcal K}(v-h)-v,
\]
where $\mathrm{proj}_{\mathcal K}$ denotes Euclidean projection.
A local feedback controller is
\begin{subequations}
\label{eq:two_vpp_local_controller}
\begin{align}
 \dot u_i&=\varepsilon\mathsf R_{\mathcal U_i}
 \bigl(u_i,\gamma_i[\nabla f_i(u_i)+\nu_i\mathbf1]\bigr),\\
 \dot\nu_i&=\varepsilon\gamma_{\nu,i}
 \bigl(p_i^{\rm vpp}(x)-\bar p_i^{\rm vpp}\bigr),
\end{align}
\end{subequations}
where $p_i^{\rm vpp}(x)$ is the measured aggregate injection.
The local multiplier integrates the error between delivered and
requested power. During transients, $-\nu_i$ is a coordinating
signal; its interpretation as an exact value-function subgradient
requires local optimality.

\begin{figure}[t]
    \centering
    \hspace{-.5cm}
        \resizebox{0.48\textwidth}{!}{%
\includegraphics[width=249.4119pt,height=211.67844pt]{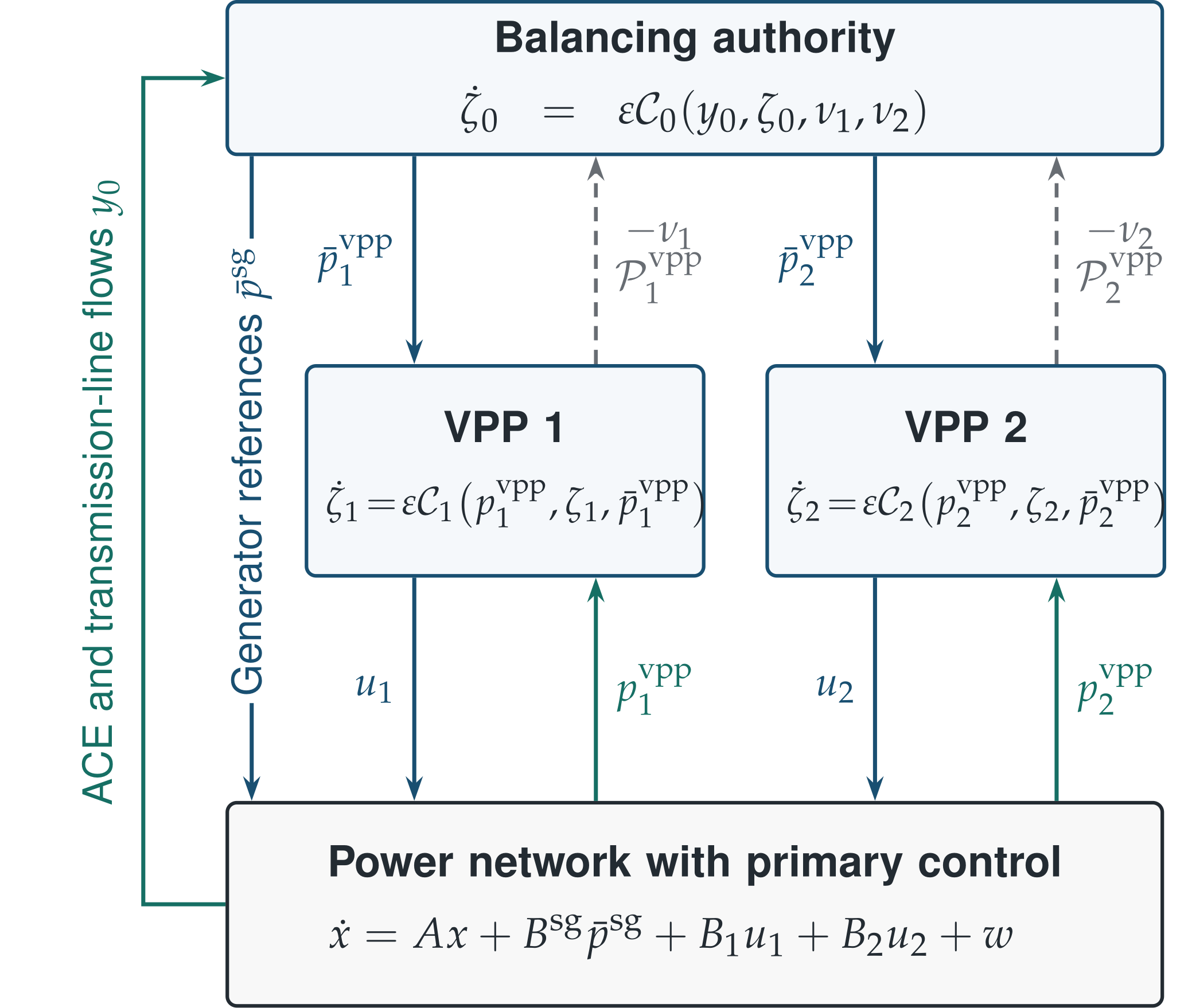}%
    }
    \caption{Example of architecture using feedback optimization methods to solve a  bi-level problem for dispatching generators and two VPPs. }    \label{fig:F_bilevel_feedback_optimization}
\end{figure}

For the balancing authority, define
\[
 \widehat c(\varrho,\nu_1,\nu_2)
 =\begin{bmatrix}
   \nabla C^{\rm sg}(\bar p^{\rm sg})\\-\nu_1\\-\nu_2
  \end{bmatrix}.
\]
Let $\lambda$ be the multiplier of the zero-ACE constraint and
$\mu^\pm\in\mathbb R_{\geq0}^{n_\ell}$ the line-limit multipliers.
The balancing authority's controller is
\begin{subequations}
\label{eq:two_vpp_authority_controller}
\begin{align}
 \dot \varrho
 &=\varepsilon\mathsf R_{\mathcal W}
 \left(\varrho,\gamma_0
 \left[\widehat c
 +G_{{\rm ace},k}^{\top}\lambda
 +H^\top(\mu^+-\mu^-)\right]\right),\\
 \dot\lambda
 &=\varepsilon\gamma_\lambda\,{\rm ACE}_k,\\
 \dot\mu^\pm
 &=\varepsilon\left[
 \mathsf P_{\mathbb R_+^{n_\ell}}
 \bigl(\mu^\pm+\gamma_\mu[
       \pm\ell^{\rm m}(x)-\bar\ell]\bigr)
 -\mu^\pm\right],
\end{align}
\end{subequations}
where $\ell^{\rm m}(x)$ contains measured transmission flows,
$\varepsilon>0$, and all controller gains are positive.
The multiplier update uses the measured ${\rm ACE}_k$, while the reference
update uses the sensitivity of its steady-state map.  Define $\zeta_0=(\varrho,\lambda,\mu^+,\mu^-)$,
 $\zeta_i=(u_i,\nu_i)$, $i\in\{1,2\}$, and let $y_0(x)=\operatorname{col}
 \bigl({\rm ACE}_k(x),\ell^{\rm m}(x)\bigr)$,
 $\zeta=\operatorname{col}(\zeta_0,\zeta_1,\zeta_2)$.
The feedback optimization methods in \eqref{eq:feedback_architecture}
can be compactly written as
\begin{equation}
 \mathcal C(x,\zeta)=
 \begin{bmatrix}
  \mathcal C_0(y_0(x),\zeta_0,\nu_1,\nu_2)\\
  \mathcal C_1(p_1^{\rm vpp}(x),\zeta_1,\bar p_1^{\rm vpp})\\
  \mathcal C_2(p_2^{\rm vpp}(x),\zeta_2,\bar p_2^{\rm vpp})
 \end{bmatrix}.
 \label{eq:two_vpp_feedback}
\end{equation}
The balancing authority updates generator references, aggregate VPP requests,
and the multipliers associated with ACE and line-limit constraints. Each VPP updates its
internal references and its request-tracking multiplier, as
illustrated in Fig.~\ref{fig:F_bilevel_feedback_optimization}.

For fixed disturbances and dispatch schedules, a closed-loop equilibrium
satisfies the local allocation optimality conditions and, under
the assumption stated above, and based on the optimality conditions of
\eqref{eq:two_vpp_upper}. An interesting research direction is to merge Example \#3 and this Example \#4 to develop coordination mechanisms across balancing areas.

\section{The grand research challenge in power systems  control architectures}

Power systems are evolving rapidly: renewable generation, inverter-based resources, and large data center loads all challenge the existing control architecture. The readers of this magazine---the systems and control community---can certainly envision many potential solutions. Before jumping on them, however, we want to stress a few aspects of power systems research that are less prominent and quite distinct to control theory research.

\begin{pullquote}
    Power systems research faces a tension between envisioning new solutions and the requirement that every
step remains compatible with a legacy system.
\end{pullquote}

Power systems research faces a tension between envisioning new solutions and the requirement that every step remains compatible with a legacy system that can never be switched off and will never be rebuilt. As a result, the deployed system changes only incrementally at its interfaces, even when the underlying technology, such as inverter-based generation, changes radically. Moreover, power systems research demands a holistic view: a change to one element of the control architecture ripples through generation, markets, grid codes, protection, and operation. Innovations must therefore be co-designed across these layers rather than optimized in isolation. To a control theorist, power systems research may thus appear to be inching forward painfully slowly.

Of course, one needs to envision a future Utopian grid in order to set our direction. The current control architecture is largely a product of history, the result of organic growth, and continuous integration of new technologies. If one were to rebuild the grid today, would one again opt for a vertically and horizontally layered architecture with time-scale separation and centralized error signals broadcast to many controlled subsystems? Or would one tailor it to today's technologies, such as battery storage and renewable sources interfaced through inverters and coordinated via high-bandwidth communication? Would it even be AC? Researchers are exploring these directions, and isolated microgrids offer testbeds where new architectures can mature before considering migration into the bulk grid.

The central grand challenge is not imagining a future Utopian grid, but charting a feasible path towards it. The authors therefore recommend embracing the backward-compatibility constraint, adopting a holistic view, and treating the transition path as a research problem in its own right; one in which stability, performance, and robustness must be assured at every intermediate stage. 

To end on a constructive note, and in addition to the research opportunities explained earlier in the article, the authors propose below a few meta-level research directions for the community to consider:
\begin{itemize}
    \item Time-scale separation arose organically as a patch whenever new layers were added to the system or simply due to slow and fast actuation. Nowadays one should rethink whether it is still needed.
    \item (Dynamic) virtual power plant solutions emerged to pool devices with different capabilities (in power, energy, and response times) so that they can contribute to the existing ancillary services. Conversely, one may also rethink the  ancillary service schemes to fit modern sources, e.g., energy neutrality for batteries.
    \item The microgrid community extensively dabbled with entirely new power system balancing architectures, such as distributed secondary control, replacing secondary control by fast redispatch, fully decentralized architectures, and feedback optimization solutions. Which of these are suited for the bulk power system?
    \item  Transmission and distribution are physically coupled, but their control, scheduling, and market processes are often organized separately. What information exchanges and market arrangements would enable coordinated operation, and how should responsibility for network constraints and service delivery be allocated across operators?
    \item  Individually stable controllers need not yield a stable interconnected power system. What locally verifiable requirements on resources and aggregators would allow system-level stability and performance to be certified without disclosing every internal model and control law?
    \item  New architectures must coexist with existing controllers and operational practices. What are systematic methodologies for exploring the space of backwards-compatible designs? How can they be introduced incrementally, and what must be re-validated when devices, controller settings, software, or operating conditions change? 
\end{itemize}

\section{ACKNOWLEDGMENTS}

The authors would like to thank Damola Ajeyemi (Boston University) for his assistance in the EMT numerical experiments for the power system models.

Use of AI disclosure: authors utilized AI to produce some of the graphics and part of the code for simulations. Text and mathematical developments are from the authors.

\section{Author Information}

\begin{IEEEbiography}{{E}miliano Dall'Anese}{\,}(edallane@bu.edu) received the Ph.D. degree from the University of Padova in 2011. He is an Associate Professor in the Department of Electrical and Computer Engineering at Boston University, where he is also with the Division of Systems Engineering. His focus is on optimization, control, and systems theory, with primary applications in power systems and autonomous systems. He is a Senior Member of IEEE. 
\end{IEEEbiography}

\begin{IEEEbiography}{{J}ohn~{W}.~{S}impson-{P}orco}{\,}(jw.simpson@utoronto.ca) received the  Ph.D. degree in mechanical engineering from the University of California at Santa Barbara in 2015. He is an Associate Professor and Edward S. Rogers Sr. Chair in Engineering in the Edward S. Rogers Sr. Department of Electrical and Computer Engineering at the University of Toronto. His research focuses on control, systems theory, and applications of control in power systems. He is a Senior Member of the IEEE.
\end{IEEEbiography}

\begin{IEEEbiography}{{F}lorian {D}\"orfler}{\,}(dorfler@ethz.ch)
received the Ph.D. degree in mechanical engineering from the
University of California at Santa Barbara in 2013. He is a
Professor at the Automatic Control Laboratory at ETH Z\"urich.
His research focuses on control, systems theory, optimization,
and learning, with particular interests in networked systems,
data-driven control, and power systems. He is a Fellow of IEEE.
\end{IEEEbiography}

\begin{IEEEbiography}{{S}airaj  {D}hople}{\,}(sdhople@umn.edu)
received the Ph.D. degree in electrical engineering from the
University of Illinois at Urbana-Champaign in 2012. He is the
Oscar A. Schott Professor in the Department of Electrical and
Computer Engineering at the University of Minnesota. His research
focuses on modeling, analysis, and control of power electronics
and power systems. He is a Senior Member of IEEE.
\end{IEEEbiography}

\begin{IEEEbiography}{{S}averio {B}olognani}{\,}(bsaverio@ethz.ch)
received the Ph.D. degree in information engineering from the
University of Padova in 2011. He is a Senior Scientist at the
Automatic Control Laboratory at ETH Z\"urich. His research
focuses on networked control, optimization, and game theory,
with applications to power systems and autonomous mobility.
He is a Member of IEEE.
\end{IEEEbiography}

\bibliographystyle{IEEEtran}
\bibliography{biblio_acgofo, biblio_modeling,biblio_books}

\end{document}